\documentclass[aps,twocolumn,pra,groupedaddress,amsmath,amssymb,superscriptaddress]{revtex4-2}
\usepackage{dcolumn}
\usepackage{bm}
\usepackage{amsmath}
\usepackage{txfonts}
\usepackage[T1]{fontenc}
\usepackage{xspace}
\usepackage{ulem}
\usepackage{comment}
\usepackage{braket}
\ifx\pdfoutput\undefined
\usepackage[dvipdfmx]{graphicx}
\usepackage[dvipdfmx]{hyperref}
\usepackage[dvipdfmx]{color}
\else
\usepackage{graphicx}
\usepackage{hyperref}
\usepackage{color}
\fi

\begin{document}
\let\emph\textit

\newcommand{\cred}[1]{\textcolor{black}{#1}}
\newcommand{\cblue}[1]{\textcolor{blue}{#1}}

\title{Measurement-induced spatially nonuniform fluctuations of the local particle number and their crossover in a quasiperiodic free-fermion chain}
\author{Toranosuke Matsubara}
\email{matsubara.t.6125@m.isct.ac.jp}
\affiliation{
  Department of Physics, Institute of Science Tokyo,
  Meguro, Tokyo 152-8551, Japan
}

\author{Kazuki Yamamoto}
\email{kazuki-yamamoto@omu.ac.jp}
\affiliation{
  Research Institute for Innovation and Co-Creation, Osaka Metropolitan University,
  Sakai, Osaka 599-8531, Japan
}
\affiliation{
  Department of Physics, Osaka Metropolitan University,
  Sumiyoshi, Osaka 558-8585, Japan
}

\affiliation{
  Nambu Yoichiro Institute of Theoretical and Experimental Physics (NITEP),
  Osaka Metropolitan University, Sumiyoshi, Osaka 558-8585, Japan
}

\date{\today}
\begin{abstract}

We study continuously monitored dynamics of a quasiperiodic free-fermion chain defined on a Fibonacci lattice.
We focus on fluctuations of the local particle number, which do not exhibit quasiperiodic spatial modulation in the unitary limit.
Remarkably, we demonstrate that
they exhibit a nonuniform spatial pattern originating from the quasiperiodic long-range order under continuous measurement.
Furthermore, employing both physical- and perpendicular-space analyses, we elucidate that a measurement-induced crossover emerges in fluctuations due to the interplay between
the incommensurate modulation and the continuous measurement.
While weak measurement yields a distribution reflecting the long-range spatial structure of the quasiperiodic
system, an increase in measurement strength alters the distribution into one dominated by the local
environment of each site.
\cred{A structure-factor analysis further reveals that the weak-measurement regime is characterized by two dominant Bragg peaks corresponding to the golden-mean modulation, whereas the strong-measurement regime exhibits the characteristic Bragg-peak hierarchy of the Fibonacci pattern.}
We also elucidate that the measurement-induced crossover emerges in other physical quantities such as connected correlation functions.
These findings offer insights into nonequilibrium quasiperiodic phenomena emerging in continuously monitored dynamics.

\end{abstract}
\maketitle

\section{Introduction}
Monitored quantum dynamics has attracted considerable attention as a novel platform for investigating open quantum systems~\cite{Harrington22, Fisher23rev, Potter_2022}, e.g.,
when a system is subjected to continuous or discrete measurements,
a measurement-induced entanglement phase transition can emerge~\cite{Li_2018, Chan_2019, Skinner_2019}.
While such transitions have been analyzed predominantly in quantum circuits~\cite{Szyniszewski_2019, Bao_2020, Choi_2020, Gullans_2020a, Gullans_2020b, Jian_2020, Zabalo_2020, Iaconis_2020, Turkeshi_2020, Zhang_2020, Szyniszewski_2020, Nahum_2021, Ippoliti_2021a, Ippoliti_2021b, Lavasani_2021a, Lavasani_2021b, Sang_2021, Fisher23rev, Block_2022, Sharma_2022, Agrawal_2022, Barratt_2022, Kelly_2023, Ha_2024},
investigations have extended the research arena to quantum many-body systems motivated by experimental advances in ultracold atoms \cite{Muller_2012}, including
interacting~\cite{Tang_2020, Goto_2020, Fuji_2020, Doggen_2022, Doggen_2023, Lunt_2020, De_2024, Patrick_2024}
and free-fermion systems~\cite{Cao_2019, Alberton_2021, Chen_2020, Tang_2021, Coppola_2022, Ladewig_2022, Carollo_2022, Yang_2022, Buchhold_2021, VanRegemortel_2021, Youenn_2023, Loio_2023, Turkeshi_2022, Kells_2023, Yu_2025, Fava_2023,Piccitto_2022,Piccitto_2023,Russomanno_2023,Poboiko_2023, Fava_2024,Starchl_2025, Chahine_2024, Poboiko_2024, Jin_2024, Minato_2022, Muller_2022}.
Experimentally, measurement-induced phase transitions have been observed in superconducting qubits~\cite{Koh_2023, Google_2023,Kamakari_2025} and trapped ions~\cite{Noel_2022, Agrawal_2024}, but there are still limitations to fully controllable realizations.
In light of these challenges, recent studies have focused on the fact that nontrivial measurement-induced many-body phenomena arise not only in entanglement entropy, but also in other physical observables, such as spatiotemporal correlation functions \cite{Cech25}, waiting-time distributions~\cite{Yamamoto_2026}, current fluctuations~\cite{Landi_2024, Yamamoto_2025}, and bipartite fluctuations~\cite{Poboiko_2023, Moghaddam_2023, Qiu_2025, Oshima_2023}.

On another front, monitored quantum dynamics in disordered systems has been actively studied, revealing that structural properties of the system play crucial roles~\cite{Yamamoto_2023, Szyniszewski_2023, Popperl_2023, Szyniszewski_2024, Liao_2026}.
These studies naturally raise the question of how monitored dynamics is influenced by aperiodic yet nonrandom structures with long-range order. In this context, quasiperiodic systems have attracted considerable interest as prototypical examples of deterministic aperiodic structures, characterized by distinctive lattice modulations that differ from those of random disorder.
Importantly, quasiperiodicity can be realized in cold-atom experiments by superimposing optical lattices with incommensurate periods or angles in one-dimensional~\cite{Fallani_2007, Roati_2008, DErrico_2014, Schreiber_2015, Bordia_2017, An_2021, Rajagopal_2019,Shimasaki_2024}, pentagonal~\cite{Guidoni_1997, Guidoni_1999, Corcovilos_2019}, and octagonal systems~\cite{Viebahn_2019, Sbroscia_2020, Yu_2024}.
So far, monitored quantum dynamics with quasiperiodic structures have been analyzed in both quantum circuits~\cite{Li_2019, Lu_2021, Shkolnik_2023, Zabalo_2023} and quantum many-body systems~\cite{Matsubara_AAH, Zhao_2026, Singha_2026, Yin_2026}.
Nevertheless, since many of the aforementioned phenomena, such as anomalous entanglement scalings and localization properties~\cite{Matsubara_AAH, Zhao_2026, Singha_2026, Yin_2026}, are also known to occur in homogeneous disordered systems,  measurement-induced phenomena that are unique or intrinsic to quasiperiodicity remain largely unexplored.
Therefore, it is significant to identify which physical observables reflect the properties of quasiperiodic structures unique to monitored quantum dynamics and to elucidate nontrivial phenomena that emerge from the interplay between measurements and quasiperiodic structures.

In this paper, we analyze free fermions in a quasiperiodic chain under continuous measurement and clarify that anomalous nonuniformity originating from the quasiperiodic structure, which is absent in the unitary limit, appears in the spatial distribution of fluctuations of the local particle number.
Moreover, employing both physical- and perpendicular-space analyses in quasiperiodic systems, we demonstrate that the property of the spatial distribution of fluctuations exhibits a crossover induced by measurement: the spatial profile captures the aperiodic long-range ordered lattice structure for weak measurement, while it is dominated by the local lattice environment for strong measurement.
\cred{We further characterize this measurement-induced crossover through a structure-factor analysis, finding that the weak-measurement regime exhibits two dominant Bragg peaks associated with the golden-mean modulation, whereas the strong-measurement regime shows a characteristic hierarchy of Bragg peaks reflecting the Fibonacci structure.}
In addition, we analyze connected correlation functions and reveal that the measurement-induced crossover occurs beyond fluctuations of the local particle number. 
Our results are relevant to ultracold atoms loaded into optical lattices and can be tested by introducing continuous measurement with the use of off-resonant probe light and quantum-gas microscopes.

The rest of this paper is organized as follows. In Sec.~\ref{sec: model}, we define a tight-binding model on the Fibonacci chain and introduce the stochastic Schr\"odinger equation to describe the continuously monitored dynamics. Section~\ref{sec: results} is devoted to numerical results, including the dynamics and steady-state properties of fluctuations of the local particle number and correlation functions \cred{studied by physical-space, perpendicular-space, and structure-factor analyses}. Finally, we summarize our findings in Sec.~\ref{sec: summary}.

\section{Model}
\label{sec: model}
\subsection{Fibonacci chain}
\begin{figure}[tb]
  \begin{center}
    \includegraphics[width=\linewidth]{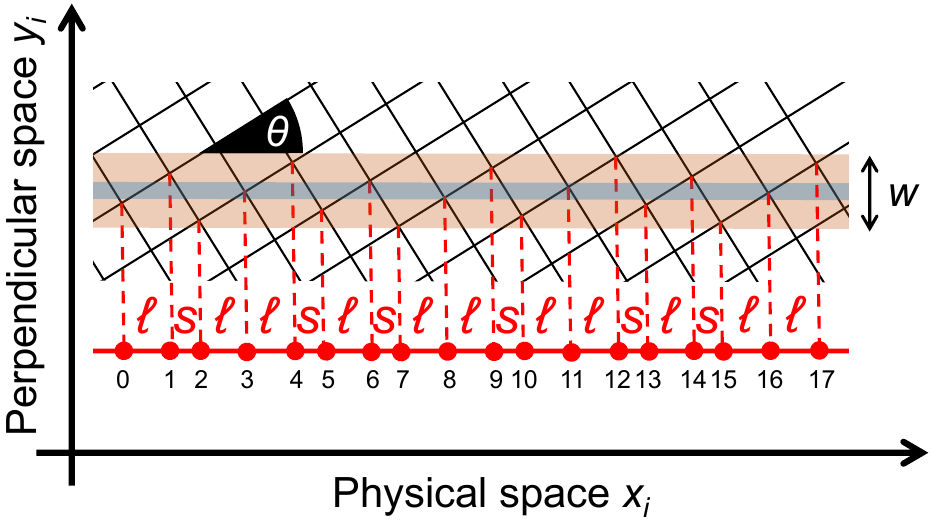}
    \caption{The Fibonacci chain obtained by cutting and projecting a square lattice.
    The number below each site indicates the site index $i$.
    The coordinates $x_i$ and $y_i$ of the $i$-th site are for physical and perpendicular spaces, respectively. The width of a blue strip (two red strips) is determined by the fraction of number of sites located between two $\ell$-spacings ($\ell$- and $s$-spacings) relative to the total number of sites, where the fraction is denoted as $r_{\ell \ell}$ ($r_{\ell s}$). These satisfy $r_{\ell s} / r_{\ell \ell} = 2 \tau$ with $r_{\ell \ell} + r_{\ell s} = 1$, which leads to the widths of the red and blue strips as $r_{\ell s} w \approx 1.05$ and $r_{\ell \ell} w \approx 0.32$, respectively. \cred{Here, $\tau$ denotes the golden mean.}}
    \label{fig: Fibonacci}
  \end{center}
\end{figure}

In this study, we consider continuously monitored dynamics of
free fermions governed by a tight-binding model on a Fibonacci chain~\cite{Jagannathan_Rev}. The Fibonacci chain is composed of two types of lengths, $\ell$ and $s$ ($\ell>s$), which we refer to as $\ell$- and $s$-spacings. 
The two different lengths are arranged in an aperiodic pattern.
One of the methods to construct this chain is the cut-and-project method.
As shown in Fig.~\ref{fig: Fibonacci}, a square lattice with a lattice constant of length 1 is cut by a strip of width $w$ tilted at an angle $\theta$, where
\begin{eqnarray}
\theta &=& \arctan \left( \frac{1}{\tau} \right),\label{Eq: theta}\\
w &=& \frac{1+\tau}{\sqrt{2+\tau}},\\
\frac{\ell}{s} &=& \tau,\\
\tau &=& \frac{\sqrt{5}+1}{2}\label{Eq: GM}.
\end{eqnarray}
The direction parallel to the strip is called a physical space, while the transverse direction is called a perpendicular space.
Since the slope of the cutting angle $\theta$ is the irrational number as given in Eqs.~\eqref{Eq: theta} and \eqref{Eq: GM},
spatial periodicity is lacking in physical space.
Note that each site is located either between $\ell$- and $s$-spacings, or between two $\ell$-spacings; a site is never located between two $s$-spacings.
The lattice generated by this method is quasiperiodic, namely, the lattice structure exhibits long-range order despite lacking spatial periodicity.

We introduce several properties of the perpendicular space.
When an $i$-th site has a coordinate $x_i$ in physical space,
the corresponding point on the square lattice can be projected onto the perpendicular space with a coordinate $y_i$.
Thus, there is a one-to-one correspondence between a coordinate in physical space and that in perpendicular space.
A key feature is that sites with similar local lattice environments in physical space are mapped to nearby points in perpendicular space.
As a simple example,
in Fig.~\ref{fig: Fibonacci}, sites located between two $\ell$-spacings are mapped to the central region of the strip with width $w$ shown by the blue region,
while those located between $\ell$- and $s$-spacings are mapped to the two red regions of the strip.
For convenience, we define a map $\Phi (i, j)$ for any two sites $i$ and $j$ ($0 \leq i < j$),
which yields a segment of the Fibonacci word between the two sites.
For example, $\Phi (0, 1) = (\ell)$, $\Phi (0, 5) = (\ell, s, \ell, \ell, s)$, and $\Phi (2, 5) = (\ell, \ell, s)$ (see Fig.~\ref{fig: Fibonacci}).
An $i$-th site mapped to the upper (lower) red regions relative to the blue region in Fig.~\ref{fig: Fibonacci} satisfies $\Phi (i-1, i+1) = (\ell, s)$ [$\Phi (i-1, i+1) = (s, \ell)$].

\subsection{Setup}
We introduce a tight-binding model defined on the Fibonacci lattice~\cite{Jagannathan_Rev}
\begin{eqnarray}
  H =-\sum_{j=0}^{L-1} J_{j,j+1} \left[  c_j^{\dagger} c_{j+1} + {\rm H.c.}\right].
  \label{eq: Hamiltonian}
\end{eqnarray}
Here, $c_j$ ($c_j^\dagger$) is the annihilation (creation) operator
of a fermion at the $j$-th site, and $L$ is the system size.
The hopping amplitude $J_{j, j+1}$ takes a value $J_\ell$ ($J_s$)
when the length between the $j$-th and $(j+1)$-th sites is $\ell$ ($s$).
When $J_\ell = J_s$, the model reduces to a standard tight-binding model with spatial periodicity. 
The system is quasiperiodic when $J_\ell \neq J_s$.
We impose periodic boundary conditions $c_L = c_0$.

\begin{figure}[tb]
  \begin{center}
    \includegraphics[width=\linewidth]{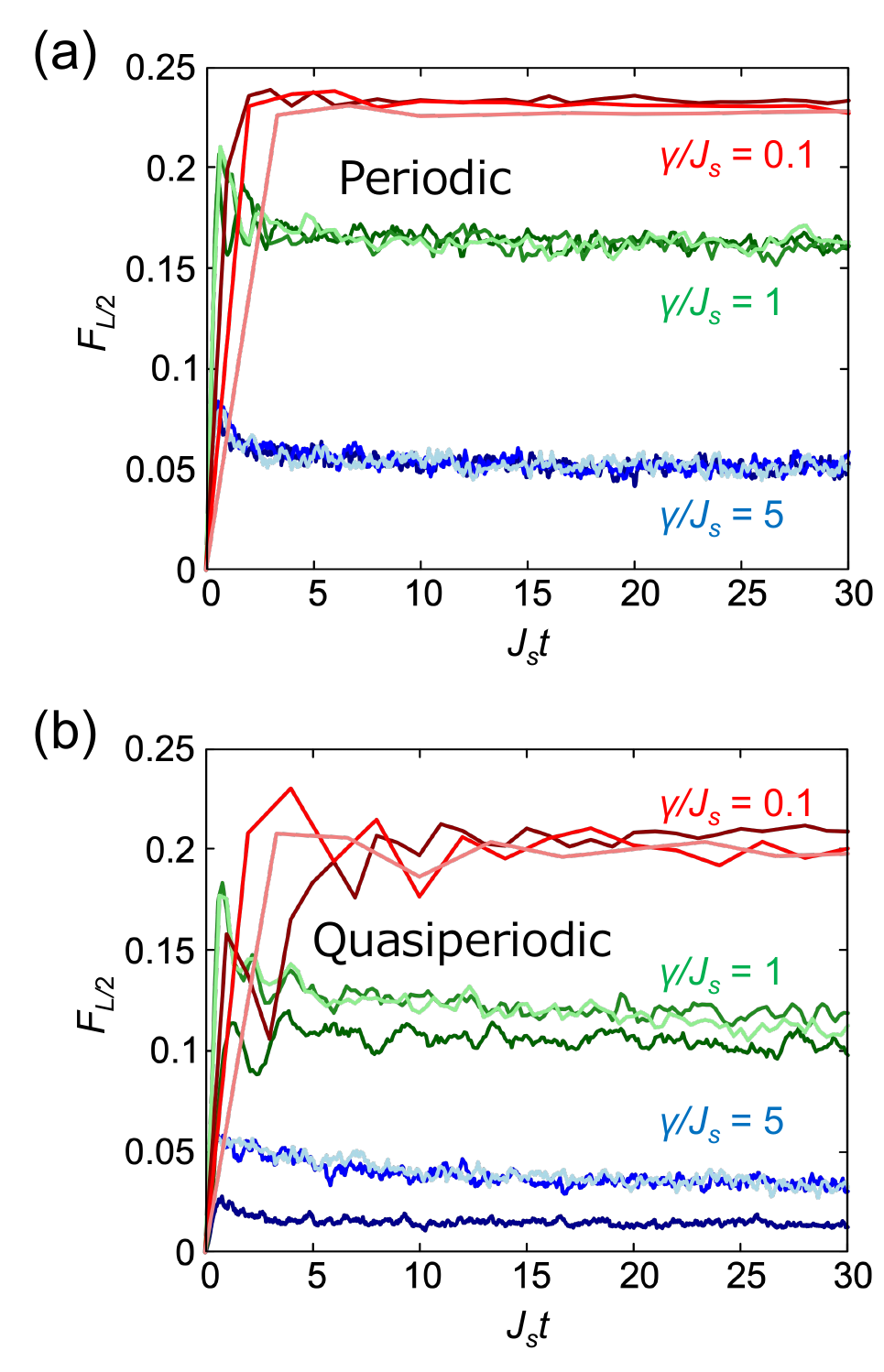}
    \caption{Numerical results for the dynamics of fluctuations of the local particle number $F_{L/2}$ at the $(L/2)$-th site. We set (a) $J_\ell / J_s = 1$ in the periodic system and (b) $J_\ell / J_s = 0.4$ in the quasiperiodic system. Light, medium, and dark colors correspond to $L=30$, $50$, and $100$, respectively. The number of quantum trajectories is set to 500.}
    \label{fig: dynamics}
  \end{center}
\end{figure}

To consider continuously monitored dynamics, we employ the quantum trajectory method \cite{Daley_2014}. When the local particle number is continuously monitored, the time evolution of the quantum state over a time interval $[t, t+dt]$ is governed by the stochastic Schr\"odinger equation
\begin{equation}
  d\left|\Psi\left\{\xi_{j, t}\right\}\right\rangle = \left[-i H \, dt + \sum_{j} \xi_{j, t}\left(\frac{n_j}{\sqrt{\left\langle n_j\right\rangle}}-1\right)\right] \left|\Psi\left\{\xi_{j, t}\right\}\right\rangle,
  \label{eq: QJ}
\end{equation}
where $\langle\cdots\rangle$ represents the expectation value with respect to the state $|\Psi\rangle$. The variable $\xi_{j,t} = 0,1$ is a discrete random number satisfying $\xi_{j, t}^2 = \xi_{j, t}$ and $\overline{\xi_{j, t}} = \gamma \braket{n_j} dt$ \cite{Daley_2014, Dalibard_1992, Wiseman_1993, Fuji_2020, Yamamoto_2023, Yamamoto_2025}.
Here, $\gamma$ is the measurement strength, and $\overline{X}$ denotes an ensemble average of $X$ over stochastic processes.
In free-fermion systems, the state $|\Psi\rangle$ can be represented by a Gaussian state, which enables a simulation of large system sizes~\cite{Alberton_2021, Muller_2022, Minato_2022, Szyniszewski_2023}. This even holds in quasiperiodic systems \cite{Matsubara_AAH}, and we conduct our calculations using Gaussian-state-based algorithms.
In the following calculations, we take the initial state as the half-filled N{\'e}el state $\ket{\Psi}|_{t=0} = \prod_j c_{2j}^\dagger \ket{\rm vac}$,
where $\ket{\rm vac}$ denotes the fermionic vacuum state satisfying
$c_j\ket{\rm vac}=0$ for all sites $j$.

\begin{figure}[tb]
  \begin{center}
    \includegraphics[width=\linewidth]{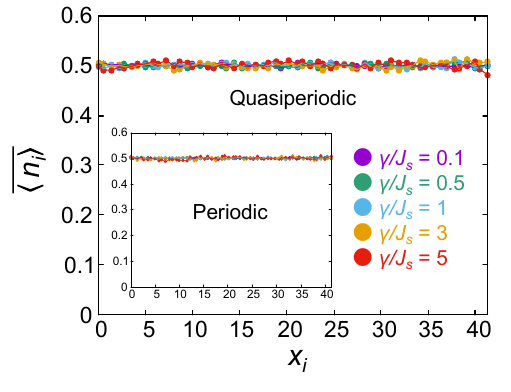}
    \caption{Spatial distribution of the local particle number with respect to $x_i$ in physical space. We set $J_\ell/J_s=0.4$ in the quasiperiodic system (main panel) and $J_\ell/J_s=1$ in the periodic system (inset). The number of trajectories is set to 1000 for $L=60$.}
    \label{fig: ni_phys}
  \end{center}
\end{figure}

\section{Results}
\label{sec: results}
\subsection{Fluctuations of the local particle number}
In this study, we focus on fluctuations of the local particle number
\begin{equation}
F_i = \overline{\braket{n_i^2} - \braket{n_i}^2},
\label{eq_Fi}
\end{equation}
where $n_i = c_i^\dagger c_i$.
\cred{Since our main focus is on the steady-state properties, we first plot the time evolution of the fluctuations of the local particle number to confirm that they have relaxed to their stationary values.
We find that $F_i$ starts from zero at $t=0$ for the N\'eel state, initially increases, and then approaches a stationary value after a sufficiently long-time evolution, irrespective of whether the system is periodic [Fig.~\ref{fig: dynamics}(a)] or quasiperiodic [Fig.~\ref{fig: dynamics}(b)].}
Since $n_i^2 = n_i$ for fermions, a sufficient number of quantum-trajectory realizations yields $\overline{\braket{n_i}} \approx 0.5$ in the steady state, and we can approximate Eq.~\eqref{eq_Fi} as $F_i \approx 0.5 - \overline{\braket{n_i}^2}$. Therefore, $\overline{\braket{n_i}^2}$ plays a primary role in determining the behavior of fluctuations.
Note that in the unitary limit $\gamma/J_s = 0$, the quantum trajectory is unique, and we have numerically checked that $\braket{n_i} = 0.5$ and $F_i = 0.25$ in the thermodynamic limit.
However, for finite measurement strengths $\gamma/J_s\neq0$ shown in Fig.~\ref{fig: dynamics}, $F_i$ significantly deviates from $0.25$, indicating that nontrivial local fluctuations are induced by measurements.
Moreover, in what follows, we will show that the spatial distribution of $F_i$
exhibits an anomalous nonuniformity when the continuous measurement is introduced
in the quasiperiodic system.
Hereafter, we focus on the property in the steady state.

Before going to the main results, we confirm that the local particle number takes a trivial value $\overline{\braket{n_i}} \approx 0.5$ in the steady state under continuous measurement.
In Fig.~\ref{fig: ni_phys}, we see that the local particle number is uniformly distributed around $\overline{\braket{n_i}} \approx 0.5$ in both quasiperiodic (main panel) and periodic systems (inset). This result indicates that a sufficient number of quantum trajectories is sampled in the numerical simulation.

\begin{figure}[tb]
  \begin{center}
    \includegraphics[width=\linewidth]{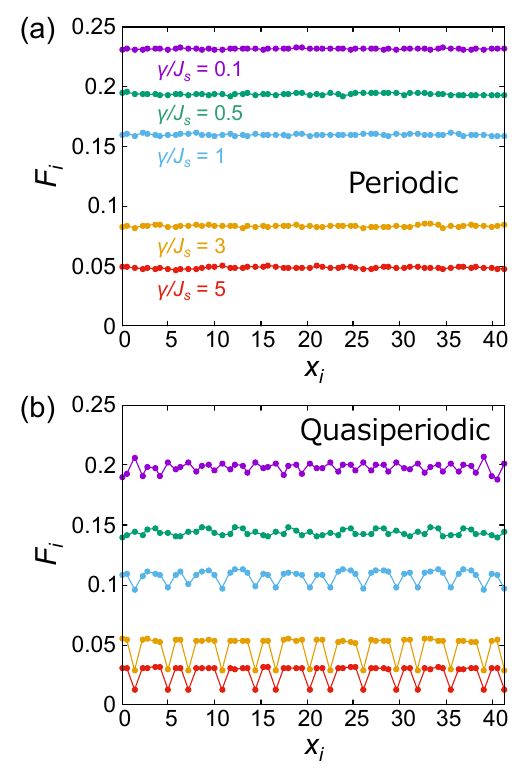}
    \caption{Spatial distribution of fluctuations of the local particle number with respect to $x_i$ in physical space. We take (a) $J_\ell/J_s=1$ in the periodic system and (b) $J_\ell/J_s=0.4$ in the quasiperiodic system. The number of trajectories is set to 1000.
    The system size is set to $L=120$, whose enlarged view is shown.}
    \label{fig: Fi_phys}
  \end{center}
\end{figure}

\subsection{Measurement-induced spatial nonuniformity and crossover}
\label{sec_Fiphysical}
We study the spatial distribution of $F_i$ in a free-fermion chain under continuous monitoring.
To focus on the spatial distribution of $F_i$, we average the numerical results over ten equally spaced time points in the stationary regime.
When the system is periodic as shown in Fig.~\ref{fig: Fi_phys}(a), $F_i$ is spatially uniform regardless of the measurement strength.
Furthermore, as the measurement strength increases, the value of $F_i$ is significantly suppressed.
This is attributed to the localization of the wavefunction under strong measurements~\cite{Szyniszewski_2023, Szyniszewski_2024, Matsubara_AAH, Singha_2026}.

On the other hand, when the system is quasiperiodic as shown in Fig.~\ref{fig: Fi_phys}(b),
we find that the spatial structure clearly becomes nonuniform under continuous monitoring.
In particular, for weak measurement strength $\gamma / J_s = 0.1$, we see that measurement-induced fluctuations exhibit a spatially nonuniform profile reflecting the underlying incommensurate modulation.
Such a fluctuation is distributed within the range $F_i \in [0.187,  0.204]$ for $\gamma / J_s = 0.1$.
\cred{Importantly, as demonstrated by the perpendicular-space analysis in Sec.~\ref{sec: perp} and structure-factor analysis in Sec.~\ref{sec: bragg}, this weak-measurement behavior cannot be understood solely from the nearest-neighbor local bond structure, reflecting the nontrivial role of long-range quasiperiodic order.}

Furthermore, increasing the measurement strength up to $\gamma / J_s = 5$,
a spatially nonuniform pattern distinct from that in the weak measurement regime appears;
the values of $F_i$ bifurcate into two different values $F_i \approx 0.01$ and $0.03$
corresponding to the sites between two $\ell$-spacings, and between $\ell$- and $s$-spacings, respectively.
Clearly, this behavior is governed by the local lattice environments.
Thus, while both weak and strong measurement regimes exhibit spatial nonuniformity, their characteristics are qualitatively different from each other.
\cred{Since these two regimes are smoothly connected as the measurement strength is increased, we refer to this phenomenon as a measurement-induced crossover, describing a smooth change between regimes with qualitatively different spatial characteristics.
Although previous studies have reported the absence of measurement-induced entanglement transitions in one-dimensional monitored free fermions~\cite{Poboiko_2023,Yin_2026}, the possible relationship between such transitions and the spatial crossover observed here remains unclear.
Our large system-size analysis [see Fig.~\ref{fig:size_physical} in Appendix~\ref{app_systemsize}] suggests that the characteristic spatial structures of both regimes and the crossover between them persist toward the thermodynamic limit, although a definitive conclusion cannot be drawn from finite-size calculations alone.}
This crossover will be further elucidated using perpendicular-space analysis in Sec.~\ref{sec: perp} \cred{and structure-factor analysis in Sec.~\ref{sec: bragg}}.

\subsection{Measurement-induced crossover in perpendicular space}\label{sec: perp}
\begin{figure}[tb]
  \begin{center}
    \includegraphics[width=\linewidth]{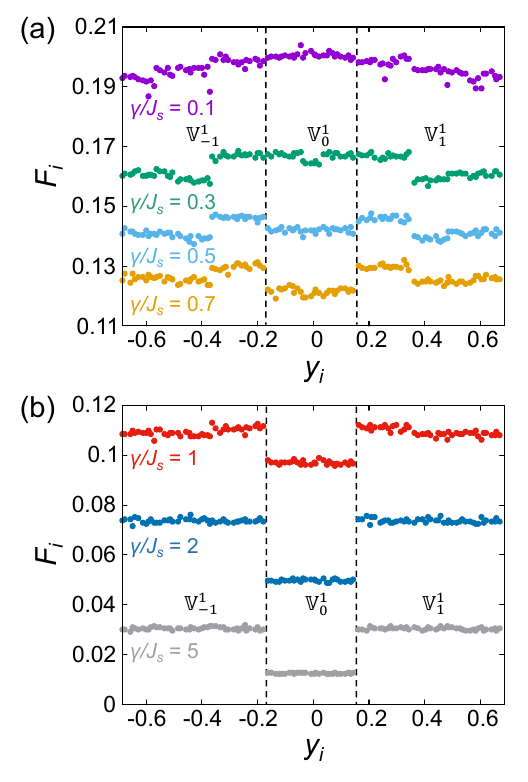}
    \caption{Spatial distribution of fluctuations of the local particle number in the quasiperiodic system against $y_i$ in perpendicular space. Results are shown for (a) $\gamma/J_s =$ 0.1, 0.3, 0.5, and 0.7, and for (b) $\gamma/J_s =$ \cred{1, 2, and 5}. The parameters are set to $J_\ell / J_s = 0.4$ and $L=120$.}
    \label{fig: Fi_perp}
  \end{center}
\end{figure}

We use perpendicular space analysis in a quasiperiodic lattice
to elucidate how the aperiodic structure affects the spatial distribution of fluctuations.
Perpendicular space analysis has been widely employed as a powerful technique for analyzing electronic properties in quasicrystals~\cite{Jagannathan_2004, Attila_2008, Koga_2017, Koga_2020, Matsubara_3GMT, Koga_honey, Takemori_2020, Sakai_2022, Murod_2020, Oktel_2022, Akif_2022, Hori_TSC, Hori_Bose, Yamamoto_2024}, and here we extend this approach to continuously monitored dynamics.
In Fig.~\ref{fig: Fi_perp}, fluctuations of the local particle number are plotted against the coordinate $y_i$ in perpendicular space. The range of $y_i$ corresponds to the width $w$ of the strip shown in Fig.~\ref{fig: Fibonacci}, spanning $y_i \in [-w/2, w/2]$ where $w/2 \approx 0.688$.
Here, we define a region in perpendicular space as
\begin{eqnarray}
{\mathbb V}^R_n &=& \{y_i \, | \, \Phi (i-R, i+R) = W^R_n\}
\end{eqnarray}
where $W^R_n$ is a sequence of length $2R$ composed of $\ell$- and $s$-spacings in the Fibonacci lattice, and $n=-R,-R+1,\cdots,R-1,R$ labels the $2R+1$ possible local environments.
Note that $n$ is assigned in ascending order for the coordinates in perpendicular space.
We emphasize that the values of $F_i$ remain unchanged even after performing the coordinate transformation from physical space to perpendicular space as $(x_i, F_i) \rightarrow (y_i,F_i)$.
Importantly, the regions ${\mathbb V}^R_n$ are determined independently of $F_i$, which is not affected by the choice of $n$ and \cred{$R$} for classifying the perpendicular space.

Focusing on three consecutive sites in the Fibonacci lattice, we divide the perpendicular space into the following regions
\begin{eqnarray}
{\mathbb V}^1_n &=& \{y_i \, | \, \Phi (i-1, i+1) = W^1_n\},
\end{eqnarray}
where $n=-1, 0, 1$, and \cred{$W_n^R$} is given by
\begin{eqnarray}
W^1_{-1} &=& (\ell, s),\\
W^1_0 &=& (\ell, \ell),\\
W^1_1 &=& (s, \ell).
\end{eqnarray}
In Fig.~\ref{fig: Fi_perp}, $F_i$ is displayed in perpendicular space, where we find three regions denoted as ${\mathbb V}^1_{-1}$, ${\mathbb V}^1_0$, and ${\mathbb V}^1_1$.
Note that $i$-th and $i^\prime$-th sites satisfying $\Phi (i-1,i+1)=W_{-1}^1=(\ell, s)$ and $\Phi (i^\prime-1,i^\prime+1)=W_1^1=(s, \ell)$ possess equivalent local lattice structures, and thereby the distributions of $F_i$ for ${\mathbb V}^1_{-1}$ and ${\mathbb V}^1_{1}$ are symmetric with respect to $y_i=0$.
Although the present numerical data may be slightly asymmetric due to finite-size effects, they are expected to be perfectly symmetric in the thermodynamic limit.
For weak measurement strength $\gamma / J_s = 0.1$ [Fig.~\ref{fig: Fi_perp}(a)], we see that the value of $F_i$ is modulated within the region $\mathbb V_n^1$ as $y_i$ is varied.
\cred{Importantly, this spatial modulation is observed even among sites sharing the same nearest-neighbor local lattice environment and therefore cannot be explained solely by the local bond configuration.
This demonstrates that, under weak measurement, $F_i$ captures the incommensurate lattice structure, reflecting the nontrivial long-range quasiperiodic order of the underlying lattice.}
As the measurement strength increases, the distribution of $F_i$ begins
to separate into several clusters.
\cred{For $0.3 \lesssim \gamma / J_s \lesssim 0.7$, $F_i$ splits into five clusters, as shown in Fig.~\ref{fig: Fi_perp}(a).
We note that, in the thermodynamic limit, the reflection symmetry of the ideal Fibonacci chain implies a symmetric distribution in perpendicular space with respect to $y_i=0$.}
In contrast to the weak measurement case,
discontinuous behavior with a jump emerges at the boundaries of the clusters.
In detail, the values of $F_i$ within the regions ${\mathbb V}_{-1}^1$ and ${\mathbb V}_1^1$ split into two distinct branches [see Fig.~\ref{fig: Fi_perp}(a) for $\gamma/J_s = 0.3, 0.5$, and $0.7$].
However, as the measurement strength increases, these branches merge into a single value [see Fig.~\ref{fig: Fi_perp}(b)],
and $F_i$ forms three clusters with two different values.
Notably, in the strong measurement regime, the value of $F_i$ becomes constant within the region $\mathbb V_n^1$.
This means that the fluctuation is dominated by local environments, namely, the types of two adjacent lattice spacings around each site.
Such splitting and merging of clusters manifest the nontrivial crossover induced by measurements:
the distribution of fluctuations of the local particle number reflects incommensurate modulation of the quasiperiodic lattice spacings for weak measurement, while it changes to the one governed by the local lattice environment for strong measurement.
\cred{These characteristic features become more prominent as the system size is increased [see Fig.~\ref{fig:size_physical}(b) in Appendix~\ref{app_systemsize}].}
It should be noted that the quasiperiodic pattern found in $F_i$ for weak measurement is absent in the unitary limit, and is characteristic of the quasiperiodic systems under continuous monitoring.

\begin{figure}[tb]
  \begin{center}
    \includegraphics[width=\linewidth]{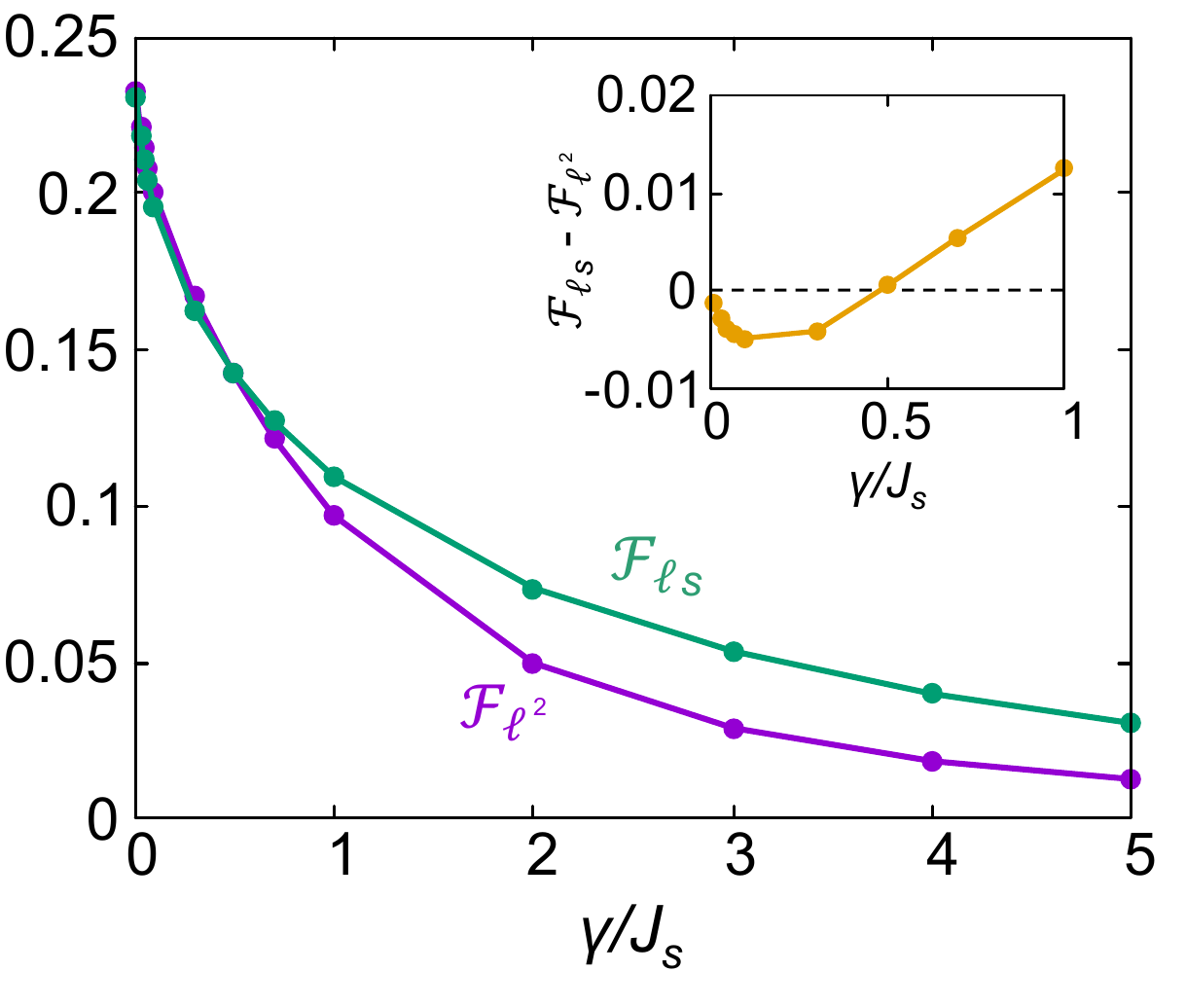}
    \caption{Spatial average of fluctuations of the local particle number in the quasiperiodic system resolved by perpendicular space $\mathbb V_n^1$ with respect to measurement strengths.
    The inset shows the difference $\mathcal F_{\ell s}-\mathcal F_{\ell^2}$ between fluctuations corresponding to two different local environments.
    We take $L=120$ and $J_\ell/J_s=0.4$, and the number of trajectories is set to 1000.}
    \label{fig: Fi_ave}
  \end{center}
\end{figure}

Another intriguing phenomenon is a measurement-induced sign reversal of $\mathcal{F}_{\ell s} -\mathcal{F}_{\ell^2}$ shown in Fig.~\ref{fig: Fi_ave},
where $\mathcal{F}_{\ell^2}$ and $\mathcal{F}_{\ell s}$ are the spatial averages of $F_i$ in regions ${\mathbb V}^1_{0}$ and ${\mathbb V}^1_{-1} \cup {\mathbb V}^1_{1}$, respectively.
When $\gamma / J_s \lesssim 0.5$, we find that $\mathcal{F}_{\ell^2} > \mathcal{F}_{\ell s}$, but for strong measurement, we see that $\mathcal{F}_{\ell s} > \mathcal{F}_{\ell^2}$,
which signals that the dominant contribution from the local lattice environments changes as the measurement strength is increased.
However, we note that this sign reversal is also observed in periodic approximants (see Appendix~\ref{app: AC1}) and is distinct from the measurement-induced crossover observed in $F_i$; it is solely determined by the local lattice structure independent of the quasiperiodic structure.

\subsection{Structure-factor analysis}
\label{sec: bragg}

To quantitatively characterize the spatial distribution of fluctuations of the local particle number, we introduce the structure factor
\begin{eqnarray}
S(q)
&=&
\frac{1}{L}
\left|
\sum_{j=0}^{L-1}
(F_j- F_{\rm space})e^{iqj}
\right|^2,\label{eq_Sq}\\
F_{\rm space}
&=&
\frac{1}{L}
\sum_{j=0}^{L-1}F_j.
\end{eqnarray}
Here, the spatial average $F_{\rm space}$ is subtracted from $F_j$ to highlight the spatial modulation of $F_j$.
We use the site index $j$ rather than the physical position $x_j$ in the Fourier transform, to characterize the quasiperiodic modulation of $F_j$ itself. Using $x_j$ would also incorporate the quasiperiodic arrangement of the lattice sites, making it difficult to distinguish its contribution from that of the spatial modulation of $F_j$.

\begin{figure*}[htb]
    \centering
    \includegraphics[width=\linewidth]{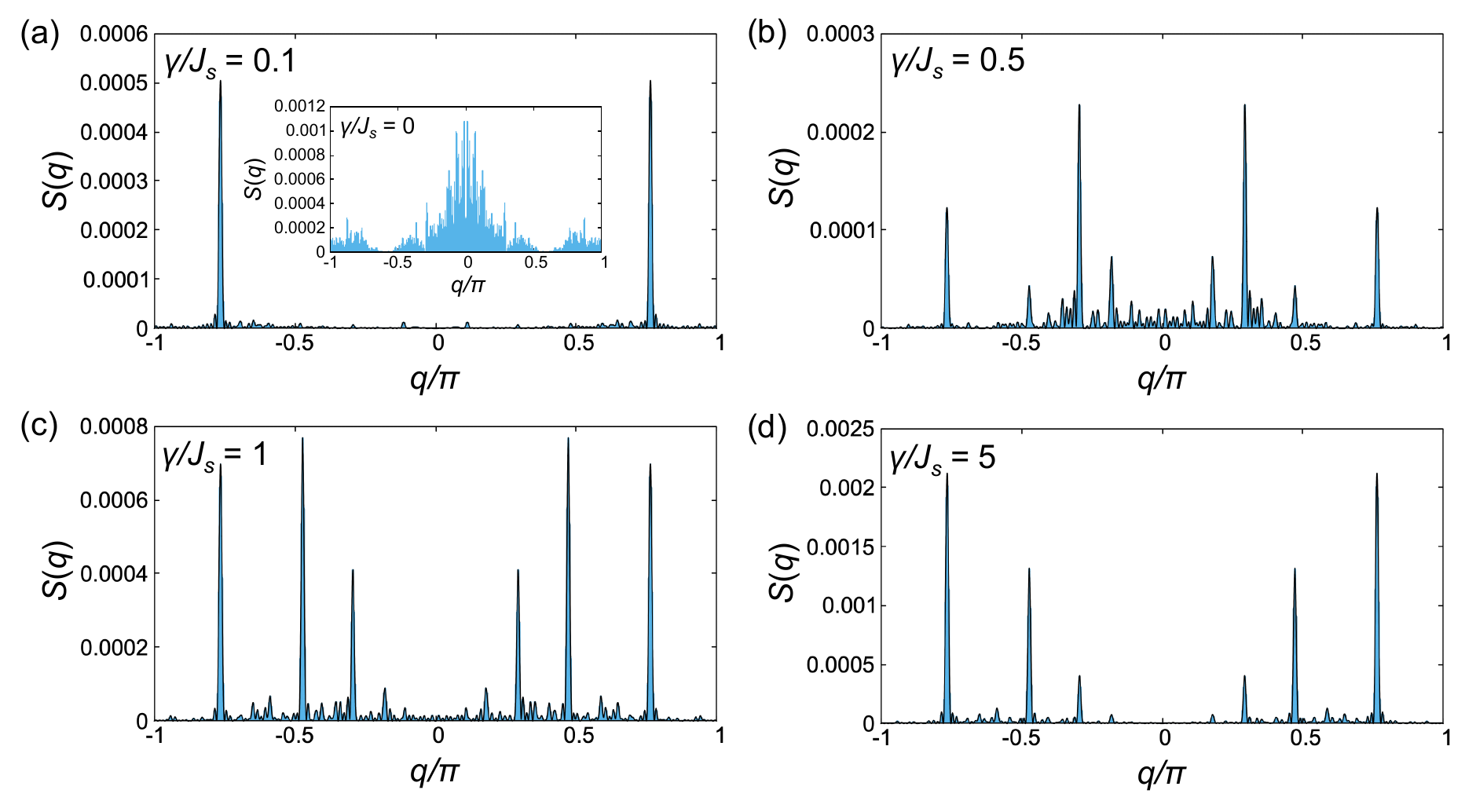}
    \caption{\cred{
    Structure factor of fluctuations of the local particle number $F_i$ at
    (a) $\gamma/J_s=0.1$,
    (b) $0.5$,
    (c) $1$, and
    (d) $5$.
    The weak-measurement result in (a) is dominated by two Bragg peaks at $q=\pm 2\pi/\tau^2$, characteristic of the golden-mean modulation, whereas the strong-measurement result in (d) exhibits additional peaks at $q=\pm 2\pi/\tau^3$ and $\pm 2\pi/\tau^4$, reflecting the characteristic Bragg-peak hierarchy of the Fibonacci pattern.
    The inset in panel (a) shows the result for $\gamma=0$.
    We take $L=150$ and $J_\ell/J_s=0.4$, and the number of trajectories is set to 1000 for finite measurement strengths.}
    }
    \label{fig:structure_F}
\end{figure*}

Figure~\ref{fig:structure_F} shows $S(q)$ for $J_\ell/J_s=0.4$ at several measurement strengths.
Since $S(q)=S(q+2\pi)$, we restrict the plot to $-\pi\leq q\leq\pi$.
For weak measurement $\gamma/J_s=0.1$ shown in Fig.~\ref{fig:structure_F}(a), two dominant Bragg peaks appear at $q=\pm 2\pi/\tau^2$.
This indicates that the dominant spatial modulation of $F_j$ can be approximately expressed as
\begin{equation}
F_j-F_{\rm space}
\propto
\cos\left(\frac{2\pi j}{\tau^2}+\phi\right)
=
\cos\left(\frac{2\pi j}{\tau}-\phi\right),
\end{equation}
where $\phi$ is a phase offset, and the equality holds for integer $j$ because $\tau^2=\tau+1$.
This peak structure is also reproduced by the single golden-mean modulation discussed in Appendix~\ref{app:reference_structure}.
We remark that our numerical results for $\gamma=0$ in the inset of Fig.~\ref{fig:structure_F}(a) show no sharp peak structures in $S(q)$, but rather a broad distribution in reciprocal space.
Thus, the emergence of Bragg peaks under weak measurement indicates that continuous monitoring induces a nontrivial spatial modulation of $F_j$ that reflects the underlying quasiperiodic order.
Furthermore, the finite-size scaling of the normalized Bragg-peak intensities $S(2\pi/\tau^2)/L$ [see Fig.~\ref{fig:size_physical}(c) in Appendix~\ref{app_systemsize}] shows approximately size-independent behavior for $L\geq90$, suggesting that these Bragg peaks persist in the thermodynamic limit.

In the strong-measurement regime, for $\gamma/J_s=5$ shown in Fig.~\ref{fig:structure_F}(d), additional Bragg peaks at $\pm 2\pi/\tau^3$ and $\pm 2\pi/\tau^4$ become prominent.
This behavior can be understood in terms of the quantum Zeno effect, which suppresses particle hopping under strong measurement, making the fluctuations predominantly dependent on the local bond configuration.
In this regime, the fluctuations are approximately classified according to the local environment as
\begin{equation}
F_j
\approx
\begin{cases}
0.01, & \text{(between $\ell$- and $\ell$-spacings)},\\
0.03, & \text{(between $\ell$- and $s$-spacings)}.
\end{cases}
\end{equation}
Consequently, the structure factor approaches that of the local-environment-dependent Fibonacci pattern shown in Appendix~\ref{app:reference_structure}, exhibiting a characteristic hierarchy of Bragg peaks that reflects the self-similar structure of the underlying lattice.
Note that, at intermediate measurement strengths $\gamma/J_s=0.5$ and $1$ shown in Figs.~\ref{fig:structure_F}(b) and \ref{fig:structure_F}(c), the structure factor exhibits many smaller closely spaced peaks.
However, our finite-size results do not allow us to determine whether these peaks develop into well-defined Bragg peaks or form a broad distribution in the thermodynamic limit.

The structure factor of $F_j$ provides a reciprocal-space characterization of the measurement-induced crossover.
While the weak-measurement regime is characterized by a dominant golden-mean modulation beyond the local bond configuration, the strong-measurement regime exhibits a Bragg-peak hierarchy reflecting the increasing dominance of the local lattice environment.
Interestingly, the disordered hopping model discussed in Appendix~\ref{app: disorder} exhibits a broad distribution in the structure factor without well-defined Bragg peaks.
This qualitative difference demonstrates that the Bragg-peak structures observed here are not simply a consequence of spatially inhomogeneous hopping, but reflect the underlying quasiperiodic order.
Thus, the fluctuations under weak measurement exhibit a characteristic incommensurate modulation reflecting the long-range quasiperiodic structure, which cannot be explained solely by the nearest-neighbor bond configuration.

\subsection{Correlation functions}
\begin{figure}[tb]
  \begin{center}
    \includegraphics[width=\linewidth]{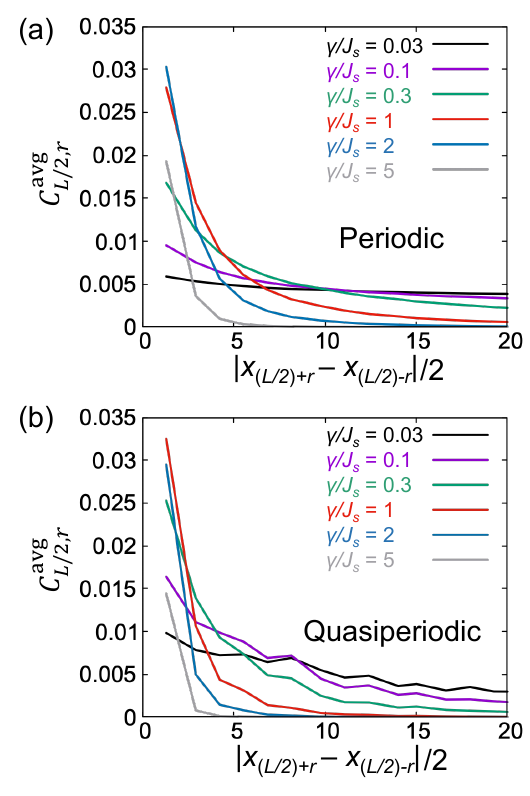}
    \caption{Correlation function $C^{\rm avg}_{L/2,r}$ as a function of distance $|x_{(L/2)+r}-x_{(L/2)-r}|/2$ in (a) the periodic system $J_\ell / J_s = 1$ and in (b) the quasiperiodic system $J_\ell / J_s = 0.4$. The number of trajectories is set to 1000.
    The system size used in the calculation is $L = 60$.}
    \label{fig: correlation}
  \end{center}
\end{figure}

To explore how the quasiperiodic structure affects physical quantities beyond fluctuations, we study the connected correlation function as
a nonlocal extension of fluctuations of the local particle number, given by
\begin{equation}
C_{i,j}= \overline{\braket{n_i}\braket{n_j} - \braket{n_i n_j}},
\end{equation}
which reduces to the fluctuation $-F_i$ when $i=j$.
To investigate how the lattice structure around each site influences the connected correlation function, we introduce an average of $C_{i,j}$ with $j=i-r$ and $j=i+r$, which correspond to two sites located at a site separation $r$ from the $i$-th site, as
\begin{equation}
C^{\rm avg}_{i,r}= \frac{C_{i,i+r} + C_{i,i-r}}{2}.
\end{equation}
Figure~\ref{fig: correlation} shows $C^{\rm avg}_{L/2,r}$ as a function of distance $|x_{(L/2)+r}-x_{(L/2)-r}|/2$. 
While the periodic system exhibits a monotonic decay of the correlation function irrespective of the measurement strength [see Fig.~\ref{fig: correlation}(a)], the quasiperiodic system exhibits an oscillatory decay for weak measurement strengths  [see Fig.~\ref{fig: correlation}(b) for $\gamma/J_s = 0.03$ and $0.1$]. 
\cred{Moreover, as the measurement strength is increased, the magnitude of $C^{\rm avg}_{L/2,r}$ at short distance initially increases, reaches a maximum around $\gamma / J_s \approx 2$ in the periodic system [Fig.~\ref{fig: correlation}(a)] and $\gamma / J_s \approx 1$ in the quasiperiodic system [Fig.~\ref{fig: correlation}(b)], and then decreases.
A similar nonmonotonic dependence of the magnitude of short-distance correlations on the measurement strength has also been observed in continuously monitored free-fermion systems, including periodic, random, and Aubry-Andr\'e-Harper models~\cite{Alberton_2021, Szyniszewski_2023, Matsubara_AAH}.}
On the other hand, at large distance $|x_{(L/2)+r} - x_{(L/2)-r}| \sim O(L)$, the correlation function decreases monotonically against the measurement strengths in both periodic and quasiperiodic systems.

\begin{figure*}[tb]
  \begin{center}
    \includegraphics[width=\linewidth]{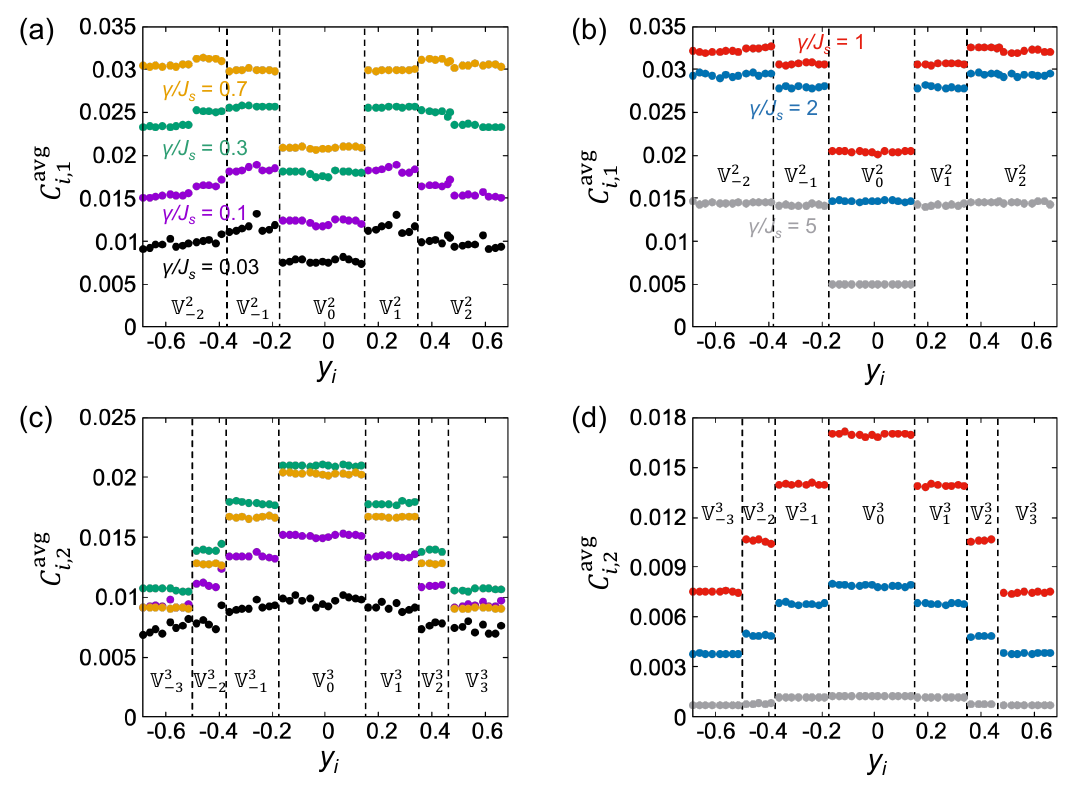}
    \caption{Distribution of correlation functions in the quasiperiodic chain with respect to $y_i$ in perpendicular space.
     $C_{i,1}^\mathrm{avg}$ ($C_{i,2}^\mathrm{avg}$) for (a) [(c)] $\gamma/J_s$ = \cred{0.03, 0.1, 0.3, and 0.7}, and (b) [(d)] for $\gamma/J_s$ = \cred{1, 2, and 5}.
     The parameters are set to $J_\ell / J_s = 0.4$ and $L=60$, and the number of trajectories is set to 1000.}
    \label{fig: correlation0}
  \end{center}
\end{figure*}

\begin{figure}[htb]
  \begin{center}
    \includegraphics[width=\linewidth]{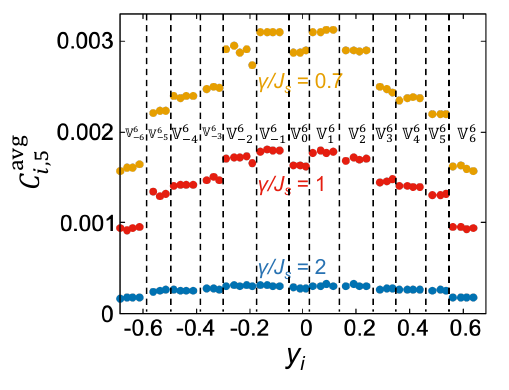}
    \caption{
    Distribution of the correlation function $C_{i,5}^\mathrm{avg}$ in the quasiperiodic chain with respect to $y_i$ in perpendicular space.
    The parameters are set to $J_\ell / J_s = 0.4$ and $L=60$ and the number of trajectories is set to 1000.}
    \label{fig: correlation1}
  \end{center}
\end{figure}

To clarify how the quasiperiodic structure affects the correlation functions,
we perform the perpendicular space analysis.
We emphasize that, in contrast to fluctuations studied in Sec.~\ref{sec: perp}, the correlation function $C_{i,r}^\mathrm{avg}$ captures local lattice structures over extended length scales, and thereby controls how the effect of the local environments of each site appears in itself by changing the distance $|x_{(L/2)+r}-x_{(L/2)-r}|/2$.
This indicates that distribution patterns of $C_{i,r}^\mathrm{avg}$ \cred{change substantially} in perpendicular space as $r$ is varied.
To see this, we first plot in Figs.~\ref{fig: correlation0}(a) and \ref{fig: correlation0}(b) the correlation function $C^{\rm avg}_{i,1}$ against $y_i$ in perpendicular space.
In this case, we can focus on five consecutive sites in physical space and divide the perpendicular space into the following regions
\begin{eqnarray}
{\mathbb V}^2_n &=& \{y_i \, | \, \Phi (i-2, i+2) = W^2_n\},
\end{eqnarray}
where $-2\leq n \leq 2$, and \cred{$W_n^R$} is given by
\begin{eqnarray}
W^2_0 &=& (s, \ell, \ell, s),\\
W^2_1 &=& (\ell, s, \ell, s),\\
W^2_2 &=& (\ell, s, \ell, \ell).
\end{eqnarray}
Here, we have omitted the definition for $W_{-2}^2$ and $W_{-1}^2$ since $W_{-n}^2$ is reproduced from $W_n^2=(a,b,c,d)$ as $W_{-n}^2=(d,c,b,a)$.
The classification of $\mathbb V_n^2$ corresponds to a decomposition of $\mathbb V_n^1$: ${\mathbb V}^1_0 = {\mathbb V}^2_0$, ${\mathbb V}^1_1 = {\mathbb V}^2_1 \cup {\mathbb V}^2_2$, and ${\mathbb V}^1_{-1} = {\mathbb V}^2_{-1} \cup {\mathbb V}^2_{-2}$.
For weak measurement strength $\gamma / J_s \lesssim 0.1$ shown in Fig.~\ref{fig: correlation0}(a), the values of correlation functions are modulated within each region $\mathbb V_n^2$ in perpendicular space, reflecting the long-range quasiperiodic lattice structure.
On the other hand, for strong measurement strength $\gamma / J_s \gtrsim 1$ shown in Fig.~\ref{fig: correlation0}(b), $C^\mathrm{avg}_{i,1}$ is almost fixed in each region $\mathbb V_n^2$,
demonstrating that the values of correlation functions $C^\mathrm{avg}_{i,1}$ are determined by the local lattice environment within the site separation $2r$.

By changing $r$, we find a distinct distribution pattern in perpendicular space. Figures~\ref{fig: correlation0}(c) and \ref{fig: correlation0}(d) display the correlation function $C^\mathrm{avg}_{i,r}$ for $r=2$. In this case, the perpendicular space is further classified into
\begin{eqnarray}
{\mathbb V}^3_n &=& \{y_i \, | \, \Phi (i-3, i+3) = W^3_n\},
\end{eqnarray}
where $-3\leq n\leq3$, and \cred{$W_n^R$} is given by
\begin{eqnarray}
W^3_0 &=& (\ell, s, \ell, \ell, s, \ell),\\
W^3_1 &=& (\ell, \ell, s, \ell, s, \ell),\\
W^3_2 &=& (\ell, \ell, s, \ell, \ell, s),\\
W^3_3 &=& (s, \ell, s, \ell, \ell, s).
\end{eqnarray}
Again, we have omitted $W_{-3}^3$, $W_{-2}^3$ and $W_{-1}^3$ since $W_{-n}^3$ is reproduced from \cred{$W_n^3=(a,b,c,d,e,f)$ as $W_{-n}^3=(f,e,d,c,b,a)$}.
In the weak measurement regime, the distribution of $C^\mathrm{avg}_{i,2}$ is again modulated in perpendicular space within each region $\mathbb V_n^3$.
However, as the measurement strength increases,
the distribution pattern changes to one, whose value of $C^\mathrm{avg}_{i,2}$ is fixed within $\mathbb V_n^3$ and determined by the local lattice structure within the site separation $2r$.
Clearly, we find that the spatial profiles of $C^\mathrm{avg}_{i,1}$ and $C^\mathrm{avg}_{i,2}$ are distinct from each other; the latter for strong measurement strength are further split into highly fragmented clusters [see Figs.~\ref{fig: correlation0}(b) and \ref{fig: correlation0}(d)].
We note that, for $\gamma/J_s=5$, although the correlation function is governed by the local lattice environments, the differences in $C^{\mathrm{avg}}_{i,2}$ among different regions of $\mathbb{V}_n^3$ become very small because the correlations are strongly suppressed by measurement.

We also briefly comment on several properties found in correlation functions.
For $C^{\rm avg}_{i,1}$ and $C^{\rm avg}_{i,2}$ in the weak-measurement regime, increasing the measurement strength initially enhances the correlation functions.
They reach their peak values, $C^{\rm avg}_{i,1}\approx 0.03$ and $C^{\rm avg}_{i,2}\approx 0.02$ around $\gamma/J_s\sim 1$ (see Fig.~\ref{fig: correlation0}), and are then suppressed.
We also find that, for strong measurement, the correlation function $C^{\rm avg}_{i,1}$ takes a minimum value at the central region ${\mathbb V}^2_0$, while $C^{\rm avg}_{i,2}$ takes a maximum value at the central region ${\mathbb V}^3_0$, which indicates that varying $r$ induces a qualitative difference of the spatial patterns in perpendicular space.

Finally, we show the numerical results of correlation functions for large $r$.
As an example, Fig.~\ref{fig: correlation1} shows the spatial distribution of $C_{i,5}^\mathrm{avg}$, where the perpendicular space is divided into
\begin{eqnarray}
{\mathbb V}^6_n &=& \{y_i \, | \, \Phi (i-6, i+6) = W^6_n\},
\end{eqnarray}
where $-6\leq n \leq 6$, and \cred{$W_n^R$} is given by
\begin{eqnarray}
W^6_0 &=& (\ell, \ell, s, \ell, s, \ell, \ell, s, \ell, s, \ell, \ell),\allowdisplaybreaks\\
W^6_1 &=& (\ell, \ell, s, \ell, s, \ell, \ell, s, \ell, \ell, s, \ell),\allowdisplaybreaks\\
W^6_2 &=& (\ell, \ell, s, \ell, \ell, s, \ell, s, \ell, \ell, s, \ell),\allowdisplaybreaks\\
W^6_3 &=& (s, \ell, s, \ell, \ell, s, \ell, s, \ell, \ell, s, \ell),\allowdisplaybreaks\\
W^6_4 &=& (s, \ell, s, \ell, \ell, s, \ell, \ell, s, \ell, s, \ell),\allowdisplaybreaks\\
W^6_5 &=& (s, \ell, \ell, s, \ell, s, \ell, \ell, s, \ell, s, \ell),\allowdisplaybreaks\\
W^6_6 &=& (s, \ell, \ell, s, \ell, s, \ell, \ell, s, \ell, \ell, s),
\end{eqnarray}
where $W_{-n}^6$ ($1\leq n \leq 6$) is given by the reversed sequence of $W_n^6$. 
We find in Fig.~\ref{fig: correlation1} that, for strong measurement, the correlation function is determined by these local lattice environments, where $C_{i,5}^\mathrm{avg}$ forms 13 fragmented clusters, many more than those for $C_{i,1}^\mathrm{avg}$ and $C_{i,2}^\mathrm{avg}$ shown in Fig.~\ref{fig: correlation0}.

\section{Summary}\label{sec: summary}
We have studied the measurement-induced dynamics of free fermions in the quasiperiodic Fibonacci lattice and revealed the novel aspects of quasiperiodic phenomena under continuous measurement. 
Our key findings are summarized as follows. First, we have found that the spatial distribution of fluctuations of the local particle number exhibits the measurement-induced nonuniformity originating from the underlying quasiperiodic structures. 
Second, we have demonstrated its nontrivial crossover driven by measurement: for weak measurement strength, the spatial pattern shows spatial modulation reflecting the long-range lattice structure, while for strong measurement strength, the distribution collapses into discrete constant values determined by the local lattice environment.
We have also elucidated that the measurement-induced crossover emerges in connected correlation functions.
These findings highlight that fluctuations of the local particle number and correlation functions serve as signatures for probing the measurement-induced phenomena unique to quasiperiodic systems.

Since the crossover from the spatial distribution reflecting the long-range lattice structure to the one determined by the local lattice structure is also reported in the behavior of order parameters in superconducting or magnetic quasiperiodic systems, where strong interactions lead to localization and local-structure-dependent ordering~~\cite{Jagannathan_2004, Attila_2008, Sakai_2017, Koga_2017, Koga_2020, Takemori_2020, Matsubara_3GMT, Koga_honey, Hori_TSC, Hori_Bose, Yamamoto_2024},
it is interesting to investigate the impact of strong interactions on the measurement-induced crossover obtained in our study.
Exploring various types of quasiperiodic lattices is essential to distinguish model-specific features from universal behavior. Additionally, investigating higher-dimensional systems remains an intriguing open question, as it clarifies whether these crossover phenomena survive in two- or three-dimensional systems in ultracold atoms and how the underlying dimensionality characterizes spatial distributions of fluctuations.

\begin{acknowledgments}
We would like to thank Akihisa Koga, Shuta Nakajima, and Shiro Sakai for fruitful discussions.
Parts of the numerical calculations were performed in the supercomputing systems in ISSP, the University of Tokyo. 
T.M. was supported by JSPS KAKENHI Grant
No. JP25KJ1247.
K.Y. was supported by JSPS Program for Forming Japan's Peak Research Universities (J-PEAKS) Grant No.\ JPJS00420230008,
JSPS KAKENHI Grant No. JP25K17327,
Hirose Foundation, Fujikura Foundation, Toyota Riken Scholar Program, and Support Center for Advanced Telecommunications Technology Research. 
\end{acknowledgments}

\section*{Data availability}
The data that support the findings of this paper are openly available~\cite{zenodo202607}.

\appendix

\section{Numerical results on system-size dependence up to $L=150$}
\label{app_systemsize}

\begin{figure*}[htb]
    \centering
    \includegraphics[width=\linewidth]{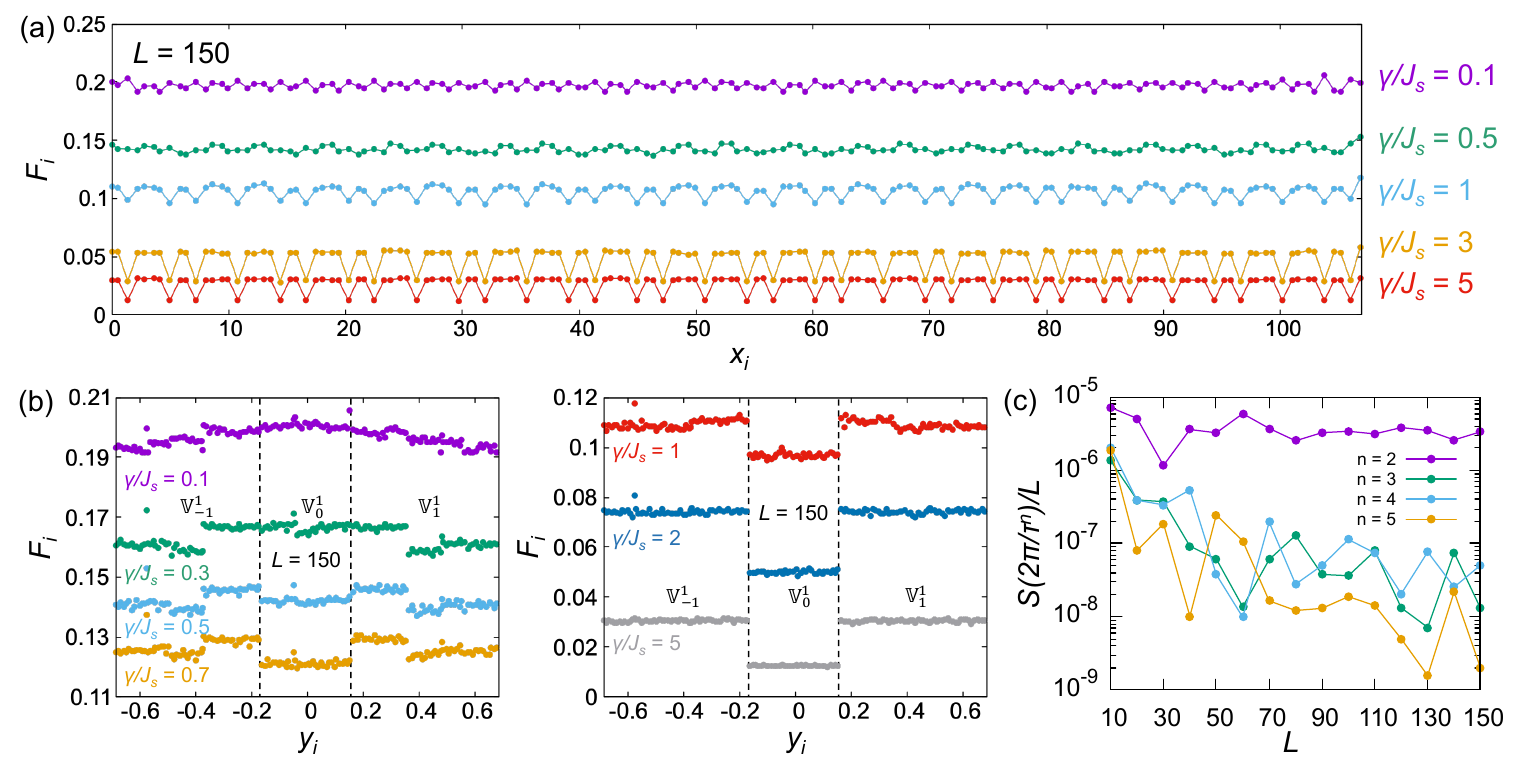}
    \caption{
    (a) [(b)] Distribution of fluctuations of the local particle number in physical (perpendicular) space. The system size is $L=150$.
    (c) System-size dependence of $S(2\pi/\tau^n)/L$ at $\gamma/J_s=0.1$, which becomes approximately independent of $L$ for $L\ge90$ and $n=2$.
    In all panels, we take $J_{\ell}/J_s=0.4$, and the number of trajectories is set to $1000$.
    Note that, in (a), the entire spatial distribution is shown, and the maximum value of $x_j$ is different from $L$.
    }
    \label{fig:size_physical}
\end{figure*}

In Fig.~\ref{fig:size_physical}(a), for $L=150$, we further examine the spatial distribution of fluctuations of the local particle number discussed in Sec.~\ref{sec_Fiphysical}.
For all system sizes considered, the spatial distributions of $F_i$ in both the weak- and strong-measurement regimes exhibit the same qualitative behavior as those shown in Fig.~\ref{fig: Fi_phys}(b), even when the system size is increased to $L=150$.
This tendency becomes clearer in perpendicular space.
Figure~\ref{fig:size_physical}(b) shows the corresponding distributions in perpendicular space for $L=150$.
The spatial distributions of $F_i$ in perpendicular space remain qualitatively unchanged as the system size increases, while their characteristic features become more clearly resolved.
This improved resolution is consistent with the denser sampling of perpendicular-space coordinates at larger system sizes.
In addition, we have explicitly examined the Fibonacci-number system size $L=34$ and found no anomalous suppression of the fluctuations compared with nearby system sizes.
Given the robustness of the observed behavior against changes in system size, we expect that no specific suppression occurs for other Fibonacci-number system sizes.
In Fig.~\ref{fig:size_physical}(c), we show the structure factor $S(2\pi/\tau^n)/L$ for several values of $n$ at $\gamma/J_s=0.1$.
For $n=2$, we obtain an approximately $L$-independent value for $L\ge90$, while the values for $n=3,$ $4,$ and $5$ approach zero as $L$ is increased.
This result supports the persistence of nontrivial Bragg peaks at $q=\pm 2\pi/\tau^2$ in large system sizes.
These results suggest that the distinct spatial patterns in the two regimes and the associated measurement-induced crossover persist toward the thermodynamic limit.

\section{Periodic approximants}\label{app: AC1}
\begin{figure}[tb]
  \begin{center}
    \includegraphics[width=\linewidth]{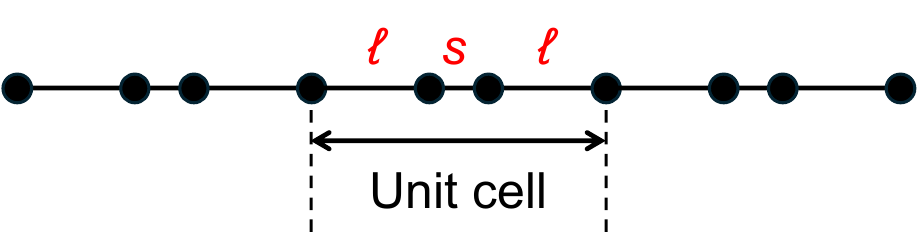}
    \caption{Schematic illustration of the $2/1$ approximant lattice.}
    \label{fig: approximants}
  \end{center}
\end{figure}

We consider a periodic approximant of the quasiperiodic Fibonacci chain~\cite{Goldman_1993}, and study the fluctuation of the local particle number under \cred{continuous} monitoring. We then compare the results with those of the quasiperiodic system to confirm that the measurement-induced crossover is intrinsic to quasiperiodic structures.
The periodic approximant of the Fibonacci lattice is obtained by approximating the golden mean $\tau$ by a rational number in the cutting angle \eqref{Eq: theta}. Using Fibonacci numbers $\{f_n \, | \, 1, 1, 2, 3, 5, 8, 13, \cdots\}$, we set $\tau \approx f_{n+1}/f_n$.
It is known that larger $n$ yields a better approximation. The resulting lattice is called the $f_{n+1}/f_n$ approximant lattice and possesses spatial periodicity as a result of the approximation using a rational number.
We focus on the simplest case that has $\ell$- and $s$-spacings, i.e., we set $\tau \approx f_2 / f_1 = 2/1$.
The unit cell of the $2/1$ approximant consists of two $\ell$-spacings and one $s$-spacing
as shown in Fig.~\ref{fig: approximants}.
As in the Fibonacci lattice, the local lattice environments can be classified into two types: those with neighboring spacings $\ell$ and $s$, and those with two neighboring $\ell$-spacings.
We remark that the lattice structure no longer possesses quasiperiodic long-range order.

\begin{figure}[tb]
  \begin{center}
    \includegraphics[width=\linewidth]{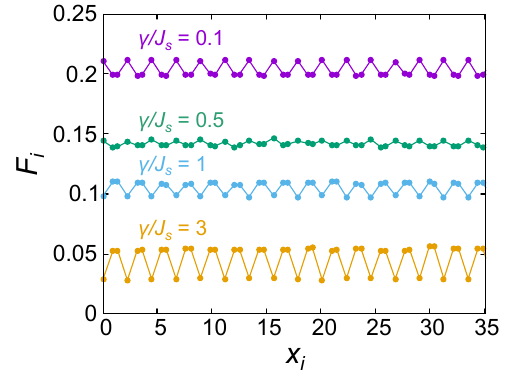}
    \caption{Spatial profile of fluctuations of the local particle number on the 2/1 approximant lattice under continuous monitoring with respect to $x_i$ in physical space. We take $L=60$ and $J_\ell/J_s=0.4$, and the number of trajectories is set to 1000.}
    \label{fig: Fi_phys_1AC}
  \end{center}
\end{figure}

\begin{figure}[tb]
  \begin{center}
  \includegraphics[width=0.85\linewidth]{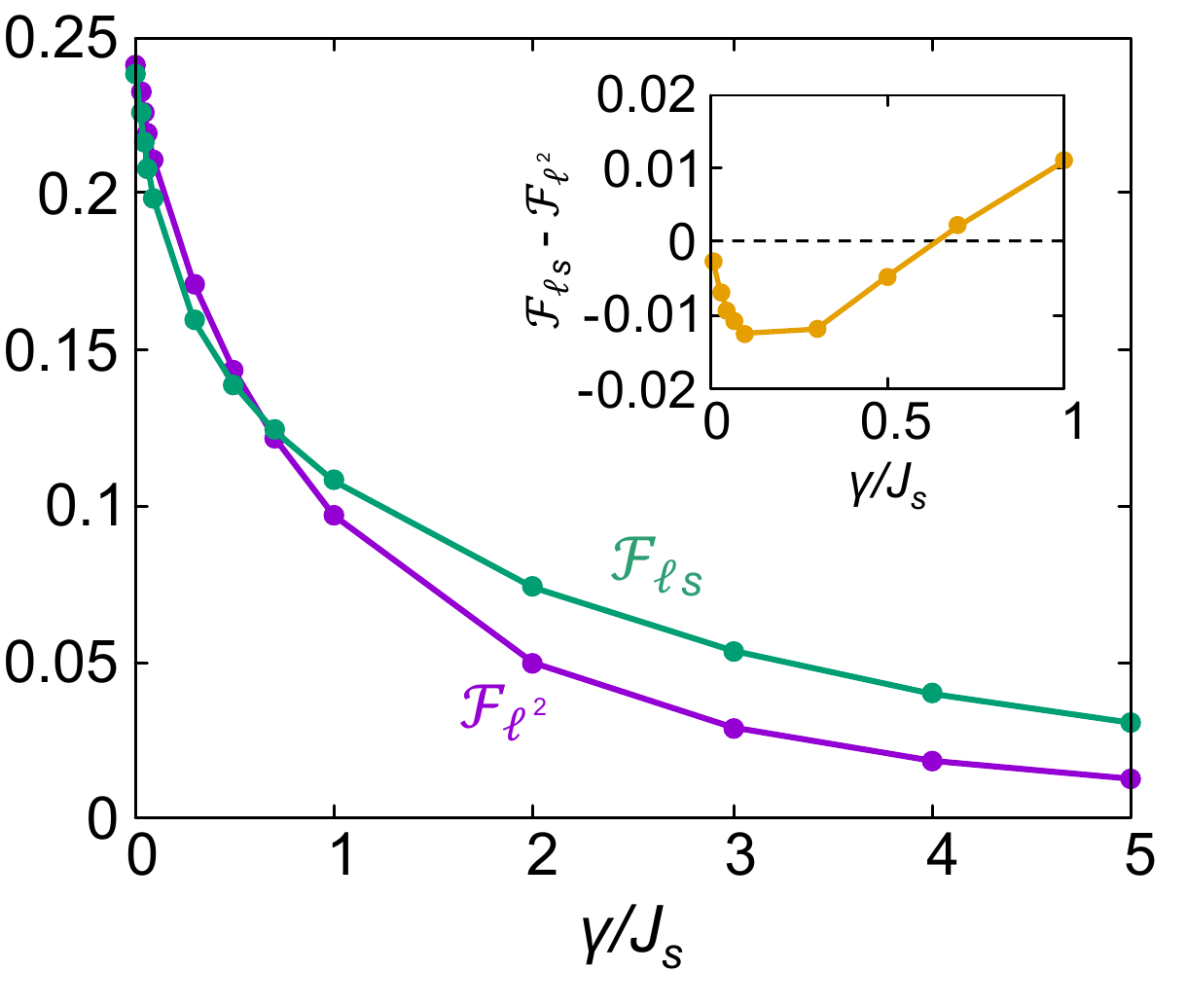}
    \caption{Spatial averages of fluctuations of the local particle number, $\mathcal{F}_{\ell^2}$ and $\mathcal{F}_{\ell s}$, against measurement strength on the $2/1$ approximant lattice. We take $L=120$ and $J_\ell/J_s=0.4$, and the number of trajectories is set to 1000.
    The inset shows \cred{$\mathcal{F}_{\ell s} - \mathcal{F}_{\ell^2}$}.}
    \label{fig: Fi_ave_1AC}
  \end{center}
\end{figure}

Figure~\ref{fig: Fi_phys_1AC} shows the spatial profile of fluctuations of the local particle number on the $2/1$ approximant lattice.
Unlike the Fibonacci chain, this periodic approximant lacks long-range quasiperiodic order, and the spatial distribution of $F_i$ is determined by the local bond configuration even in the weak-measurement regime.
Therefore, importantly, measurement-induced crossover of fluctuations of the local particle number found in Sec.~\ref{sec_Fiphysical} does not appear for periodic approximants, confirming that measurement-induced spatial nonuniformity for weak measurement on the Fibonacci chain is originating from the underlying quasiperiodic structure.
We also note that the spatial averages of $F_i$ in regions ${\mathbb V}^1_{0}$ and ${\mathbb V}^1_{-1} \cup {\mathbb V}^1_{1}$, namely $\mathcal F_{\ell^2}$ and $\mathcal F_{\ell s}$ introduced in Sec.~\ref{sec: perp}, exhibit characteristic behavior in Fig.~\ref{fig: Fi_ave_1AC}.
For weak measurement strengths $\gamma / J_s = 0.1$ and $0.5$, we see that $\mathcal{F}_{\ell^2} > \mathcal{F}_{\ell s}$, whereas $\mathcal{F}_{\ell s} > \mathcal{F}_{\ell^2}$ for strong measurement strengths $\gamma / J_s = 1$ and $3$.
Indeed, as shown in the inset in Fig.~\ref{fig: Fi_ave_1AC}, the relative magnitudes of $\mathcal F_{\ell^2}$ and $\mathcal F_{\ell s}$ are reversed around $\gamma/J_s \approx 0.7$.
Thus, though such a sign reversal of $\mathcal{F}_{\ell^2} - \mathcal{F}_{\ell s}$ was also observed in quasiperiodic systems in \cred{Fig.~\ref{fig: Fi_ave}}, we consider that this is driven purely by the local lattice structure rather than the quasiperiodicity.

\section{Reference structure factors of Fibonacci patterns}
\label{app:reference_structure}
\begin{figure}[htb]
    \centering
    \includegraphics[width=\linewidth]{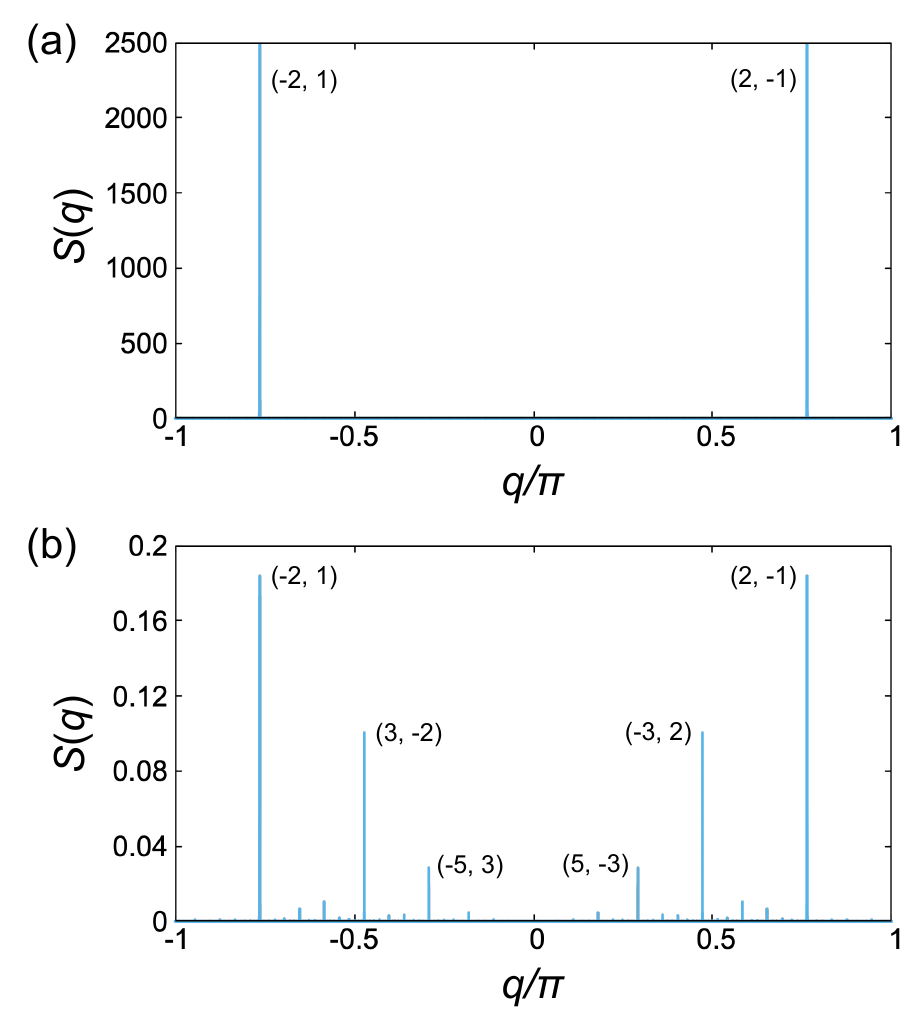}
    \caption{\cred{
    Reference structure factors of
    (a) the golden-mean modulation $d_j=\cos(2\pi j/\tau)$ and
    (b) a Fibonacci pattern in which $d_j=\alpha=0.01$ between two $\ell$-spacings and $d_j=\beta=0.03$ between an $\ell$-spacing and an $s$-spacing.
    The labels $(m,n)$ indicate the Bragg-peak positions $q_{m,n}=2\pi(m+n\tau)$ (mod $2\pi$).
    The system size is $L=10000$ for both panels.}
    }
    \label{fig:structure_reference}
\end{figure}

To facilitate the interpretation of the Bragg-peak structures discussed in Sec.~\ref{sec: bragg}, we consider two reference spatial patterns $\{d_j\}$.
Their structure factors are calculated by replacing $F_j$ and $F_{\rm space}$ in Eq.~\eqref{eq_Sq} with $d_j$ and $d_{\rm space}=L^{-1}\sum_{j=0}^{L-1}d_j$, respectively.
The first reference pattern is the golden-mean modulation
\begin{equation}
d_j=\cos\left(\frac{2\pi j}{\tau}\right).
\end{equation}
As shown in Fig.~\ref{fig:structure_reference}(a), its structure factor exhibits two Bragg peaks at $q=\pm 2\pi/\tau^2$.
The second reference pattern is determined solely by the nearest-neighbor lattice spacings of the Fibonacci chain.
For two distinct constants $\alpha$ and $\beta$, we define
\begin{equation}
d_j=
\begin{cases}
\alpha, & \text{between two $\ell$-spacings},\\
\beta, & \text{between $\ell$- and $s$-spacings}.
\end{cases}
\end{equation}
The Fibonacci structure is known to exhibit a characteristic hierarchy of Bragg peaks in reciprocal space~\cite{Godreche_1990}.
Since the Fourier transform is performed with respect to the site index $j$, the Bragg-peak positions are given by
\begin{equation}
q_{m,n}=2\pi(m+n\tau)\quad (\mathrm{mod}\ 2\pi),
\end{equation}
where $m$ and $n$ are integers.
The two integer indices reflect the two-dimensional parent lattice underlying the cut-and-project construction of the one-dimensional Fibonacci chain.
Figure~\ref{fig:structure_reference}(b) shows the corresponding structure factor for $\alpha=0.01$ and $\beta=0.03$.
The Bragg-peak hierarchy reflects the self-similar structure of the Fibonacci pattern.
For example, a sequence of prominent peaks appears at
\begin{eqnarray}
\frac{q_{2,-1}}{\pi}
&=&
\frac{2}{\tau^2}
\approx 0.76,\\
\frac{q_{-3,2}}{\pi}
&=&
\frac{2}{\tau^3}
\approx 0.47,\\
\frac{q_{5,-3}}{\pi}
&=&
\frac{2}{\tau^4}
\approx 0.29,
\end{eqnarray}
whose positions are successively related by a factor of $\tau$.
These reference patterns illustrate the difference between a single golden-mean modulation and the hierarchical Bragg-peak structure arising from the local lattice environments of the Fibonacci chain, discussed in Sec.~\ref{sec: bragg}.

\section{Comparison with the disordered system}
\label{app: disorder}
To clarify the qualitative difference between quasiperiodic and disordered systems, we consider a continuously monitored disordered hopping model described by
\begin{eqnarray}
  H =-\sum_{j=0}^{L-1} J^\prime_{j,j+1} \left[c_j^{\dagger} c_{j+1} + {\rm H.c.}\right],
  \label{eq: Hamiltonian_random}
\end{eqnarray}
where each hopping amplitude $J^\prime_{j,j+1}$ independently takes either $J_\ell$ or $J_s$ with equal probability.
For each system size, we generate a single realization of the random hopping configuration and use the same configuration throughout the calculations.
In particular, the random pattern is kept fixed when varying the measurement strength and quantum-trajectory realization.

\begin{figure*}[htb]
    \centering
    \includegraphics[width=\linewidth]{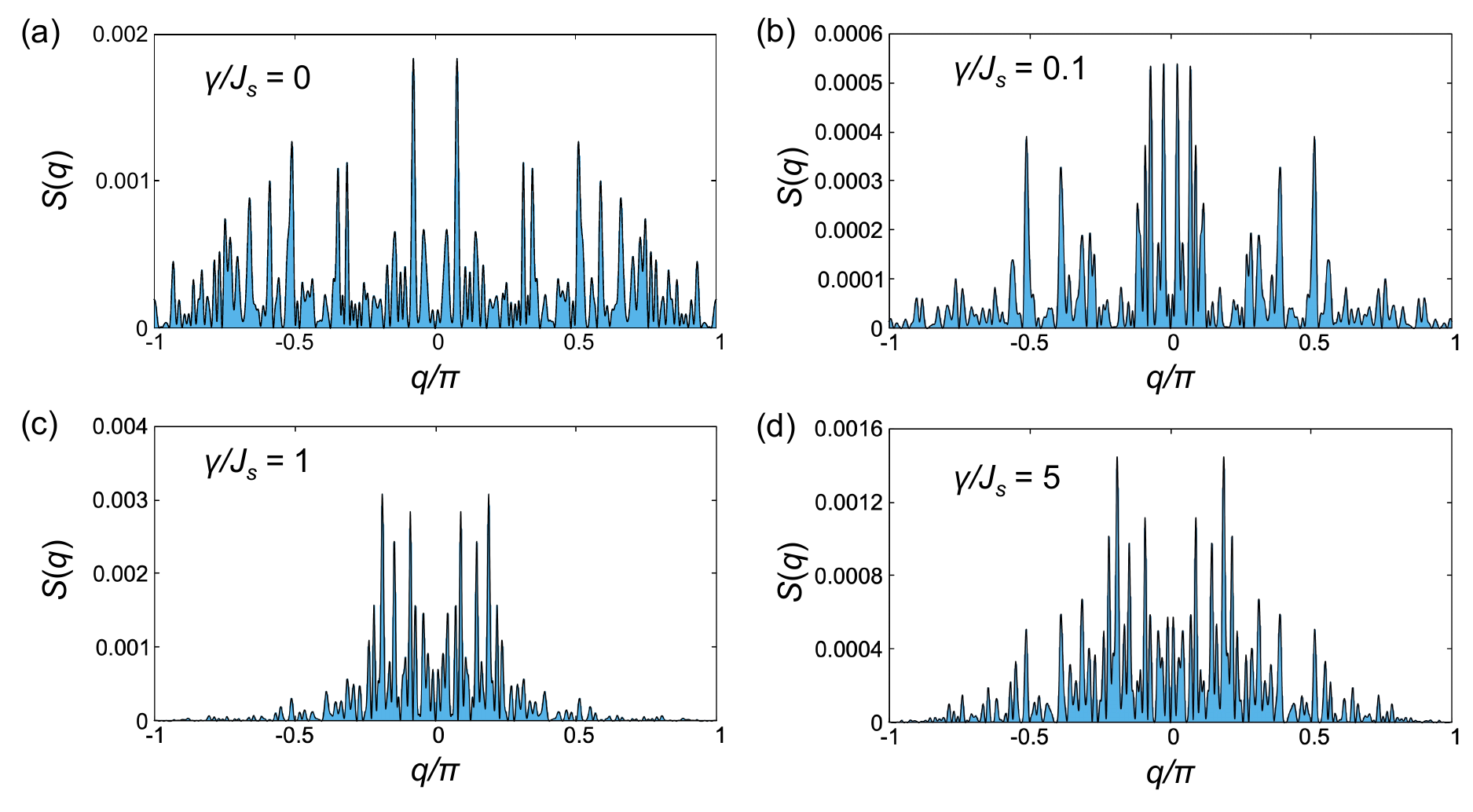}
    \caption{
    Structure factor of fluctuations of the local particle number in the disordered system at measurement strengths
    (a) $\gamma/J_s=0$,
    (b) $0.1$,
    (c) $1$, and
    (d) $5$.
    We take $J_\ell/J_s=0.4$ and $L=150$.
    The number of trajectories is set to 1000.
    }
    \label{fig:random_structure_factor}
\end{figure*}

We examine the structure factor of the spatial distribution of $F_i$ to distinguish the quasiperiodic and disordered systems.
Figure~\ref{fig:random_structure_factor} shows $S(q)$ in the disordered system for several measurement strengths.
Remarkably, no well-defined Bragg peaks are observed and the distribution is broadened for any of the measurement strengths.
This is in sharp contrast to the quasiperiodic system, where characteristic Bragg peaks appear under both weak and strong measurement, as shown in Fig.~\ref{fig:structure_F}.
In particular, under weak measurement, the quasiperiodic system exhibits dominant peaks at $q=\pm 2\pi/\tau^2$, reflecting the nontrivial long-range quasiperiodic order of the fluctuations.
Thus, the structure-factor analysis reveals a clear qualitative difference between the quasiperiodic and disordered systems, which persists even in the strong-measurement regime.

\bibliographystyle{apsrev4-2}
\bibliography{./refs}

\begin{thebibliography}{126}%
\makeatletter
\providecommand \@ifxundefined [1]{%
 \@ifx{#1\undefined}
}%
\providecommand \@ifnum [1]{%
 \ifnum #1\expandafter \@firstoftwo
 \else \expandafter \@secondoftwo
 \fi
}%
\providecommand \@ifx [1]{%
 \ifx #1\expandafter \@firstoftwo
 \else \expandafter \@secondoftwo
 \fi
}%
\providecommand \natexlab [1]{#1}%
\providecommand \enquote  [1]{``#1''}%
\providecommand \bibnamefont  [1]{#1}%
\providecommand \bibfnamefont [1]{#1}%
\providecommand \citenamefont [1]{#1}%
\providecommand \href@noop [0]{\@secondoftwo}%
\providecommand \href [0]{\begingroup \@sanitize@url \@href}%
\providecommand \@href[1]{\@@startlink{#1}\@@href}%
\providecommand \@@href[1]{\endgroup#1\@@endlink}%
\providecommand \@sanitize@url [0]{\catcode `\\12\catcode `\$12\catcode
  `\&12\catcode `\#12\catcode `\^12\catcode `\_12\catcode `\%12\relax}%
\providecommand \@@startlink[1]{}%
\providecommand \@@endlink[0]{}%
\providecommand \url  [0]{\begingroup\@sanitize@url \@url }%
\providecommand \@url [1]{\endgroup\@href {#1}{\urlprefix }}%
\providecommand \urlprefix  [0]{URL }%
\providecommand \Eprint [0]{\href }%
\providecommand \doibase [0]{https://doi.org/}%
\providecommand \selectlanguage [0]{\@gobble}%
\providecommand \bibinfo  [0]{\@secondoftwo}%
\providecommand \bibfield  [0]{\@secondoftwo}%
\providecommand \translation [1]{[#1]}%
\providecommand \BibitemOpen [0]{}%
\providecommand \bibitemStop [0]{}%
\providecommand \bibitemNoStop [0]{.\EOS\space}%
\providecommand \EOS [0]{\spacefactor3000\relax}%
\providecommand \BibitemShut  [1]{\csname bibitem#1\endcsname}%
\let\auto@bib@innerbib\@empty
\bibitem [{\citenamefont {Harrington}\ \emph {et~al.}(2022)\citenamefont
  {Harrington}, \citenamefont {Mueller},\ and\ \citenamefont
  {Murch}}]{Harrington22}%
  \BibitemOpen
  \bibfield  {author} {\bibinfo {author} {\bibfnamefont {P.~M.}\ \bibnamefont
  {Harrington}}, \bibinfo {author} {\bibfnamefont {E.~J.}\ \bibnamefont
  {Mueller}},\ and\ \bibinfo {author} {\bibfnamefont {K.~W.}\ \bibnamefont
  {Murch}},\ }\bibfield  {title} {\bibinfo {title} {Engineered dissipation for
  quantum information science},\ }\href
  {https://doi.org/10.1038/s42254-022-00494-8} {\bibfield  {journal} {\bibinfo
  {journal} {Nat. Rev. Phys.}\ }\textbf {\bibinfo {volume} {4}},\ \bibinfo
  {pages} {660} (\bibinfo {year} {2022})}\BibitemShut {NoStop}%
\bibitem [{\citenamefont {Fisher}\ \emph {et~al.}(2023)\citenamefont {Fisher},
  \citenamefont {Khemani}, \citenamefont {Nahum},\ and\ \citenamefont
  {Vijay}}]{Fisher23rev}%
  \BibitemOpen
  \bibfield  {author} {\bibinfo {author} {\bibfnamefont {M.~P.}\ \bibnamefont
  {Fisher}}, \bibinfo {author} {\bibfnamefont {V.}~\bibnamefont {Khemani}},
  \bibinfo {author} {\bibfnamefont {A.}~\bibnamefont {Nahum}},\ and\ \bibinfo
  {author} {\bibfnamefont {S.}~\bibnamefont {Vijay}},\ }\bibfield  {title}
  {\bibinfo {title} {Random quantum circuits},\ }\href
  {https://doi.org/10.1146/annurev-conmatphys-031720-030658} {\bibfield
  {journal} {\bibinfo  {journal} {Annu. Rev. Condens. Matter Phys.}\ }\textbf
  {\bibinfo {volume} {14}},\ \bibinfo {pages} {335} (\bibinfo {year}
  {2023})}\BibitemShut {NoStop}%
\bibitem [{\citenamefont {Potter}\ and\ \citenamefont
  {Vasseur}(2022)}]{Potter_2022}%
  \BibitemOpen
  \bibfield  {author} {\bibinfo {author} {\bibfnamefont {A.~C.}\ \bibnamefont
  {Potter}}\ and\ \bibinfo {author} {\bibfnamefont {R.}~\bibnamefont
  {Vasseur}},\ }\bibinfo {title} {Entanglement dynamics in hybrid quantum
  circuits},\ in\ \href {https://doi.org/10.1007/978-3-031-03998-0_9} {\emph
  {\bibinfo {booktitle} {Entanglement in Spin Chains: From Theory to Quantum
  Technology Applications}}},\ \bibinfo {editor} {edited by\ \bibinfo {editor}
  {\bibfnamefont {A.}~\bibnamefont {Bayat}}, \bibinfo {editor} {\bibfnamefont
  {S.}~\bibnamefont {Bose}},\ and\ \bibinfo {editor} {\bibfnamefont
  {H.}~\bibnamefont {Johannesson}}}\ (\bibinfo  {publisher} {Springer
  International Publishing},\ \bibinfo {address} {Cham},\ \bibinfo {year}
  {2022})\ p.\ \bibinfo {pages} {211}\BibitemShut {NoStop}%
\bibitem [{\citenamefont {Li}\ \emph {et~al.}(2018)\citenamefont {Li},
  \citenamefont {Chen},\ and\ \citenamefont {Fisher}}]{Li_2018}%
  \BibitemOpen
  \bibfield  {author} {\bibinfo {author} {\bibfnamefont {Y.}~\bibnamefont
  {Li}}, \bibinfo {author} {\bibfnamefont {X.}~\bibnamefont {Chen}},\ and\
  \bibinfo {author} {\bibfnamefont {M.~P.}\ \bibnamefont {Fisher}},\ }\bibfield
   {title} {\bibinfo {title} {Quantum zeno effect and the many-body
  entanglement transition},\ }\href
  {https://doi.org/10.1103/PhysRevB.98.205136} {\bibfield  {journal} {\bibinfo
  {journal} {Phys. Rev. B}\ }\textbf {\bibinfo {volume} {98}},\ \bibinfo
  {pages} {205136} (\bibinfo {year} {2018})}\BibitemShut {NoStop}%
\bibitem [{\citenamefont {Chan}\ \emph {et~al.}(2019)\citenamefont {Chan},
  \citenamefont {Nandkishore}, \citenamefont {Pretko},\ and\ \citenamefont
  {Smith}}]{Chan_2019}%
  \BibitemOpen
  \bibfield  {author} {\bibinfo {author} {\bibfnamefont {A.}~\bibnamefont
  {Chan}}, \bibinfo {author} {\bibfnamefont {R.~M.}\ \bibnamefont
  {Nandkishore}}, \bibinfo {author} {\bibfnamefont {M.}~\bibnamefont
  {Pretko}},\ and\ \bibinfo {author} {\bibfnamefont {G.}~\bibnamefont
  {Smith}},\ }\bibfield  {title} {\bibinfo {title} {Unitary-projective
  entanglement dynamics},\ }\href {https://doi.org/10.1103/PhysRevB.99.224307}
  {\bibfield  {journal} {\bibinfo  {journal} {Phys. Rev. B}\ }\textbf {\bibinfo
  {volume} {99}},\ \bibinfo {pages} {224307} (\bibinfo {year}
  {2019})}\BibitemShut {NoStop}%
\bibitem [{\citenamefont {Skinner}\ \emph {et~al.}(2019)\citenamefont
  {Skinner}, \citenamefont {Ruhman},\ and\ \citenamefont
  {Nahum}}]{Skinner_2019}%
  \BibitemOpen
  \bibfield  {author} {\bibinfo {author} {\bibfnamefont {B.}~\bibnamefont
  {Skinner}}, \bibinfo {author} {\bibfnamefont {J.}~\bibnamefont {Ruhman}},\
  and\ \bibinfo {author} {\bibfnamefont {A.}~\bibnamefont {Nahum}},\ }\bibfield
   {title} {\bibinfo {title} {Measurement-induced phase transitions in the
  dynamics of entanglement},\ }\href
  {https://doi.org/10.1103/PhysRevX.9.031009} {\bibfield  {journal} {\bibinfo
  {journal} {Phys. Rev. X}\ }\textbf {\bibinfo {volume} {9}},\ \bibinfo {pages}
  {031009} (\bibinfo {year} {2019})}\BibitemShut {NoStop}%
\bibitem [{\citenamefont {Szyniszewski}\ \emph {et~al.}(2019)\citenamefont
  {Szyniszewski}, \citenamefont {Romito},\ and\ \citenamefont
  {Schomerus}}]{Szyniszewski_2019}%
  \BibitemOpen
  \bibfield  {author} {\bibinfo {author} {\bibfnamefont {M.}~\bibnamefont
  {Szyniszewski}}, \bibinfo {author} {\bibfnamefont {A.}~\bibnamefont
  {Romito}},\ and\ \bibinfo {author} {\bibfnamefont {H.}~\bibnamefont
  {Schomerus}},\ }\bibfield  {title} {\bibinfo {title} {Entanglement transition
  from variable-strength weak measurements},\ }\href
  {https://doi.org/10.1103/PhysRevB.100.064204} {\bibfield  {journal} {\bibinfo
   {journal} {Phys. Rev. B}\ }\textbf {\bibinfo {volume} {100}},\ \bibinfo
  {pages} {064204} (\bibinfo {year} {2019})}\BibitemShut {NoStop}%
\bibitem [{\citenamefont {Bao}\ \emph {et~al.}(2020)\citenamefont {Bao},
  \citenamefont {Choi},\ and\ \citenamefont {Altman}}]{Bao_2020}%
  \BibitemOpen
  \bibfield  {author} {\bibinfo {author} {\bibfnamefont {Y.}~\bibnamefont
  {Bao}}, \bibinfo {author} {\bibfnamefont {S.}~\bibnamefont {Choi}},\ and\
  \bibinfo {author} {\bibfnamefont {E.}~\bibnamefont {Altman}},\ }\bibfield
  {title} {\bibinfo {title} {Theory of the phase transition in random unitary
  circuits with measurements},\ }\href
  {https://doi.org/10.1103/PhysRevB.101.104301} {\bibfield  {journal} {\bibinfo
   {journal} {Phys. Rev. B}\ }\textbf {\bibinfo {volume} {101}},\ \bibinfo
  {pages} {104301} (\bibinfo {year} {2020})}\BibitemShut {NoStop}%
\bibitem [{\citenamefont {Choi}\ \emph {et~al.}(2020)\citenamefont {Choi},
  \citenamefont {Bao}, \citenamefont {Qi},\ and\ \citenamefont
  {Altman}}]{Choi_2020}%
  \BibitemOpen
  \bibfield  {author} {\bibinfo {author} {\bibfnamefont {S.}~\bibnamefont
  {Choi}}, \bibinfo {author} {\bibfnamefont {Y.}~\bibnamefont {Bao}}, \bibinfo
  {author} {\bibfnamefont {X.-L.}\ \bibnamefont {Qi}},\ and\ \bibinfo {author}
  {\bibfnamefont {E.}~\bibnamefont {Altman}},\ }\bibfield  {title} {\bibinfo
  {title} {Quantum error correction in scrambling dynamics and
  measurement-induced phase transition},\ }\href
  {https://doi.org/10.1103/PhysRevLett.125.030505} {\bibfield  {journal}
  {\bibinfo  {journal} {Phys. Rev. Lett.}\ }\textbf {\bibinfo {volume} {125}},\
  \bibinfo {pages} {030505} (\bibinfo {year} {2020})}\BibitemShut {NoStop}%
\bibitem [{\citenamefont {Gullans}\ and\ \citenamefont
  {Huse}(2020{\natexlab{a}})}]{Gullans_2020a}%
  \BibitemOpen
  \bibfield  {author} {\bibinfo {author} {\bibfnamefont {M.~J.}\ \bibnamefont
  {Gullans}}\ and\ \bibinfo {author} {\bibfnamefont {D.~A.}\ \bibnamefont
  {Huse}},\ }\bibfield  {title} {\bibinfo {title} {Dynamical purification phase
  transition induced by quantum measurements},\ }\href
  {https://doi.org/10.1103/PhysRevX.10.041020} {\bibfield  {journal} {\bibinfo
  {journal} {Phys. Rev. X}\ }\textbf {\bibinfo {volume} {10}},\ \bibinfo
  {pages} {041020} (\bibinfo {year} {2020}{\natexlab{a}})}\BibitemShut
  {NoStop}%
\bibitem [{\citenamefont {Gullans}\ and\ \citenamefont
  {Huse}(2020{\natexlab{b}})}]{Gullans_2020b}%
  \BibitemOpen
  \bibfield  {author} {\bibinfo {author} {\bibfnamefont {M.~J.}\ \bibnamefont
  {Gullans}}\ and\ \bibinfo {author} {\bibfnamefont {D.~A.}\ \bibnamefont
  {Huse}},\ }\bibfield  {title} {\bibinfo {title} {Scalable probes of
  measurement-induced criticality},\ }\href
  {https://doi.org/10.1103/PhysRevLett.125.070606} {\bibfield  {journal}
  {\bibinfo  {journal} {Phys. Rev. Lett.}\ }\textbf {\bibinfo {volume} {125}},\
  \bibinfo {pages} {070606} (\bibinfo {year} {2020}{\natexlab{b}})}\BibitemShut
  {NoStop}%
\bibitem [{\citenamefont {Jian}\ \emph {et~al.}(2020)\citenamefont {Jian},
  \citenamefont {You}, \citenamefont {Vasseur},\ and\ \citenamefont
  {Ludwig}}]{Jian_2020}%
  \BibitemOpen
  \bibfield  {author} {\bibinfo {author} {\bibfnamefont {C.-M.}\ \bibnamefont
  {Jian}}, \bibinfo {author} {\bibfnamefont {Y.-Z.}\ \bibnamefont {You}},
  \bibinfo {author} {\bibfnamefont {R.}~\bibnamefont {Vasseur}},\ and\ \bibinfo
  {author} {\bibfnamefont {A.~W.~W.}\ \bibnamefont {Ludwig}},\ }\bibfield
  {title} {\bibinfo {title} {Measurement-induced criticality in random quantum
  circuits},\ }\href {https://doi.org/10.1103/PhysRevB.101.104302} {\bibfield
  {journal} {\bibinfo  {journal} {Phys. Rev. B}\ }\textbf {\bibinfo {volume}
  {101}},\ \bibinfo {pages} {104302} (\bibinfo {year} {2020})}\BibitemShut
  {NoStop}%
\bibitem [{\citenamefont {Zabalo}\ \emph {et~al.}(2020)\citenamefont {Zabalo},
  \citenamefont {Gullans}, \citenamefont {Wilson}, \citenamefont
  {Gopalakrishnan}, \citenamefont {Huse},\ and\ \citenamefont
  {Pixley}}]{Zabalo_2020}%
  \BibitemOpen
  \bibfield  {author} {\bibinfo {author} {\bibfnamefont {A.}~\bibnamefont
  {Zabalo}}, \bibinfo {author} {\bibfnamefont {M.~J.}\ \bibnamefont {Gullans}},
  \bibinfo {author} {\bibfnamefont {J.~H.}\ \bibnamefont {Wilson}}, \bibinfo
  {author} {\bibfnamefont {S.}~\bibnamefont {Gopalakrishnan}}, \bibinfo
  {author} {\bibfnamefont {D.~A.}\ \bibnamefont {Huse}},\ and\ \bibinfo
  {author} {\bibfnamefont {J.~H.}\ \bibnamefont {Pixley}},\ }\bibfield  {title}
  {\bibinfo {title} {Critical properties of the measurement-induced transition
  in random quantum circuits},\ }\href
  {https://doi.org/10.1103/PhysRevB.101.060301} {\bibfield  {journal} {\bibinfo
   {journal} {Phys. Rev. B}\ }\textbf {\bibinfo {volume} {101}},\ \bibinfo
  {pages} {060301(R)} (\bibinfo {year} {2020})}\BibitemShut {NoStop}%
\bibitem [{\citenamefont {Iaconis}\ \emph {et~al.}(2020)\citenamefont
  {Iaconis}, \citenamefont {Lucas},\ and\ \citenamefont {Chen}}]{Iaconis_2020}%
  \BibitemOpen
  \bibfield  {author} {\bibinfo {author} {\bibfnamefont {J.}~\bibnamefont
  {Iaconis}}, \bibinfo {author} {\bibfnamefont {A.}~\bibnamefont {Lucas}},\
  and\ \bibinfo {author} {\bibfnamefont {X.}~\bibnamefont {Chen}},\ }\bibfield
  {title} {\bibinfo {title} {Measurement-induced phase transitions in quantum
  automaton circuits},\ }\href {https://doi.org/10.1103/PhysRevB.102.224311}
  {\bibfield  {journal} {\bibinfo  {journal} {Phys. Rev. B}\ }\textbf {\bibinfo
  {volume} {102}},\ \bibinfo {pages} {224311} (\bibinfo {year}
  {2020})}\BibitemShut {NoStop}%
\bibitem [{\citenamefont {Turkeshi}\ \emph {et~al.}(2020)\citenamefont
  {Turkeshi}, \citenamefont {Fazio},\ and\ \citenamefont
  {Dalmonte}}]{Turkeshi_2020}%
  \BibitemOpen
  \bibfield  {author} {\bibinfo {author} {\bibfnamefont {X.}~\bibnamefont
  {Turkeshi}}, \bibinfo {author} {\bibfnamefont {R.}~\bibnamefont {Fazio}},\
  and\ \bibinfo {author} {\bibfnamefont {M.}~\bibnamefont {Dalmonte}},\
  }\bibfield  {title} {\bibinfo {title} {Measurement-induced criticality in
  $(2+1)$-dimensional hybrid quantum circuits},\ }\href
  {https://doi.org/10.1103/PhysRevB.102.014315} {\bibfield  {journal} {\bibinfo
   {journal} {Phys. Rev. B}\ }\textbf {\bibinfo {volume} {102}},\ \bibinfo
  {pages} {014315} (\bibinfo {year} {2020})}\BibitemShut {NoStop}%
\bibitem [{\citenamefont {Zhang}\ \emph {et~al.}(2020)\citenamefont {Zhang},
  \citenamefont {Reyes}, \citenamefont {Kourtis}, \citenamefont {Chamon},
  \citenamefont {Mucciolo},\ and\ \citenamefont {Ruckenstein}}]{Zhang_2020}%
  \BibitemOpen
  \bibfield  {author} {\bibinfo {author} {\bibfnamefont {L.}~\bibnamefont
  {Zhang}}, \bibinfo {author} {\bibfnamefont {J.~A.}\ \bibnamefont {Reyes}},
  \bibinfo {author} {\bibfnamefont {S.}~\bibnamefont {Kourtis}}, \bibinfo
  {author} {\bibfnamefont {C.}~\bibnamefont {Chamon}}, \bibinfo {author}
  {\bibfnamefont {E.~R.}\ \bibnamefont {Mucciolo}},\ and\ \bibinfo {author}
  {\bibfnamefont {A.~E.}\ \bibnamefont {Ruckenstein}},\ }\bibfield  {title}
  {\bibinfo {title} {Nonuniversal entanglement level statistics in
  projection-driven quantum circuits},\ }\href
  {https://doi.org/10.1103/PhysRevB.101.235104} {\bibfield  {journal} {\bibinfo
   {journal} {Phys. Rev. B}\ }\textbf {\bibinfo {volume} {101}},\ \bibinfo
  {pages} {235104} (\bibinfo {year} {2020})}\BibitemShut {NoStop}%
\bibitem [{\citenamefont {Szyniszewski}\ \emph {et~al.}(2020)\citenamefont
  {Szyniszewski}, \citenamefont {Romito},\ and\ \citenamefont
  {Schomerus}}]{Szyniszewski_2020}%
  \BibitemOpen
  \bibfield  {author} {\bibinfo {author} {\bibfnamefont {M.}~\bibnamefont
  {Szyniszewski}}, \bibinfo {author} {\bibfnamefont {A.}~\bibnamefont
  {Romito}},\ and\ \bibinfo {author} {\bibfnamefont {H.}~\bibnamefont
  {Schomerus}},\ }\bibfield  {title} {\bibinfo {title} {Universality of
  entanglement transitions from stroboscopic to continuous measurements},\
  }\href {https://doi.org/10.1103/PhysRevLett.125.210602} {\bibfield  {journal}
  {\bibinfo  {journal} {Phys. Rev. Lett.}\ }\textbf {\bibinfo {volume} {125}},\
  \bibinfo {pages} {210602} (\bibinfo {year} {2020})}\BibitemShut {NoStop}%
\bibitem [{\citenamefont {Nahum}\ \emph {et~al.}(2021)\citenamefont {Nahum},
  \citenamefont {Roy}, \citenamefont {Skinner},\ and\ \citenamefont
  {Ruhman}}]{Nahum_2021}%
  \BibitemOpen
  \bibfield  {author} {\bibinfo {author} {\bibfnamefont {A.}~\bibnamefont
  {Nahum}}, \bibinfo {author} {\bibfnamefont {S.}~\bibnamefont {Roy}}, \bibinfo
  {author} {\bibfnamefont {B.}~\bibnamefont {Skinner}},\ and\ \bibinfo {author}
  {\bibfnamefont {J.}~\bibnamefont {Ruhman}},\ }\bibfield  {title} {\bibinfo
  {title} {Measurement and entanglement phase transitions in all-to-all quantum
  circuits, on quantum trees, and in {Landau-Ginsburg} theory},\ }\href
  {https://doi.org/10.1103/PRXQuantum.2.010352} {\bibfield  {journal} {\bibinfo
   {journal} {PRX Quantum}\ }\textbf {\bibinfo {volume} {2}},\ \bibinfo {pages}
  {010352} (\bibinfo {year} {2021})}\BibitemShut {NoStop}%
\bibitem [{\citenamefont {Ippoliti}\ \emph {et~al.}(2021)\citenamefont
  {Ippoliti}, \citenamefont {Gullans}, \citenamefont {Gopalakrishnan},
  \citenamefont {Huse},\ and\ \citenamefont {Khemani}}]{Ippoliti_2021a}%
  \BibitemOpen
  \bibfield  {author} {\bibinfo {author} {\bibfnamefont {M.}~\bibnamefont
  {Ippoliti}}, \bibinfo {author} {\bibfnamefont {M.~J.}\ \bibnamefont
  {Gullans}}, \bibinfo {author} {\bibfnamefont {S.}~\bibnamefont
  {Gopalakrishnan}}, \bibinfo {author} {\bibfnamefont {D.~A.}\ \bibnamefont
  {Huse}},\ and\ \bibinfo {author} {\bibfnamefont {V.}~\bibnamefont
  {Khemani}},\ }\bibfield  {title} {\bibinfo {title} {Entanglement phase
  transitions in measurement-only dynamics},\ }\href
  {https://doi.org/10.1103/PhysRevX.11.011030} {\bibfield  {journal} {\bibinfo
  {journal} {Phys. Rev. X}\ }\textbf {\bibinfo {volume} {11}},\ \bibinfo
  {pages} {011030} (\bibinfo {year} {2021})}\BibitemShut {NoStop}%
\bibitem [{\citenamefont {Ippoliti}\ and\ \citenamefont
  {Khemani}(2021)}]{Ippoliti_2021b}%
  \BibitemOpen
  \bibfield  {author} {\bibinfo {author} {\bibfnamefont {M.}~\bibnamefont
  {Ippoliti}}\ and\ \bibinfo {author} {\bibfnamefont {V.}~\bibnamefont
  {Khemani}},\ }\bibfield  {title} {\bibinfo {title} {Postselection-free
  entanglement dynamics via spacetime duality},\ }\href
  {https://doi.org/10.1103/PhysRevLett.126.060501} {\bibfield  {journal}
  {\bibinfo  {journal} {Phys. Rev. Lett.}\ }\textbf {\bibinfo {volume} {126}},\
  \bibinfo {pages} {060501} (\bibinfo {year} {2021})}\BibitemShut {NoStop}%
\bibitem [{\citenamefont {Lavasani}\ \emph
  {et~al.}(2021{\natexlab{a}})\citenamefont {Lavasani}, \citenamefont
  {Alavirad},\ and\ \citenamefont {Barkeshli}}]{Lavasani_2021a}%
  \BibitemOpen
  \bibfield  {author} {\bibinfo {author} {\bibfnamefont {A.}~\bibnamefont
  {Lavasani}}, \bibinfo {author} {\bibfnamefont {Y.}~\bibnamefont {Alavirad}},\
  and\ \bibinfo {author} {\bibfnamefont {M.}~\bibnamefont {Barkeshli}},\
  }\bibfield  {title} {\bibinfo {title} {Measurement-induced topological
  entanglement transitions in symmetric random quantum circuits},\ }\href
  {https://doi.org/10.1038/s41567-020-01112-z} {\bibfield  {journal} {\bibinfo
  {journal} {Nat. Phys.}\ }\textbf {\bibinfo {volume} {17}},\ \bibinfo {pages}
  {342} (\bibinfo {year} {2021}{\natexlab{a}})}\BibitemShut {NoStop}%
\bibitem [{\citenamefont {Lavasani}\ \emph
  {et~al.}(2021{\natexlab{b}})\citenamefont {Lavasani}, \citenamefont
  {Alavirad},\ and\ \citenamefont {Barkeshli}}]{Lavasani_2021b}%
  \BibitemOpen
  \bibfield  {author} {\bibinfo {author} {\bibfnamefont {A.}~\bibnamefont
  {Lavasani}}, \bibinfo {author} {\bibfnamefont {Y.}~\bibnamefont {Alavirad}},\
  and\ \bibinfo {author} {\bibfnamefont {M.}~\bibnamefont {Barkeshli}},\
  }\bibfield  {title} {\bibinfo {title} {Topological order and criticality in
  $(2+1)\mathrm{D}$ monitored random quantum circuits},\ }\href
  {https://doi.org/10.1103/PhysRevLett.127.235701} {\bibfield  {journal}
  {\bibinfo  {journal} {Phys. Rev. Lett.}\ }\textbf {\bibinfo {volume} {127}},\
  \bibinfo {pages} {235701} (\bibinfo {year} {2021}{\natexlab{b}})}\BibitemShut
  {NoStop}%
\bibitem [{\citenamefont {Sang}\ and\ \citenamefont {Hsieh}(2021)}]{Sang_2021}%
  \BibitemOpen
  \bibfield  {author} {\bibinfo {author} {\bibfnamefont {S.}~\bibnamefont
  {Sang}}\ and\ \bibinfo {author} {\bibfnamefont {T.~H.}\ \bibnamefont
  {Hsieh}},\ }\bibfield  {title} {\bibinfo {title} {Measurement-protected
  quantum phases},\ }\href {https://doi.org/10.1103/PhysRevResearch.3.023200}
  {\bibfield  {journal} {\bibinfo  {journal} {Phys. Rev. Research}\ }\textbf
  {\bibinfo {volume} {3}},\ \bibinfo {pages} {023200} (\bibinfo {year}
  {2021})}\BibitemShut {NoStop}%
\bibitem [{\citenamefont {Block}\ \emph {et~al.}(2022)\citenamefont {Block},
  \citenamefont {Bao}, \citenamefont {Choi}, \citenamefont {Altman},\ and\
  \citenamefont {Yao}}]{Block_2022}%
  \BibitemOpen
  \bibfield  {author} {\bibinfo {author} {\bibfnamefont {M.}~\bibnamefont
  {Block}}, \bibinfo {author} {\bibfnamefont {Y.}~\bibnamefont {Bao}}, \bibinfo
  {author} {\bibfnamefont {S.}~\bibnamefont {Choi}}, \bibinfo {author}
  {\bibfnamefont {E.}~\bibnamefont {Altman}},\ and\ \bibinfo {author}
  {\bibfnamefont {N.~Y.}\ \bibnamefont {Yao}},\ }\bibfield  {title} {\bibinfo
  {title} {Measurement-induced transition in long-range interacting quantum
  circuits},\ }\href {https://doi.org/10.1103/PhysRevLett.128.010604}
  {\bibfield  {journal} {\bibinfo  {journal} {Phys. Rev. Lett.}\ }\textbf
  {\bibinfo {volume} {128}},\ \bibinfo {pages} {010604} (\bibinfo {year}
  {2022})}\BibitemShut {NoStop}%
\bibitem [{\citenamefont {Sharma}\ \emph {et~al.}(2022)\citenamefont {Sharma},
  \citenamefont {Turkeshi}, \citenamefont {Fazio},\ and\ \citenamefont
  {Dalmonte}}]{Sharma_2022}%
  \BibitemOpen
  \bibfield  {author} {\bibinfo {author} {\bibfnamefont {S.}~\bibnamefont
  {Sharma}}, \bibinfo {author} {\bibfnamefont {X.}~\bibnamefont {Turkeshi}},
  \bibinfo {author} {\bibfnamefont {R.}~\bibnamefont {Fazio}},\ and\ \bibinfo
  {author} {\bibfnamefont {M.}~\bibnamefont {Dalmonte}},\ }\bibfield  {title}
  {\bibinfo {title} {{Measurement-induced criticality in extended and
  long-range unitary circuits}},\ }\href
  {https://doi.org/10.21468/SciPostPhysCore.5.2.023} {\bibfield  {journal}
  {\bibinfo  {journal} {SciPost Phys. Core}\ }\textbf {\bibinfo {volume} {5}},\
  \bibinfo {pages} {023} (\bibinfo {year} {2022})}\BibitemShut {NoStop}%
\bibitem [{\citenamefont {Agrawal}\ \emph {et~al.}(2022)\citenamefont
  {Agrawal}, \citenamefont {Zabalo}, \citenamefont {Chen}, \citenamefont
  {Wilson}, \citenamefont {Potter}, \citenamefont {Pixley}, \citenamefont
  {Gopalakrishnan},\ and\ \citenamefont {Vasseur}}]{Agrawal_2022}%
  \BibitemOpen
  \bibfield  {author} {\bibinfo {author} {\bibfnamefont {U.}~\bibnamefont
  {Agrawal}}, \bibinfo {author} {\bibfnamefont {A.}~\bibnamefont {Zabalo}},
  \bibinfo {author} {\bibfnamefont {K.}~\bibnamefont {Chen}}, \bibinfo {author}
  {\bibfnamefont {J.~H.}\ \bibnamefont {Wilson}}, \bibinfo {author}
  {\bibfnamefont {A.~C.}\ \bibnamefont {Potter}}, \bibinfo {author}
  {\bibfnamefont {J.~H.}\ \bibnamefont {Pixley}}, \bibinfo {author}
  {\bibfnamefont {S.}~\bibnamefont {Gopalakrishnan}},\ and\ \bibinfo {author}
  {\bibfnamefont {R.}~\bibnamefont {Vasseur}},\ }\bibfield  {title} {\bibinfo
  {title} {Entanglement and charge-sharpening transitions in {U(1)} symmetric
  monitored quantum circuits},\ }\href
  {https://doi.org/10.1103/PhysRevX.12.041002} {\bibfield  {journal} {\bibinfo
  {journal} {Phys. Rev. X}\ }\textbf {\bibinfo {volume} {12}},\ \bibinfo
  {pages} {041002} (\bibinfo {year} {2022})}\BibitemShut {NoStop}%
\bibitem [{\citenamefont {Barratt}\ \emph {et~al.}(2022)\citenamefont
  {Barratt}, \citenamefont {Agrawal}, \citenamefont {Gopalakrishnan},
  \citenamefont {Huse}, \citenamefont {Vasseur},\ and\ \citenamefont
  {Potter}}]{Barratt_2022}%
  \BibitemOpen
  \bibfield  {author} {\bibinfo {author} {\bibfnamefont {F.}~\bibnamefont
  {Barratt}}, \bibinfo {author} {\bibfnamefont {U.}~\bibnamefont {Agrawal}},
  \bibinfo {author} {\bibfnamefont {S.}~\bibnamefont {Gopalakrishnan}},
  \bibinfo {author} {\bibfnamefont {D.~A.}\ \bibnamefont {Huse}}, \bibinfo
  {author} {\bibfnamefont {R.}~\bibnamefont {Vasseur}},\ and\ \bibinfo {author}
  {\bibfnamefont {A.~C.}\ \bibnamefont {Potter}},\ }\bibfield  {title}
  {\bibinfo {title} {Field theory of charge sharpening in symmetric monitored
  quantum circuits},\ }\href {https://doi.org/10.1103/PhysRevLett.129.120604}
  {\bibfield  {journal} {\bibinfo  {journal} {Phys. Rev. Lett.}\ }\textbf
  {\bibinfo {volume} {129}},\ \bibinfo {pages} {120604} (\bibinfo {year}
  {2022})}\BibitemShut {NoStop}%
\bibitem [{\citenamefont {Kelly}\ \emph {et~al.}(2023)\citenamefont {Kelly},
  \citenamefont {Poschinger}, \citenamefont {Schmidt-Kaler}, \citenamefont
  {Fisher},\ and\ \citenamefont {Marino}}]{Kelly_2023}%
  \BibitemOpen
  \bibfield  {author} {\bibinfo {author} {\bibfnamefont {S.~P.}\ \bibnamefont
  {Kelly}}, \bibinfo {author} {\bibfnamefont {U.}~\bibnamefont {Poschinger}},
  \bibinfo {author} {\bibfnamefont {F.}~\bibnamefont {Schmidt-Kaler}}, \bibinfo
  {author} {\bibfnamefont {M.~P.~A.}\ \bibnamefont {Fisher}},\ and\ \bibinfo
  {author} {\bibfnamefont {J.}~\bibnamefont {Marino}},\ }\bibfield  {title}
  {\bibinfo {title} {{Coherence requirements for quantum communication from
  hybrid circuit dynamics}},\ }\href
  {https://doi.org/10.21468/SciPostPhys.15.6.250} {\bibfield  {journal}
  {\bibinfo  {journal} {SciPost Phys.}\ }\textbf {\bibinfo {volume} {15}},\
  \bibinfo {pages} {250} (\bibinfo {year} {2023})}\BibitemShut {NoStop}%
\bibitem [{\citenamefont {Ha}\ \emph {et~al.}(2024)\citenamefont {Ha},
  \citenamefont {Pandey}, \citenamefont {Gopalakrishnan},\ and\ \citenamefont
  {Huse}}]{Ha_2024}%
  \BibitemOpen
  \bibfield  {author} {\bibinfo {author} {\bibfnamefont {H.}~\bibnamefont
  {Ha}}, \bibinfo {author} {\bibfnamefont {A.}~\bibnamefont {Pandey}}, \bibinfo
  {author} {\bibfnamefont {S.}~\bibnamefont {Gopalakrishnan}},\ and\ \bibinfo
  {author} {\bibfnamefont {D.~A.}\ \bibnamefont {Huse}},\ }\bibfield  {title}
  {\bibinfo {title} {Measurement-induced phase transitions in systems with
  diffusive dynamics},\ }\href {https://doi.org/10.1103/PhysRevB.110.L140301}
  {\bibfield  {journal} {\bibinfo  {journal} {Phys. Rev. B}\ }\textbf {\bibinfo
  {volume} {110}},\ \bibinfo {pages} {L140301} (\bibinfo {year}
  {2024})}\BibitemShut {NoStop}%
\bibitem [{\citenamefont {M{\"u}ller}\ \emph {et~al.}(2012)\citenamefont
  {M{\"u}ller}, \citenamefont {Diehl}, \citenamefont {Pupillo},\ and\
  \citenamefont {Zoller}}]{Muller_2012}%
  \BibitemOpen
  \bibfield  {author} {\bibinfo {author} {\bibfnamefont {M.}~\bibnamefont
  {M{\"u}ller}}, \bibinfo {author} {\bibfnamefont {S.}~\bibnamefont {Diehl}},
  \bibinfo {author} {\bibfnamefont {G.}~\bibnamefont {Pupillo}},\ and\ \bibinfo
  {author} {\bibfnamefont {P.}~\bibnamefont {Zoller}},\ }\bibfield  {title}
  {\bibinfo {title} {Engineered open systems and quantum simulations with atoms
  and ions},\ }\href
  {https://doi.org/https://doi.org/10.1016/B978-0-12-396482-3.00001-6}
  {\bibfield  {journal} {\bibinfo  {journal} {Adv. Atom. Mol. Opt. Phys.}\
  }\textbf {\bibinfo {volume} {61}},\ \bibinfo {pages} {1} (\bibinfo {year}
  {2012})}\BibitemShut {NoStop}%
\bibitem [{\citenamefont {Tang}\ and\ \citenamefont {Zhu}(2020)}]{Tang_2020}%
  \BibitemOpen
  \bibfield  {author} {\bibinfo {author} {\bibfnamefont {Q.}~\bibnamefont
  {Tang}}\ and\ \bibinfo {author} {\bibfnamefont {W.}~\bibnamefont {Zhu}},\
  }\bibfield  {title} {\bibinfo {title} {Measurement-induced phase transition:
  A case study in the nonintegrable model by density-matrix renormalization
  group calculations},\ }\href
  {https://doi.org/10.1103/PhysRevResearch.2.013022} {\bibfield  {journal}
  {\bibinfo  {journal} {Phys. Rev. Res.}\ }\textbf {\bibinfo {volume} {2}},\
  \bibinfo {pages} {013022} (\bibinfo {year} {2020})}\BibitemShut {NoStop}%
\bibitem [{\citenamefont {Goto}\ and\ \citenamefont
  {Danshita}(2020)}]{Goto_2020}%
  \BibitemOpen
  \bibfield  {author} {\bibinfo {author} {\bibfnamefont {S.}~\bibnamefont
  {Goto}}\ and\ \bibinfo {author} {\bibfnamefont {I.}~\bibnamefont
  {Danshita}},\ }\bibfield  {title} {\bibinfo {title} {Measurement-induced
  transitions of the entanglement scaling law in ultracold gases with
  controllable dissipation},\ }\href
  {https://doi.org/10.1103/PhysRevA.102.033316} {\bibfield  {journal} {\bibinfo
   {journal} {Phys. Rev. A}\ }\textbf {\bibinfo {volume} {102}},\ \bibinfo
  {pages} {033316} (\bibinfo {year} {2020})}\BibitemShut {NoStop}%
\bibitem [{\citenamefont {Fuji}\ and\ \citenamefont
  {Ashida}(2020)}]{Fuji_2020}%
  \BibitemOpen
  \bibfield  {author} {\bibinfo {author} {\bibfnamefont {Y.}~\bibnamefont
  {Fuji}}\ and\ \bibinfo {author} {\bibfnamefont {Y.}~\bibnamefont {Ashida}},\
  }\bibfield  {title} {\bibinfo {title} {Measurement-induced quantum
  criticality under continuous monitoring},\ }\href
  {https://doi.org/10.1103/PhysRevB.102.054302} {\bibfield  {journal} {\bibinfo
   {journal} {Phys. Rev. B}\ }\textbf {\bibinfo {volume} {102}},\ \bibinfo
  {pages} {054302} (\bibinfo {year} {2020})}\BibitemShut {NoStop}%
\bibitem [{\citenamefont {Doggen}\ \emph {et~al.}(2022)\citenamefont {Doggen},
  \citenamefont {Gefen}, \citenamefont {Gornyi}, \citenamefont {Mirlin},\ and\
  \citenamefont {Polyakov}}]{Doggen_2022}%
  \BibitemOpen
  \bibfield  {author} {\bibinfo {author} {\bibfnamefont {E.~V.~H.}\
  \bibnamefont {Doggen}}, \bibinfo {author} {\bibfnamefont {Y.}~\bibnamefont
  {Gefen}}, \bibinfo {author} {\bibfnamefont {I.~V.}\ \bibnamefont {Gornyi}},
  \bibinfo {author} {\bibfnamefont {A.~D.}\ \bibnamefont {Mirlin}},\ and\
  \bibinfo {author} {\bibfnamefont {D.~G.}\ \bibnamefont {Polyakov}},\
  }\bibfield  {title} {\bibinfo {title} {Generalized quantum measurements with
  matrix product states: Entanglement phase transition and clusterization},\
  }\href {https://doi.org/10.1103/PhysRevResearch.4.023146} {\bibfield
  {journal} {\bibinfo  {journal} {Phys. Rev. Res.}\ }\textbf {\bibinfo {volume}
  {4}},\ \bibinfo {pages} {023146} (\bibinfo {year} {2022})}\BibitemShut
  {NoStop}%
\bibitem [{\citenamefont {Doggen}\ \emph {et~al.}(2023)\citenamefont {Doggen},
  \citenamefont {Gefen}, \citenamefont {Gornyi}, \citenamefont {Mirlin},\ and\
  \citenamefont {Polyakov}}]{Doggen_2023}%
  \BibitemOpen
  \bibfield  {author} {\bibinfo {author} {\bibfnamefont {E.~V.~H.}\
  \bibnamefont {Doggen}}, \bibinfo {author} {\bibfnamefont {Y.}~\bibnamefont
  {Gefen}}, \bibinfo {author} {\bibfnamefont {I.~V.}\ \bibnamefont {Gornyi}},
  \bibinfo {author} {\bibfnamefont {A.~D.}\ \bibnamefont {Mirlin}},\ and\
  \bibinfo {author} {\bibfnamefont {D.~G.}\ \bibnamefont {Polyakov}},\
  }\bibfield  {title} {\bibinfo {title} {Evolution of many-body systems under
  ancilla quantum measurements},\ }\href
  {https://doi.org/10.1103/PhysRevB.107.214203} {\bibfield  {journal} {\bibinfo
   {journal} {Phys. Rev. B}\ }\textbf {\bibinfo {volume} {107}},\ \bibinfo
  {pages} {214203} (\bibinfo {year} {2023})}\BibitemShut {NoStop}%
\bibitem [{\citenamefont {Lunt}\ and\ \citenamefont {Pal}(2020)}]{Lunt_2020}%
  \BibitemOpen
  \bibfield  {author} {\bibinfo {author} {\bibfnamefont {O.}~\bibnamefont
  {Lunt}}\ and\ \bibinfo {author} {\bibfnamefont {A.}~\bibnamefont {Pal}},\
  }\bibfield  {title} {\bibinfo {title} {Measurement-induced entanglement
  transitions in many-body localized systems},\ }\href
  {https://doi.org/10.1103/PhysRevResearch.2.043072} {\bibfield  {journal}
  {\bibinfo  {journal} {Phys. Rev. Res.}\ }\textbf {\bibinfo {volume} {2}},\
  \bibinfo {pages} {043072} (\bibinfo {year} {2020})}\BibitemShut {NoStop}%
\bibitem [{\citenamefont {De~Tomasi}\ and\ \citenamefont
  {Khaymovich}(2024)}]{De_2024}%
  \BibitemOpen
  \bibfield  {author} {\bibinfo {author} {\bibfnamefont {G.}~\bibnamefont
  {De~Tomasi}}\ and\ \bibinfo {author} {\bibfnamefont {I.~M.}\ \bibnamefont
  {Khaymovich}},\ }\bibfield  {title} {\bibinfo {title} {Stable many-body
  localization under random continuous measurements in the no-click limit},\
  }\href {https://doi.org/10.1103/PhysRevB.109.174205} {\bibfield  {journal}
  {\bibinfo  {journal} {Phys. Rev. B}\ }\textbf {\bibinfo {volume} {109}},\
  \bibinfo {pages} {174205} (\bibinfo {year} {2024})}\BibitemShut {NoStop}%
\bibitem [{\citenamefont {Patrick}\ \emph {et~al.}(2024)\citenamefont
  {Patrick}, \citenamefont {Yang},\ and\ \citenamefont {Liu}}]{Patrick_2024}%
  \BibitemOpen
  \bibfield  {author} {\bibinfo {author} {\bibfnamefont {K.}~\bibnamefont
  {Patrick}}, \bibinfo {author} {\bibfnamefont {Q.}~\bibnamefont {Yang}},\ and\
  \bibinfo {author} {\bibfnamefont {D.~E.}\ \bibnamefont {Liu}},\ }\bibfield
  {title} {\bibinfo {title} {Enhanced localization in the prethermal regime of
  continuously measured many-body localized systems},\ }\href
  {https://doi.org/10.1103/PhysRevB.110.184211} {\bibfield  {journal} {\bibinfo
   {journal} {Phys. Rev. B}\ }\textbf {\bibinfo {volume} {110}},\ \bibinfo
  {pages} {184211} (\bibinfo {year} {2024})}\BibitemShut {NoStop}%
\bibitem [{\citenamefont {Cao}\ \emph {et~al.}(2019)\citenamefont {Cao},
  \citenamefont {Tilloy},\ and\ \citenamefont {De~Luca}}]{Cao_2019}%
  \BibitemOpen
  \bibfield  {author} {\bibinfo {author} {\bibfnamefont {X.}~\bibnamefont
  {Cao}}, \bibinfo {author} {\bibfnamefont {A.}~\bibnamefont {Tilloy}},\ and\
  \bibinfo {author} {\bibfnamefont {A.}~\bibnamefont {De~Luca}},\ }\bibfield
  {title} {\bibinfo {title} {Entanglement in a fermion chain under continuous
  monitoring},\ }\href {https://doi.org/10.21468/SciPostPhys.7.2.024}
  {\bibfield  {journal} {\bibinfo  {journal} {SciPost Phys.}\ }\textbf
  {\bibinfo {volume} {7}},\ \bibinfo {pages} {024} (\bibinfo {year}
  {2019})}\BibitemShut {NoStop}%
\bibitem [{\citenamefont {Alberton}\ \emph {et~al.}(2021)\citenamefont
  {Alberton}, \citenamefont {Buchhold},\ and\ \citenamefont
  {Diehl}}]{Alberton_2021}%
  \BibitemOpen
  \bibfield  {author} {\bibinfo {author} {\bibfnamefont {O.}~\bibnamefont
  {Alberton}}, \bibinfo {author} {\bibfnamefont {M.}~\bibnamefont {Buchhold}},\
  and\ \bibinfo {author} {\bibfnamefont {S.}~\bibnamefont {Diehl}},\ }\bibfield
   {title} {\bibinfo {title} {Entanglement transition in a monitored
  free-fermion chain: From extended criticality to area law},\ }\href
  {https://doi.org/10.1103/PhysRevLett.126.170602} {\bibfield  {journal}
  {\bibinfo  {journal} {Phys. Rev. Lett.}\ }\textbf {\bibinfo {volume} {126}},\
  \bibinfo {pages} {170602} (\bibinfo {year} {2021})}\BibitemShut {NoStop}%
\bibitem [{\citenamefont {Chen}\ \emph {et~al.}(2020)\citenamefont {Chen},
  \citenamefont {Li}, \citenamefont {Fisher},\ and\ \citenamefont
  {Lucas}}]{Chen_2020}%
  \BibitemOpen
  \bibfield  {author} {\bibinfo {author} {\bibfnamefont {X.}~\bibnamefont
  {Chen}}, \bibinfo {author} {\bibfnamefont {Y.}~\bibnamefont {Li}}, \bibinfo
  {author} {\bibfnamefont {M.~P.~A.}\ \bibnamefont {Fisher}},\ and\ \bibinfo
  {author} {\bibfnamefont {A.}~\bibnamefont {Lucas}},\ }\bibfield  {title}
  {\bibinfo {title} {Emergent conformal symmetry in nonunitary random dynamics
  of free fermions},\ }\href {https://doi.org/10.1103/PhysRevResearch.2.033017}
  {\bibfield  {journal} {\bibinfo  {journal} {Phys. Rev. Research}\ }\textbf
  {\bibinfo {volume} {2}},\ \bibinfo {pages} {033017} (\bibinfo {year}
  {2020})}\BibitemShut {NoStop}%
\bibitem [{\citenamefont {Tang}\ \emph {et~al.}(2021)\citenamefont {Tang},
  \citenamefont {Chen},\ and\ \citenamefont {Zhu}}]{Tang_2021}%
  \BibitemOpen
  \bibfield  {author} {\bibinfo {author} {\bibfnamefont {Q.}~\bibnamefont
  {Tang}}, \bibinfo {author} {\bibfnamefont {X.}~\bibnamefont {Chen}},\ and\
  \bibinfo {author} {\bibfnamefont {W.}~\bibnamefont {Zhu}},\ }\bibfield
  {title} {\bibinfo {title} {Quantum criticality in the nonunitary dynamics of
  $(2+1)$-dimensional free fermions},\ }\href
  {https://doi.org/10.1103/PhysRevB.103.174303} {\bibfield  {journal} {\bibinfo
   {journal} {Phys. Rev. B}\ }\textbf {\bibinfo {volume} {103}},\ \bibinfo
  {pages} {174303} (\bibinfo {year} {2021})}\BibitemShut {NoStop}%
\bibitem [{\citenamefont {Coppola}\ \emph {et~al.}(2022)\citenamefont
  {Coppola}, \citenamefont {Tirrito}, \citenamefont {Karevski},\ and\
  \citenamefont {Collura}}]{Coppola_2022}%
  \BibitemOpen
  \bibfield  {author} {\bibinfo {author} {\bibfnamefont {M.}~\bibnamefont
  {Coppola}}, \bibinfo {author} {\bibfnamefont {E.}~\bibnamefont {Tirrito}},
  \bibinfo {author} {\bibfnamefont {D.}~\bibnamefont {Karevski}},\ and\
  \bibinfo {author} {\bibfnamefont {M.}~\bibnamefont {Collura}},\ }\bibfield
  {title} {\bibinfo {title} {Growth of entanglement entropy under local
  projective measurements},\ }\href
  {https://doi.org/10.1103/PhysRevB.105.094303} {\bibfield  {journal} {\bibinfo
   {journal} {Phys. Rev. B}\ }\textbf {\bibinfo {volume} {105}},\ \bibinfo
  {pages} {094303} (\bibinfo {year} {2022})}\BibitemShut {NoStop}%
\bibitem [{\citenamefont {Ladewig}\ \emph {et~al.}(2022)\citenamefont
  {Ladewig}, \citenamefont {Diehl},\ and\ \citenamefont
  {Buchhold}}]{Ladewig_2022}%
  \BibitemOpen
  \bibfield  {author} {\bibinfo {author} {\bibfnamefont {B.}~\bibnamefont
  {Ladewig}}, \bibinfo {author} {\bibfnamefont {S.}~\bibnamefont {Diehl}},\
  and\ \bibinfo {author} {\bibfnamefont {M.}~\bibnamefont {Buchhold}},\
  }\bibfield  {title} {\bibinfo {title} {Monitored open fermion dynamics:
  {E}xploring the interplay of measurement, decoherence, and free {Hamiltonian}
  evolution},\ }\href {https://doi.org/10.1103/PhysRevResearch.4.033001}
  {\bibfield  {journal} {\bibinfo  {journal} {Phys. Rev. Research}\ }\textbf
  {\bibinfo {volume} {4}},\ \bibinfo {pages} {033001} (\bibinfo {year}
  {2022})}\BibitemShut {NoStop}%
\bibitem [{\citenamefont {Carollo}\ and\ \citenamefont
  {Alba}(2022)}]{Carollo_2022}%
  \BibitemOpen
  \bibfield  {author} {\bibinfo {author} {\bibfnamefont {F.}~\bibnamefont
  {Carollo}}\ and\ \bibinfo {author} {\bibfnamefont {V.}~\bibnamefont {Alba}},\
  }\bibfield  {title} {\bibinfo {title} {Entangled multiplets and spreading of
  quantum correlations in a continuously monitored tight-binding chain},\
  }\href {https://doi.org/10.1103/PhysRevB.106.L220304} {\bibfield  {journal}
  {\bibinfo  {journal} {Phys. Rev. B}\ }\textbf {\bibinfo {volume} {106}},\
  \bibinfo {pages} {L220304} (\bibinfo {year} {2022})}\BibitemShut {NoStop}%
\bibitem [{\citenamefont {Yang}\ \emph {et~al.}(2023)\citenamefont {Yang},
  \citenamefont {Zuo},\ and\ \citenamefont {Liu}}]{Yang_2022}%
  \BibitemOpen
  \bibfield  {author} {\bibinfo {author} {\bibfnamefont {Q.}~\bibnamefont
  {Yang}}, \bibinfo {author} {\bibfnamefont {Y.}~\bibnamefont {Zuo}},\ and\
  \bibinfo {author} {\bibfnamefont {D.~E.}\ \bibnamefont {Liu}},\ }\bibfield
  {title} {\bibinfo {title} {Keldysh nonlinear sigma model for a free-fermion
  gas under continuous measurements},\ }\href
  {https://doi.org/10.1103/PhysRevResearch.5.033174} {\bibfield  {journal}
  {\bibinfo  {journal} {Phys. Rev. Res.}\ }\textbf {\bibinfo {volume} {5}},\
  \bibinfo {pages} {033174} (\bibinfo {year} {2023})}\BibitemShut {NoStop}%
\bibitem [{\citenamefont {Buchhold}\ \emph {et~al.}(2021)\citenamefont
  {Buchhold}, \citenamefont {Minoguchi}, \citenamefont {Altland},\ and\
  \citenamefont {Diehl}}]{Buchhold_2021}%
  \BibitemOpen
  \bibfield  {author} {\bibinfo {author} {\bibfnamefont {M.}~\bibnamefont
  {Buchhold}}, \bibinfo {author} {\bibfnamefont {Y.}~\bibnamefont {Minoguchi}},
  \bibinfo {author} {\bibfnamefont {A.}~\bibnamefont {Altland}},\ and\ \bibinfo
  {author} {\bibfnamefont {S.}~\bibnamefont {Diehl}},\ }\bibfield  {title}
  {\bibinfo {title} {Effective theory for the measurement-induced phase
  transition of {Dirac} fermions},\ }\href
  {https://doi.org/10.1103/PhysRevX.11.041004} {\bibfield  {journal} {\bibinfo
  {journal} {Phys. Rev. X}\ }\textbf {\bibinfo {volume} {11}},\ \bibinfo
  {pages} {041004} (\bibinfo {year} {2021})}\BibitemShut {NoStop}%
\bibitem [{\citenamefont {Van~Regemortel}\ \emph {et~al.}(2021)\citenamefont
  {Van~Regemortel}, \citenamefont {Cian}, \citenamefont {Seif}, \citenamefont
  {Dehghani},\ and\ \citenamefont {Hafezi}}]{VanRegemortel_2021}%
  \BibitemOpen
  \bibfield  {author} {\bibinfo {author} {\bibfnamefont {M.}~\bibnamefont
  {Van~Regemortel}}, \bibinfo {author} {\bibfnamefont {Z.-P.}\ \bibnamefont
  {Cian}}, \bibinfo {author} {\bibfnamefont {A.}~\bibnamefont {Seif}}, \bibinfo
  {author} {\bibfnamefont {H.}~\bibnamefont {Dehghani}},\ and\ \bibinfo
  {author} {\bibfnamefont {M.}~\bibnamefont {Hafezi}},\ }\bibfield  {title}
  {\bibinfo {title} {Entanglement entropy scaling transition under competing
  monitoring protocols},\ }\href
  {https://doi.org/10.1103/PhysRevLett.126.123604} {\bibfield  {journal}
  {\bibinfo  {journal} {Phys. Rev. Lett.}\ }\textbf {\bibinfo {volume} {126}},\
  \bibinfo {pages} {123604} (\bibinfo {year} {2021})}\BibitemShut {NoStop}%
\bibitem [{\citenamefont {Gal}\ \emph {et~al.}(2023)\citenamefont {Gal},
  \citenamefont {Turkeshi},\ and\ \citenamefont {Schirò}}]{Youenn_2023}%
  \BibitemOpen
  \bibfield  {author} {\bibinfo {author} {\bibfnamefont {Y.~L.}\ \bibnamefont
  {Gal}}, \bibinfo {author} {\bibfnamefont {X.}~\bibnamefont {Turkeshi}},\ and\
  \bibinfo {author} {\bibfnamefont {M.}~\bibnamefont {Schirò}},\ }\bibfield
  {title} {\bibinfo {title} {{Volume-to-area law entanglement transition in a
  non-Hermitian free fermionic chain}},\ }\href
  {https://doi.org/10.21468/SciPostPhys.14.5.138} {\bibfield  {journal}
  {\bibinfo  {journal} {SciPost Phys.}\ }\textbf {\bibinfo {volume} {14}},\
  \bibinfo {pages} {138} (\bibinfo {year} {2023})}\BibitemShut {NoStop}%
\bibitem [{\citenamefont {L\'oio}\ \emph {et~al.}(2023)\citenamefont {L\'oio},
  \citenamefont {De~Luca}, \citenamefont {De~Nardis},\ and\ \citenamefont
  {Turkeshi}}]{Loio_2023}%
  \BibitemOpen
  \bibfield  {author} {\bibinfo {author} {\bibfnamefont {H.}~\bibnamefont
  {L\'oio}}, \bibinfo {author} {\bibfnamefont {A.}~\bibnamefont {De~Luca}},
  \bibinfo {author} {\bibfnamefont {J.}~\bibnamefont {De~Nardis}},\ and\
  \bibinfo {author} {\bibfnamefont {X.}~\bibnamefont {Turkeshi}},\ }\bibfield
  {title} {\bibinfo {title} {Purification timescales in monitored fermions},\
  }\href {https://doi.org/10.1103/PhysRevB.108.L020306} {\bibfield  {journal}
  {\bibinfo  {journal} {Phys. Rev. B}\ }\textbf {\bibinfo {volume} {108}},\
  \bibinfo {pages} {L020306} (\bibinfo {year} {2023})}\BibitemShut {NoStop}%
\bibitem [{\citenamefont {Turkeshi}\ \emph {et~al.}(2022)\citenamefont
  {Turkeshi}, \citenamefont {Piroli},\ and\ \citenamefont
  {Schir\'o}}]{Turkeshi_2022}%
  \BibitemOpen
  \bibfield  {author} {\bibinfo {author} {\bibfnamefont {X.}~\bibnamefont
  {Turkeshi}}, \bibinfo {author} {\bibfnamefont {L.}~\bibnamefont {Piroli}},\
  and\ \bibinfo {author} {\bibfnamefont {M.}~\bibnamefont {Schir\'o}},\
  }\bibfield  {title} {\bibinfo {title} {Enhanced entanglement negativity in
  boundary-driven monitored fermionic chains},\ }\href
  {https://doi.org/10.1103/PhysRevB.106.024304} {\bibfield  {journal} {\bibinfo
   {journal} {Phys. Rev. B}\ }\textbf {\bibinfo {volume} {106}},\ \bibinfo
  {pages} {024304} (\bibinfo {year} {2022})}\BibitemShut {NoStop}%
\bibitem [{\citenamefont {Kells}\ \emph {et~al.}(2023)\citenamefont {Kells},
  \citenamefont {Meidan},\ and\ \citenamefont {Romito}}]{Kells_2023}%
  \BibitemOpen
  \bibfield  {author} {\bibinfo {author} {\bibfnamefont {G.}~\bibnamefont
  {Kells}}, \bibinfo {author} {\bibfnamefont {D.}~\bibnamefont {Meidan}},\ and\
  \bibinfo {author} {\bibfnamefont {A.}~\bibnamefont {Romito}},\ }\bibfield
  {title} {\bibinfo {title} {{Topological transitions in weakly monitored free
  fermions}},\ }\href {https://doi.org/10.21468/SciPostPhys.14.3.031}
  {\bibfield  {journal} {\bibinfo  {journal} {SciPost Phys.}\ }\textbf
  {\bibinfo {volume} {14}},\ \bibinfo {pages} {031} (\bibinfo {year}
  {2023})}\BibitemShut {NoStop}%
\bibitem [{\citenamefont {Li}\ \emph {et~al.}(2025)\citenamefont {Li},
  \citenamefont {Zhong},\ and\ \citenamefont {Yu}}]{Yu_2025}%
  \BibitemOpen
  \bibfield  {author} {\bibinfo {author} {\bibfnamefont {H.-Z.}\ \bibnamefont
  {Li}}, \bibinfo {author} {\bibfnamefont {J.-X.}\ \bibnamefont {Zhong}},\ and\
  \bibinfo {author} {\bibfnamefont {X.-J.}\ \bibnamefont {Yu}},\ }\bibfield
  {title} {\bibinfo {title} {Measurement-induced entanglement phase transition
  in free fermion systems},\ }\href {https://doi.org/10.1088/1361-648X/ade7e5}
  {\bibfield  {journal} {\bibinfo  {journal} {Journal of Physics: Condensed
  Matter}\ }\textbf {\bibinfo {volume} {37}},\ \bibinfo {pages} {273002}
  (\bibinfo {year} {2025})}\BibitemShut {NoStop}%
\bibitem [{\citenamefont {Fava}\ \emph {et~al.}(2023)\citenamefont {Fava},
  \citenamefont {Piroli}, \citenamefont {Swann}, \citenamefont {Bernard},\ and\
  \citenamefont {Nahum}}]{Fava_2023}%
  \BibitemOpen
  \bibfield  {author} {\bibinfo {author} {\bibfnamefont {M.}~\bibnamefont
  {Fava}}, \bibinfo {author} {\bibfnamefont {L.}~\bibnamefont {Piroli}},
  \bibinfo {author} {\bibfnamefont {T.}~\bibnamefont {Swann}}, \bibinfo
  {author} {\bibfnamefont {D.}~\bibnamefont {Bernard}},\ and\ \bibinfo {author}
  {\bibfnamefont {A.}~\bibnamefont {Nahum}},\ }\bibfield  {title} {\bibinfo
  {title} {Nonlinear sigma models for monitored dynamics of free fermions},\
  }\href {https://doi.org/10.1103/PhysRevX.13.041045} {\bibfield  {journal}
  {\bibinfo  {journal} {Phys. Rev. X}\ }\textbf {\bibinfo {volume} {13}},\
  \bibinfo {pages} {041045} (\bibinfo {year} {2023})}\BibitemShut {NoStop}%
\bibitem [{\citenamefont {Piccitto}\ \emph {et~al.}(2022)\citenamefont
  {Piccitto}, \citenamefont {Russomanno},\ and\ \citenamefont
  {Rossini}}]{Piccitto_2022}%
  \BibitemOpen
  \bibfield  {author} {\bibinfo {author} {\bibfnamefont {G.}~\bibnamefont
  {Piccitto}}, \bibinfo {author} {\bibfnamefont {A.}~\bibnamefont
  {Russomanno}},\ and\ \bibinfo {author} {\bibfnamefont {D.}~\bibnamefont
  {Rossini}},\ }\bibfield  {title} {\bibinfo {title} {Entanglement transitions
  in the quantum ising chain: A comparison between different unravelings of the
  same lindbladian},\ }\href {https://doi.org/10.1103/PhysRevB.105.064305}
  {\bibfield  {journal} {\bibinfo  {journal} {Phys. Rev. B}\ }\textbf {\bibinfo
  {volume} {105}},\ \bibinfo {pages} {064305} (\bibinfo {year}
  {2022})}\BibitemShut {NoStop}%
\bibitem [{\citenamefont {Piccitto}\ \emph {et~al.}(2023)\citenamefont
  {Piccitto}, \citenamefont {Russomanno},\ and\ \citenamefont
  {Rossini}}]{Piccitto_2023}%
  \BibitemOpen
  \bibfield  {author} {\bibinfo {author} {\bibfnamefont {G.}~\bibnamefont
  {Piccitto}}, \bibinfo {author} {\bibfnamefont {A.}~\bibnamefont
  {Russomanno}},\ and\ \bibinfo {author} {\bibfnamefont {D.}~\bibnamefont
  {Rossini}},\ }\bibfield  {title} {\bibinfo {title} {{Entanglement dynamics
  with string measurement operators}},\ }\href
  {https://doi.org/10.21468/SciPostPhysCore.6.4.078} {\bibfield  {journal}
  {\bibinfo  {journal} {SciPost Phys. Core}\ }\textbf {\bibinfo {volume} {6}},\
  \bibinfo {pages} {078} (\bibinfo {year} {2023})}\BibitemShut {NoStop}%
\bibitem [{\citenamefont {Russomanno}\ \emph {et~al.}(2023)\citenamefont
  {Russomanno}, \citenamefont {Piccitto},\ and\ \citenamefont
  {Rossini}}]{Russomanno_2023}%
  \BibitemOpen
  \bibfield  {author} {\bibinfo {author} {\bibfnamefont {A.}~\bibnamefont
  {Russomanno}}, \bibinfo {author} {\bibfnamefont {G.}~\bibnamefont
  {Piccitto}},\ and\ \bibinfo {author} {\bibfnamefont {D.}~\bibnamefont
  {Rossini}},\ }\bibfield  {title} {\bibinfo {title} {Entanglement transitions
  and quantum bifurcations under continuous long-range monitoring},\ }\href
  {https://doi.org/10.1103/PhysRevB.108.104313} {\bibfield  {journal} {\bibinfo
   {journal} {Phys. Rev. B}\ }\textbf {\bibinfo {volume} {108}},\ \bibinfo
  {pages} {104313} (\bibinfo {year} {2023})}\BibitemShut {NoStop}%
\bibitem [{\citenamefont {Poboiko}\ \emph {et~al.}(2023)\citenamefont
  {Poboiko}, \citenamefont {P{\"o}pperl}, \citenamefont {Gornyi},\ and\
  \citenamefont {Mirlin}}]{Poboiko_2023}%
  \BibitemOpen
  \bibfield  {author} {\bibinfo {author} {\bibfnamefont {I.}~\bibnamefont
  {Poboiko}}, \bibinfo {author} {\bibfnamefont {P.}~\bibnamefont
  {P{\"o}pperl}}, \bibinfo {author} {\bibfnamefont {I.~V.}\ \bibnamefont
  {Gornyi}},\ and\ \bibinfo {author} {\bibfnamefont {A.~D.}\ \bibnamefont
  {Mirlin}},\ }\bibfield  {title} {\bibinfo {title} {Theory of free fermions
  under random projective measurements},\ }\href
  {https://doi.org/10.1103/PhysRevX.13.041046} {\bibfield  {journal} {\bibinfo
  {journal} {Phys. Rev. X}\ }\textbf {\bibinfo {volume} {13}},\ \bibinfo
  {pages} {041046} (\bibinfo {year} {2023})}\BibitemShut {NoStop}%
\bibitem [{\citenamefont {Fava}\ \emph {et~al.}(2024)\citenamefont {Fava},
  \citenamefont {Piroli}, \citenamefont {Bernard},\ and\ \citenamefont
  {Nahum}}]{Fava_2024}%
  \BibitemOpen
  \bibfield  {author} {\bibinfo {author} {\bibfnamefont {M.}~\bibnamefont
  {Fava}}, \bibinfo {author} {\bibfnamefont {L.}~\bibnamefont {Piroli}},
  \bibinfo {author} {\bibfnamefont {D.}~\bibnamefont {Bernard}},\ and\ \bibinfo
  {author} {\bibfnamefont {A.}~\bibnamefont {Nahum}},\ }\bibfield  {title}
  {\bibinfo {title} {Monitored fermions with conserved $u(1)$ charge},\ }\href
  {https://doi.org/10.1103/PhysRevResearch.6.043246} {\bibfield  {journal}
  {\bibinfo  {journal} {Phys. Rev. Res.}\ }\textbf {\bibinfo {volume} {6}},\
  \bibinfo {pages} {043246} (\bibinfo {year} {2024})}\BibitemShut {NoStop}%
\bibitem [{\citenamefont {Starchl}\ \emph {et~al.}(2025)\citenamefont
  {Starchl}, \citenamefont {Fischer},\ and\ \citenamefont
  {Sieberer}}]{Starchl_2025}%
  \BibitemOpen
  \bibfield  {author} {\bibinfo {author} {\bibfnamefont {E.}~\bibnamefont
  {Starchl}}, \bibinfo {author} {\bibfnamefont {M.~H.}\ \bibnamefont
  {Fischer}},\ and\ \bibinfo {author} {\bibfnamefont {L.~M.}\ \bibnamefont
  {Sieberer}},\ }\bibfield  {title} {\bibinfo {title} {Generalized zeno effect
  and entanglement dynamics induced by fermion counting},\ }\href
  {https://doi.org/10.1103/jppz-vdgn} {\bibfield  {journal} {\bibinfo
  {journal} {PRX Quantum}\ }\textbf {\bibinfo {volume} {6}},\ \bibinfo {pages}
  {030302} (\bibinfo {year} {2025})}\BibitemShut {NoStop}%
\bibitem [{\citenamefont {Chahine}\ and\ \citenamefont
  {Buchhold}(2024)}]{Chahine_2024}%
  \BibitemOpen
  \bibfield  {author} {\bibinfo {author} {\bibfnamefont {K.}~\bibnamefont
  {Chahine}}\ and\ \bibinfo {author} {\bibfnamefont {M.}~\bibnamefont
  {Buchhold}},\ }\bibfield  {title} {\bibinfo {title} {Entanglement phases,
  localization, and multifractality of monitored free fermions in two
  dimensions},\ }\href {https://doi.org/10.1103/PhysRevB.110.054313} {\bibfield
   {journal} {\bibinfo  {journal} {Phys. Rev. B}\ }\textbf {\bibinfo {volume}
  {110}},\ \bibinfo {pages} {054313} (\bibinfo {year} {2024})}\BibitemShut
  {NoStop}%
\bibitem [{\citenamefont {Poboiko}\ \emph {et~al.}(2024)\citenamefont
  {Poboiko}, \citenamefont {Gornyi},\ and\ \citenamefont
  {Mirlin}}]{Poboiko_2024}%
  \BibitemOpen
  \bibfield  {author} {\bibinfo {author} {\bibfnamefont {I.}~\bibnamefont
  {Poboiko}}, \bibinfo {author} {\bibfnamefont {I.~V.}\ \bibnamefont
  {Gornyi}},\ and\ \bibinfo {author} {\bibfnamefont {A.~D.}\ \bibnamefont
  {Mirlin}},\ }\bibfield  {title} {\bibinfo {title} {Measurement-induced phase
  transition for free fermions above one dimension},\ }\href
  {https://doi.org/10.1103/PhysRevLett.132.110403} {\bibfield  {journal}
  {\bibinfo  {journal} {Phys. Rev. Lett.}\ }\textbf {\bibinfo {volume} {132}},\
  \bibinfo {pages} {110403} (\bibinfo {year} {2024})}\BibitemShut {NoStop}%
\bibitem [{\citenamefont {Jin}\ and\ \citenamefont {Martin}(2024)}]{Jin_2024}%
  \BibitemOpen
  \bibfield  {author} {\bibinfo {author} {\bibfnamefont {T.}~\bibnamefont
  {Jin}}\ and\ \bibinfo {author} {\bibfnamefont {D.~G.}\ \bibnamefont
  {Martin}},\ }\bibfield  {title} {\bibinfo {title} {Measurement-induced phase
  transition in a single-body tight-binding model},\ }\href
  {https://doi.org/10.1103/PhysRevB.110.L060202} {\bibfield  {journal}
  {\bibinfo  {journal} {Phys. Rev. B}\ }\textbf {\bibinfo {volume} {110}},\
  \bibinfo {pages} {L060202} (\bibinfo {year} {2024})}\BibitemShut {NoStop}%
\bibitem [{\citenamefont {Minato}\ \emph {et~al.}(2022)\citenamefont {Minato},
  \citenamefont {Sugimoto}, \citenamefont {Kuwahara},\ and\ \citenamefont
  {Saito}}]{Minato_2022}%
  \BibitemOpen
  \bibfield  {author} {\bibinfo {author} {\bibfnamefont {T.}~\bibnamefont
  {Minato}}, \bibinfo {author} {\bibfnamefont {K.}~\bibnamefont {Sugimoto}},
  \bibinfo {author} {\bibfnamefont {T.}~\bibnamefont {Kuwahara}},\ and\
  \bibinfo {author} {\bibfnamefont {K.}~\bibnamefont {Saito}},\ }\bibfield
  {title} {\bibinfo {title} {{Fate of Measurement-Induced Phase Transition in
  Long-Range Interactions}},\ }\href
  {https://doi.org/10.1103/PhysRevLett.128.010603} {\bibfield  {journal}
  {\bibinfo  {journal} {Phys. Rev. Lett.}\ }\textbf {\bibinfo {volume} {128}},\
  \bibinfo {pages} {010603} (\bibinfo {year} {2022})}\BibitemShut {NoStop}%
\bibitem [{\citenamefont {M\"uller}\ \emph {et~al.}(2022)\citenamefont
  {M\"uller}, \citenamefont {Diehl},\ and\ \citenamefont
  {Buchhold}}]{Muller_2022}%
  \BibitemOpen
  \bibfield  {author} {\bibinfo {author} {\bibfnamefont {T.}~\bibnamefont
  {M\"uller}}, \bibinfo {author} {\bibfnamefont {S.}~\bibnamefont {Diehl}},\
  and\ \bibinfo {author} {\bibfnamefont {M.}~\bibnamefont {Buchhold}},\
  }\bibfield  {title} {\bibinfo {title} {{Measurement-Induced Dark State Phase
  Transitions in Long-Ranged Fermion Systems}},\ }\href
  {https://doi.org/10.1103/PhysRevLett.128.010605} {\bibfield  {journal}
  {\bibinfo  {journal} {Phys. Rev. Lett.}\ }\textbf {\bibinfo {volume} {128}},\
  \bibinfo {pages} {010605} (\bibinfo {year} {2022})}\BibitemShut {NoStop}%
\bibitem [{\citenamefont {Koh}\ \emph {et~al.}(2023)\citenamefont {Koh},
  \citenamefont {Sun}, \citenamefont {Motta},\ and\ \citenamefont
  {Minnich}}]{Koh_2023}%
  \BibitemOpen
  \bibfield  {author} {\bibinfo {author} {\bibfnamefont {J.~M.}\ \bibnamefont
  {Koh}}, \bibinfo {author} {\bibfnamefont {S.-N.}\ \bibnamefont {Sun}},
  \bibinfo {author} {\bibfnamefont {M.}~\bibnamefont {Motta}},\ and\ \bibinfo
  {author} {\bibfnamefont {A.~J.}\ \bibnamefont {Minnich}},\ }\bibfield
  {title} {\bibinfo {title} {Measurement-induced entanglement phase transition
  on a superconducting quantum processor with mid-circuit readout},\ }\href
  {https://doi.org/10.1038/s41567-023-02076-6} {\bibfield  {journal} {\bibinfo
  {journal} {Nat. Phys.}\ }\textbf {\bibinfo {volume} {19}},\ \bibinfo {pages}
  {1314} (\bibinfo {year} {2023})}\BibitemShut {NoStop}%
\bibitem [{\citenamefont {{Google Quantum AI}}\ and\ \citenamefont
  {{Collaborators}}(2023)}]{Google_2023}%
  \BibitemOpen
  \bibfield  {author} {\bibinfo {author} {\bibnamefont {{Google Quantum AI}}}\
  and\ \bibinfo {author} {\bibnamefont {{Collaborators}}},\ }\bibfield  {title}
  {\bibinfo {title} {Measurement-induced entanglement and teleportation on a
  noisy quantum processor},\ }\href
  {https://doi.org/10.1038/s41586-023-06505-7} {\bibfield  {journal} {\bibinfo
  {journal} {Nature}\ }\textbf {\bibinfo {volume} {622}},\ \bibinfo {pages}
  {481} (\bibinfo {year} {2023})}\BibitemShut {NoStop}%
\bibitem [{\citenamefont {Kamakari}\ \emph {et~al.}(2025)\citenamefont
  {Kamakari}, \citenamefont {Sun}, \citenamefont {Li}, \citenamefont {Thio},
  \citenamefont {Gujarati}, \citenamefont {Fisher}, \citenamefont {Motta},\
  and\ \citenamefont {Minnich}}]{Kamakari_2025}%
  \BibitemOpen
  \bibfield  {author} {\bibinfo {author} {\bibfnamefont {H.}~\bibnamefont
  {Kamakari}}, \bibinfo {author} {\bibfnamefont {J.}~\bibnamefont {Sun}},
  \bibinfo {author} {\bibfnamefont {Y.}~\bibnamefont {Li}}, \bibinfo {author}
  {\bibfnamefont {J.~J.}\ \bibnamefont {Thio}}, \bibinfo {author}
  {\bibfnamefont {T.~P.}\ \bibnamefont {Gujarati}}, \bibinfo {author}
  {\bibfnamefont {M.~P.~A.}\ \bibnamefont {Fisher}}, \bibinfo {author}
  {\bibfnamefont {M.}~\bibnamefont {Motta}},\ and\ \bibinfo {author}
  {\bibfnamefont {A.~J.}\ \bibnamefont {Minnich}},\ }\bibfield  {title}
  {\bibinfo {title} {Experimental demonstration of scalable cross-entropy
  benchmarking to detect measurement-induced phase transitions on a
  superconducting quantum processor},\ }\href
  {https://doi.org/10.1103/PhysRevLett.134.120401} {\bibfield  {journal}
  {\bibinfo  {journal} {Phys. Rev. Lett.}\ }\textbf {\bibinfo {volume} {134}},\
  \bibinfo {pages} {120401} (\bibinfo {year} {2025})}\BibitemShut {NoStop}%
\bibitem [{\citenamefont {Noel}\ \emph {et~al.}(2022)\citenamefont {Noel},
  \citenamefont {Niroula}, \citenamefont {Zhu}, \citenamefont {Risinger},
  \citenamefont {Egan}, \citenamefont {Biswas}, \citenamefont {Cetina},
  \citenamefont {Gorshkov}, \citenamefont {Gullans}, \citenamefont {Huse} \emph
  {et~al.}}]{Noel_2022}%
  \BibitemOpen
  \bibfield  {author} {\bibinfo {author} {\bibfnamefont {C.}~\bibnamefont
  {Noel}}, \bibinfo {author} {\bibfnamefont {P.}~\bibnamefont {Niroula}},
  \bibinfo {author} {\bibfnamefont {D.}~\bibnamefont {Zhu}}, \bibinfo {author}
  {\bibfnamefont {A.}~\bibnamefont {Risinger}}, \bibinfo {author}
  {\bibfnamefont {L.}~\bibnamefont {Egan}}, \bibinfo {author} {\bibfnamefont
  {D.}~\bibnamefont {Biswas}}, \bibinfo {author} {\bibfnamefont
  {M.}~\bibnamefont {Cetina}}, \bibinfo {author} {\bibfnamefont {A.~V.}\
  \bibnamefont {Gorshkov}}, \bibinfo {author} {\bibfnamefont {M.~J.}\
  \bibnamefont {Gullans}}, \bibinfo {author} {\bibfnamefont {D.~A.}\
  \bibnamefont {Huse}}, \emph {et~al.},\ }\bibfield  {title} {\bibinfo {title}
  {Measurement-induced quantum phases realized in a trapped-ion quantum
  computer},\ }\href {https://doi.org/10.1038/s41567-022-01619-7} {\bibfield
  {journal} {\bibinfo  {journal} {Nat. Phys.}\ }\textbf {\bibinfo {volume}
  {18}},\ \bibinfo {pages} {760} (\bibinfo {year} {2022})}\BibitemShut
  {NoStop}%
\bibitem [{\citenamefont {Agrawal}\ \emph {et~al.}(2024)\citenamefont
  {Agrawal}, \citenamefont {Lopez-Piqueres}, \citenamefont {Vasseur},
  \citenamefont {Gopalakrishnan},\ and\ \citenamefont {Potter}}]{Agrawal_2024}%
  \BibitemOpen
  \bibfield  {author} {\bibinfo {author} {\bibfnamefont {U.}~\bibnamefont
  {Agrawal}}, \bibinfo {author} {\bibfnamefont {J.}~\bibnamefont
  {Lopez-Piqueres}}, \bibinfo {author} {\bibfnamefont {R.}~\bibnamefont
  {Vasseur}}, \bibinfo {author} {\bibfnamefont {S.}~\bibnamefont
  {Gopalakrishnan}},\ and\ \bibinfo {author} {\bibfnamefont {A.~C.}\
  \bibnamefont {Potter}},\ }\bibfield  {title} {\bibinfo {title} {Observing
  quantum measurement collapse as a learnability phase transition},\ }\href
  {https://doi.org/10.1103/PhysRevX.14.041012} {\bibfield  {journal} {\bibinfo
  {journal} {Phys. Rev. X}\ }\textbf {\bibinfo {volume} {14}},\ \bibinfo
  {pages} {041012} (\bibinfo {year} {2024})}\BibitemShut {NoStop}%
\bibitem [{\citenamefont {Cech}\ \emph {et~al.}(2025)\citenamefont {Cech},
  \citenamefont {Cea}, \citenamefont {Ba\~nuls}, \citenamefont {Lesanovsky},\
  and\ \citenamefont {Carollo}}]{Cech25}%
  \BibitemOpen
  \bibfield  {author} {\bibinfo {author} {\bibfnamefont {M.}~\bibnamefont
  {Cech}}, \bibinfo {author} {\bibfnamefont {M.}~\bibnamefont {Cea}}, \bibinfo
  {author} {\bibfnamefont {M.~C.}\ \bibnamefont {Ba\~nuls}}, \bibinfo {author}
  {\bibfnamefont {I.}~\bibnamefont {Lesanovsky}},\ and\ \bibinfo {author}
  {\bibfnamefont {F.}~\bibnamefont {Carollo}},\ }\bibfield  {title} {\bibinfo
  {title} {{Space-Time Correlations in Monitored Kinetically Constrained
  Discrete-Time Quantum Dynamics}},\ }\href {https://doi.org/10.1103/2lxs-wccj}
  {\bibfield  {journal} {\bibinfo  {journal} {Phys. Rev. Lett.}\ }\textbf
  {\bibinfo {volume} {134}},\ \bibinfo {pages} {230403} (\bibinfo {year}
  {2025})}\BibitemShut {NoStop}%
\bibitem [{\citenamefont {Yamamoto}\ and\ \citenamefont
  {Hamazaki}(2026{\natexlab{a}})}]{Yamamoto_2026}%
  \BibitemOpen
  \bibfield  {author} {\bibinfo {author} {\bibfnamefont {K.}~\bibnamefont
  {Yamamoto}}\ and\ \bibinfo {author} {\bibfnamefont {R.}~\bibnamefont
  {Hamazaki}},\ }\bibfield  {title} {\bibinfo {title} {Anomalous waiting-time
  distributions in postselection-free quantum many-body dynamics under
  continuous monitoring},\ }\href {https://doi.org/10.1088/2058-9565/ae9183}
  {\bibfield  {journal} {\bibinfo  {journal} {Quantum Science and Technology}\
  }\textbf {\bibinfo {volume} {11}},\ \bibinfo {pages} {045007} (\bibinfo
  {year} {2026}{\natexlab{a}})}\BibitemShut {NoStop}%
\bibitem [{\citenamefont {Landi}\ \emph {et~al.}(2024)\citenamefont {Landi},
  \citenamefont {Kewming}, \citenamefont {Mitchison},\ and\ \citenamefont
  {Potts}}]{Landi_2024}%
  \BibitemOpen
  \bibfield  {author} {\bibinfo {author} {\bibfnamefont {G.~T.}\ \bibnamefont
  {Landi}}, \bibinfo {author} {\bibfnamefont {M.~J.}\ \bibnamefont {Kewming}},
  \bibinfo {author} {\bibfnamefont {M.~T.}\ \bibnamefont {Mitchison}},\ and\
  \bibinfo {author} {\bibfnamefont {P.~P.}\ \bibnamefont {Potts}},\ }\bibfield
  {title} {\bibinfo {title} {Current fluctuations in open quantum systems:
  Bridging the gap between quantum continuous measurements and full counting
  statistics},\ }\href {https://doi.org/10.1103/PRXQuantum.5.020201} {\bibfield
   {journal} {\bibinfo  {journal} {PRX Quantum}\ }\textbf {\bibinfo {volume}
  {5}},\ \bibinfo {pages} {020201} (\bibinfo {year} {2024})}\BibitemShut
  {NoStop}%
\bibitem [{\citenamefont {Yamamoto}\ and\ \citenamefont
  {Hamazaki}(2026{\natexlab{b}})}]{Yamamoto_2025}%
  \BibitemOpen
  \bibfield  {author} {\bibinfo {author} {\bibfnamefont {K.}~\bibnamefont
  {Yamamoto}}\ and\ \bibinfo {author} {\bibfnamefont {R.}~\bibnamefont
  {Hamazaki}},\ }\bibfield  {title} {\bibinfo {title} {Measurement-induced
  crossover of quantum jump statistics in postselection-free many-body
  dynamics},\ }\bibfield  {journal} {\bibinfo  {journal} {Phys. Rev. Lett.}\
  }\href {https://doi.org/10.1103/wv5b-r6sb} {10.1103/wv5b-r6sb} (\bibinfo
  {year} {2026}{\natexlab{b}}),\ \bibinfo {note} {to be published},\ \Eprint
  {https://arxiv.org/abs/2503.02418} {arXiv:2503.02418 [cond-mat.stat-mech]}
  \BibitemShut {NoStop}%
\bibitem [{\citenamefont {Moghaddam}\ \emph {et~al.}(2023)\citenamefont
  {Moghaddam}, \citenamefont {P\"oyh\"onen},\ and\ \citenamefont
  {Ojanen}}]{Moghaddam_2023}%
  \BibitemOpen
  \bibfield  {author} {\bibinfo {author} {\bibfnamefont {A.~G.}\ \bibnamefont
  {Moghaddam}}, \bibinfo {author} {\bibfnamefont {K.}~\bibnamefont
  {P\"oyh\"onen}},\ and\ \bibinfo {author} {\bibfnamefont {T.}~\bibnamefont
  {Ojanen}},\ }\bibfield  {title} {\bibinfo {title} {Exponential shortcut to
  measurement-induced entanglement phase transitions},\ }\href
  {https://doi.org/10.1103/PhysRevLett.131.020401} {\bibfield  {journal}
  {\bibinfo  {journal} {Phys. Rev. Lett.}\ }\textbf {\bibinfo {volume} {131}},\
  \bibinfo {pages} {020401} (\bibinfo {year} {2023})}\BibitemShut {NoStop}%
\bibitem [{\citenamefont {Qiu}\ \emph {et~al.}(2025)\citenamefont {Qiu},
  \citenamefont {Yu},\ and\ \citenamefont {Cai}}]{Qiu_2025}%
  \BibitemOpen
  \bibfield  {author} {\bibinfo {author} {\bibfnamefont {W.-J.}\ \bibnamefont
  {Qiu}}, \bibinfo {author} {\bibfnamefont {Y.-C.}\ \bibnamefont {Yu}},\ and\
  \bibinfo {author} {\bibfnamefont {X.}~\bibnamefont {Cai}},\ }\bibfield
  {title} {\bibinfo {title} {Fluctuation as a probe of entanglement transition
  in a monitored free-fermion chain},\ }\href
  {https://doi.org/10.1103/ps1h-ftfw} {\bibfield  {journal} {\bibinfo
  {journal} {Phys. Rev. A}\ }\textbf {\bibinfo {volume} {112}},\ \bibinfo
  {pages} {012604} (\bibinfo {year} {2025})}\BibitemShut {NoStop}%
\bibitem [{\citenamefont {Oshima}\ and\ \citenamefont
  {Fuji}(2023)}]{Oshima_2023}%
  \BibitemOpen
  \bibfield  {author} {\bibinfo {author} {\bibfnamefont {H.}~\bibnamefont
  {Oshima}}\ and\ \bibinfo {author} {\bibfnamefont {Y.}~\bibnamefont {Fuji}},\
  }\bibfield  {title} {\bibinfo {title} {Charge fluctuation and charge-resolved
  entanglement in a monitored quantum circuit with $u(1)$ symmetry},\ }\href
  {https://doi.org/10.1103/PhysRevB.107.014308} {\bibfield  {journal} {\bibinfo
   {journal} {Phys. Rev. B}\ }\textbf {\bibinfo {volume} {107}},\ \bibinfo
  {pages} {014308} (\bibinfo {year} {2023})}\BibitemShut {NoStop}%
\bibitem [{\citenamefont {Yamamoto}\ and\ \citenamefont
  {Hamazaki}(2023)}]{Yamamoto_2023}%
  \BibitemOpen
  \bibfield  {author} {\bibinfo {author} {\bibfnamefont {K.}~\bibnamefont
  {Yamamoto}}\ and\ \bibinfo {author} {\bibfnamefont {R.}~\bibnamefont
  {Hamazaki}},\ }\bibfield  {title} {\bibinfo {title} {Localization properties
  in disordered quantum many-body dynamics under continuous measurement},\
  }\href {https://doi.org/10.1103/PhysRevB.107.L220201} {\bibfield  {journal}
  {\bibinfo  {journal} {Phys. Rev. B}\ }\textbf {\bibinfo {volume} {107}},\
  \bibinfo {pages} {L220201} (\bibinfo {year} {2023})}\BibitemShut {NoStop}%
\bibitem [{\citenamefont {Szyniszewski}\ \emph {et~al.}(2023)\citenamefont
  {Szyniszewski}, \citenamefont {Lunt},\ and\ \citenamefont
  {Pal}}]{Szyniszewski_2023}%
  \BibitemOpen
  \bibfield  {author} {\bibinfo {author} {\bibfnamefont {M.}~\bibnamefont
  {Szyniszewski}}, \bibinfo {author} {\bibfnamefont {O.}~\bibnamefont {Lunt}},\
  and\ \bibinfo {author} {\bibfnamefont {A.}~\bibnamefont {Pal}},\ }\bibfield
  {title} {\bibinfo {title} {Disordered monitored free fermions},\ }\href
  {https://doi.org/10.1103/PhysRevB.108.165126} {\bibfield  {journal} {\bibinfo
   {journal} {Phys. Rev. B}\ }\textbf {\bibinfo {volume} {108}},\ \bibinfo
  {pages} {165126} (\bibinfo {year} {2023})}\BibitemShut {NoStop}%
\bibitem [{\citenamefont {P\"opperl}\ \emph {et~al.}(2023)\citenamefont
  {P\"opperl}, \citenamefont {Gornyi},\ and\ \citenamefont
  {Gefen}}]{Popperl_2023}%
  \BibitemOpen
  \bibfield  {author} {\bibinfo {author} {\bibfnamefont {P.}~\bibnamefont
  {P\"opperl}}, \bibinfo {author} {\bibfnamefont {I.~V.}\ \bibnamefont
  {Gornyi}},\ and\ \bibinfo {author} {\bibfnamefont {Y.}~\bibnamefont
  {Gefen}},\ }\bibfield  {title} {\bibinfo {title} {{Measurements on an
  Anderson chain}},\ }\href {https://doi.org/10.1103/PhysRevB.107.174203}
  {\bibfield  {journal} {\bibinfo  {journal} {Phys. Rev. B}\ }\textbf {\bibinfo
  {volume} {107}},\ \bibinfo {pages} {174203} (\bibinfo {year}
  {2023})}\BibitemShut {NoStop}%
\bibitem [{\citenamefont {Szyniszewski}(2024)}]{Szyniszewski_2024}%
  \BibitemOpen
  \bibfield  {author} {\bibinfo {author} {\bibfnamefont {M.}~\bibnamefont
  {Szyniszewski}},\ }\bibfield  {title} {\bibinfo {title} {Unscrambling of
  single-particle wave functions in systems localized through disorder and
  monitoring},\ }\href {https://doi.org/10.1103/PhysRevB.110.024303} {\bibfield
   {journal} {\bibinfo  {journal} {Phys. Rev. B}\ }\textbf {\bibinfo {volume}
  {110}},\ \bibinfo {pages} {024303} (\bibinfo {year} {2024})}\BibitemShut
  {NoStop}%
\bibitem [{\citenamefont {Liao}\ \emph {et~al.}(2026)\citenamefont {Liao},
  \citenamefont {Matheussen},\ and\ \citenamefont {Zhang}}]{Liao_2026}%
  \BibitemOpen
  \bibfield  {author} {\bibinfo {author} {\bibfnamefont {Y.}~\bibnamefont
  {Liao}}, \bibinfo {author} {\bibfnamefont {M.}~\bibnamefont {Matheussen}},\
  and\ \bibinfo {author} {\bibfnamefont {X.}~\bibnamefont {Zhang}},\ }\bibfield
   {title} {\bibinfo {title} {Measurement-induced phase transitions in
  disordered fermions},\ }\bibfield  {journal} {\bibinfo  {journal} {arXiv
  preprint arXiv:2605.05306}\ }\href
  {https://doi.org/10.48550/arXiv.2605.05306} {10.48550/arXiv.2605.05306}
  (\bibinfo {year} {2026})\BibitemShut {NoStop}%
\bibitem [{\citenamefont {Lye}\ \emph {et~al.}(2007)\citenamefont {Lye},
  \citenamefont {Fallani}, \citenamefont {Fort}, \citenamefont {Guarrera},
  \citenamefont {Modugno}, \citenamefont {Wiersma},\ and\ \citenamefont
  {Inguscio}}]{Fallani_2007}%
  \BibitemOpen
  \bibfield  {author} {\bibinfo {author} {\bibfnamefont {J.~E.}\ \bibnamefont
  {Lye}}, \bibinfo {author} {\bibfnamefont {L.}~\bibnamefont {Fallani}},
  \bibinfo {author} {\bibfnamefont {C.}~\bibnamefont {Fort}}, \bibinfo {author}
  {\bibfnamefont {V.}~\bibnamefont {Guarrera}}, \bibinfo {author}
  {\bibfnamefont {M.}~\bibnamefont {Modugno}}, \bibinfo {author} {\bibfnamefont
  {D.~S.}\ \bibnamefont {Wiersma}},\ and\ \bibinfo {author} {\bibfnamefont
  {M.}~\bibnamefont {Inguscio}},\ }\bibfield  {title} {\bibinfo {title} {Effect
  of interactions on the localization of a bose-einstein condensate in a
  quasiperiodic lattice},\ }\href {https://doi.org/10.1103/PhysRevA.75.061603}
  {\bibfield  {journal} {\bibinfo  {journal} {Phys. Rev. A}\ }\textbf {\bibinfo
  {volume} {75}},\ \bibinfo {pages} {061603} (\bibinfo {year}
  {2007})}\BibitemShut {NoStop}%
\bibitem [{\citenamefont {Roati}\ \emph {et~al.}(2008)\citenamefont {Roati},
  \citenamefont {D’Errico}, \citenamefont {Fallani}, \citenamefont {Fattori},
  \citenamefont {Fort}, \citenamefont {Zaccanti}, \citenamefont {Modugno},
  \citenamefont {Modugno},\ and\ \citenamefont {Inguscio}}]{Roati_2008}%
  \BibitemOpen
  \bibfield  {author} {\bibinfo {author} {\bibfnamefont {G.}~\bibnamefont
  {Roati}}, \bibinfo {author} {\bibfnamefont {C.}~\bibnamefont {D’Errico}},
  \bibinfo {author} {\bibfnamefont {L.}~\bibnamefont {Fallani}}, \bibinfo
  {author} {\bibfnamefont {M.}~\bibnamefont {Fattori}}, \bibinfo {author}
  {\bibfnamefont {C.}~\bibnamefont {Fort}}, \bibinfo {author} {\bibfnamefont
  {M.}~\bibnamefont {Zaccanti}}, \bibinfo {author} {\bibfnamefont
  {G.}~\bibnamefont {Modugno}}, \bibinfo {author} {\bibfnamefont
  {M.}~\bibnamefont {Modugno}},\ and\ \bibinfo {author} {\bibfnamefont
  {M.}~\bibnamefont {Inguscio}},\ }\bibfield  {title} {\bibinfo {title}
  {Anderson localization of a non-interacting bose--einstein condensate},\
  }\href {https://doi.org/10.1038/nature07071} {\bibfield  {journal} {\bibinfo
  {journal} {Nature}\ }\textbf {\bibinfo {volume} {453}},\ \bibinfo {pages}
  {895} (\bibinfo {year} {2008})}\BibitemShut {NoStop}%
\bibitem [{\citenamefont {D'Errico}\ \emph {et~al.}(2014)\citenamefont
  {D'Errico}, \citenamefont {Lucioni}, \citenamefont {Tanzi}, \citenamefont
  {Gori}, \citenamefont {Roux}, \citenamefont {McCulloch}, \citenamefont
  {Giamarchi}, \citenamefont {Inguscio},\ and\ \citenamefont
  {Modugno}}]{DErrico_2014}%
  \BibitemOpen
  \bibfield  {author} {\bibinfo {author} {\bibfnamefont {C.}~\bibnamefont
  {D'Errico}}, \bibinfo {author} {\bibfnamefont {E.}~\bibnamefont {Lucioni}},
  \bibinfo {author} {\bibfnamefont {L.}~\bibnamefont {Tanzi}}, \bibinfo
  {author} {\bibfnamefont {L.}~\bibnamefont {Gori}}, \bibinfo {author}
  {\bibfnamefont {G.}~\bibnamefont {Roux}}, \bibinfo {author} {\bibfnamefont
  {I.~P.}\ \bibnamefont {McCulloch}}, \bibinfo {author} {\bibfnamefont
  {T.}~\bibnamefont {Giamarchi}}, \bibinfo {author} {\bibfnamefont
  {M.}~\bibnamefont {Inguscio}},\ and\ \bibinfo {author} {\bibfnamefont
  {G.}~\bibnamefont {Modugno}},\ }\bibfield  {title} {\bibinfo {title}
  {Observation of a disordered bosonic insulator from weak to strong
  interactions},\ }\href {https://doi.org/10.1103/PhysRevLett.113.095301}
  {\bibfield  {journal} {\bibinfo  {journal} {Phys. Rev. Lett.}\ }\textbf
  {\bibinfo {volume} {113}},\ \bibinfo {pages} {095301} (\bibinfo {year}
  {2014})}\BibitemShut {NoStop}%
\bibitem [{\citenamefont {Schreiber}\ \emph {et~al.}(2015)\citenamefont
  {Schreiber}, \citenamefont {Hodgman}, \citenamefont {Bordia}, \citenamefont
  {L{\"u}schen}, \citenamefont {Fischer}, \citenamefont {Vosk}, \citenamefont
  {Altman}, \citenamefont {Schneider},\ and\ \citenamefont
  {Bloch}}]{Schreiber_2015}%
  \BibitemOpen
  \bibfield  {author} {\bibinfo {author} {\bibfnamefont {M.}~\bibnamefont
  {Schreiber}}, \bibinfo {author} {\bibfnamefont {S.~S.}\ \bibnamefont
  {Hodgman}}, \bibinfo {author} {\bibfnamefont {P.}~\bibnamefont {Bordia}},
  \bibinfo {author} {\bibfnamefont {H.~P.}\ \bibnamefont {L{\"u}schen}},
  \bibinfo {author} {\bibfnamefont {M.~H.}\ \bibnamefont {Fischer}}, \bibinfo
  {author} {\bibfnamefont {R.}~\bibnamefont {Vosk}}, \bibinfo {author}
  {\bibfnamefont {E.}~\bibnamefont {Altman}}, \bibinfo {author} {\bibfnamefont
  {U.}~\bibnamefont {Schneider}},\ and\ \bibinfo {author} {\bibfnamefont
  {I.}~\bibnamefont {Bloch}},\ }\bibfield  {title} {\bibinfo {title}
  {Observation of many-body localization of interacting fermions in a
  quasirandom optical lattice},\ }\href
  {https://doi.org/10.1126/science.aaa7432} {\bibfield  {journal} {\bibinfo
  {journal} {Science}\ }\textbf {\bibinfo {volume} {349}},\ \bibinfo {pages}
  {842} (\bibinfo {year} {2015})}\BibitemShut {NoStop}%
\bibitem [{\citenamefont {Bordia}\ \emph {et~al.}(2017)\citenamefont {Bordia},
  \citenamefont {L\"uschen}, \citenamefont {Scherg}, \citenamefont
  {Gopalakrishnan}, \citenamefont {Knap}, \citenamefont {Schneider},\ and\
  \citenamefont {Bloch}}]{Bordia_2017}%
  \BibitemOpen
  \bibfield  {author} {\bibinfo {author} {\bibfnamefont {P.}~\bibnamefont
  {Bordia}}, \bibinfo {author} {\bibfnamefont {H.}~\bibnamefont {L\"uschen}},
  \bibinfo {author} {\bibfnamefont {S.}~\bibnamefont {Scherg}}, \bibinfo
  {author} {\bibfnamefont {S.}~\bibnamefont {Gopalakrishnan}}, \bibinfo
  {author} {\bibfnamefont {M.}~\bibnamefont {Knap}}, \bibinfo {author}
  {\bibfnamefont {U.}~\bibnamefont {Schneider}},\ and\ \bibinfo {author}
  {\bibfnamefont {I.}~\bibnamefont {Bloch}},\ }\bibfield  {title} {\bibinfo
  {title} {Probing slow relaxation and many-body localization in
  two-dimensional quasiperiodic systems},\ }\href
  {https://doi.org/10.1103/PhysRevX.7.041047} {\bibfield  {journal} {\bibinfo
  {journal} {Phys. Rev. X}\ }\textbf {\bibinfo {volume} {7}},\ \bibinfo {pages}
  {041047} (\bibinfo {year} {2017})}\BibitemShut {NoStop}%
\bibitem [{\citenamefont {An}\ \emph {et~al.}(2021)\citenamefont {An},
  \citenamefont {Padavi\ifmmode~\acute{c}\else \'{c}\fi{}}, \citenamefont
  {Meier}, \citenamefont {Hegde}, \citenamefont {Ganeshan}, \citenamefont
  {Pixley}, \citenamefont {Vishveshwara},\ and\ \citenamefont
  {Gadway}}]{An_2021}%
  \BibitemOpen
  \bibfield  {author} {\bibinfo {author} {\bibfnamefont {F.~A.}\ \bibnamefont
  {An}}, \bibinfo {author} {\bibfnamefont {K.}~\bibnamefont
  {Padavi\ifmmode~\acute{c}\else \'{c}\fi{}}}, \bibinfo {author} {\bibfnamefont
  {E.~J.}\ \bibnamefont {Meier}}, \bibinfo {author} {\bibfnamefont
  {S.}~\bibnamefont {Hegde}}, \bibinfo {author} {\bibfnamefont
  {S.}~\bibnamefont {Ganeshan}}, \bibinfo {author} {\bibfnamefont {J.~H.}\
  \bibnamefont {Pixley}}, \bibinfo {author} {\bibfnamefont {S.}~\bibnamefont
  {Vishveshwara}},\ and\ \bibinfo {author} {\bibfnamefont {B.}~\bibnamefont
  {Gadway}},\ }\bibfield  {title} {\bibinfo {title} {Interactions and mobility
  edges: Observing the generalized aubry-andr\'e model},\ }\href
  {https://doi.org/10.1103/PhysRevLett.126.040603} {\bibfield  {journal}
  {\bibinfo  {journal} {Phys. Rev. Lett.}\ }\textbf {\bibinfo {volume} {126}},\
  \bibinfo {pages} {040603} (\bibinfo {year} {2021})}\BibitemShut {NoStop}%
\bibitem [{\citenamefont {Rajagopal}\ \emph {et~al.}(2019)\citenamefont
  {Rajagopal}, \citenamefont {Shimasaki}, \citenamefont {Dotti}, \citenamefont
  {Ra\ifmmode \check{c}\else \v{c}\fi{}i\ifmmode~\bar{u}\else \={u}\fi{}nas},
  \citenamefont {Senaratne}, \citenamefont {Anisimovas}, \citenamefont
  {Eckardt},\ and\ \citenamefont {Weld}}]{Rajagopal_2019}%
  \BibitemOpen
  \bibfield  {author} {\bibinfo {author} {\bibfnamefont {S.~V.}\ \bibnamefont
  {Rajagopal}}, \bibinfo {author} {\bibfnamefont {T.}~\bibnamefont
  {Shimasaki}}, \bibinfo {author} {\bibfnamefont {P.}~\bibnamefont {Dotti}},
  \bibinfo {author} {\bibfnamefont {M.}~\bibnamefont {Ra\ifmmode \check{c}\else
  \v{c}\fi{}i\ifmmode~\bar{u}\else \={u}\fi{}nas}}, \bibinfo {author}
  {\bibfnamefont {R.}~\bibnamefont {Senaratne}}, \bibinfo {author}
  {\bibfnamefont {E.}~\bibnamefont {Anisimovas}}, \bibinfo {author}
  {\bibfnamefont {A.}~\bibnamefont {Eckardt}},\ and\ \bibinfo {author}
  {\bibfnamefont {D.~M.}\ \bibnamefont {Weld}},\ }\bibfield  {title} {\bibinfo
  {title} {Phasonic spectroscopy of a quantum gas in a quasicrystalline
  lattice},\ }\href {https://doi.org/10.1103/PhysRevLett.123.223201} {\bibfield
   {journal} {\bibinfo  {journal} {Phys. Rev. Lett.}\ }\textbf {\bibinfo
  {volume} {123}},\ \bibinfo {pages} {223201} (\bibinfo {year}
  {2019})}\BibitemShut {NoStop}%
\bibitem [{\citenamefont {Shimasaki}\ \emph {et~al.}(2024)\citenamefont
  {Shimasaki}, \citenamefont {Prichard}, \citenamefont {Kondakci},
  \citenamefont {Pagett}, \citenamefont {Bai}, \citenamefont {Dotti},
  \citenamefont {Cao}, \citenamefont {Dardia}, \citenamefont {Lu},
  \citenamefont {Grover} \emph {et~al.}}]{Shimasaki_2024}%
  \BibitemOpen
  \bibfield  {author} {\bibinfo {author} {\bibfnamefont {T.}~\bibnamefont
  {Shimasaki}}, \bibinfo {author} {\bibfnamefont {M.}~\bibnamefont {Prichard}},
  \bibinfo {author} {\bibfnamefont {H.~E.}\ \bibnamefont {Kondakci}}, \bibinfo
  {author} {\bibfnamefont {J.~E.}\ \bibnamefont {Pagett}}, \bibinfo {author}
  {\bibfnamefont {Y.}~\bibnamefont {Bai}}, \bibinfo {author} {\bibfnamefont
  {P.}~\bibnamefont {Dotti}}, \bibinfo {author} {\bibfnamefont
  {A.}~\bibnamefont {Cao}}, \bibinfo {author} {\bibfnamefont {A.~R.}\
  \bibnamefont {Dardia}}, \bibinfo {author} {\bibfnamefont {T.-C.}\
  \bibnamefont {Lu}}, \bibinfo {author} {\bibfnamefont {T.}~\bibnamefont
  {Grover}}, \emph {et~al.},\ }\bibfield  {title} {\bibinfo {title} {Anomalous
  localization in a kicked quasicrystal},\ }\href
  {https://doi.org/10.1038/s41567-023-02329-4} {\bibfield  {journal} {\bibinfo
  {journal} {Nat. Phys.}\ }\textbf {\bibinfo {volume} {20}},\ \bibinfo {pages}
  {409} (\bibinfo {year} {2024})}\BibitemShut {NoStop}%
\bibitem [{\citenamefont {Guidoni}\ \emph {et~al.}(1997)\citenamefont
  {Guidoni}, \citenamefont {Trich\'e}, \citenamefont {Verkerk},\ and\
  \citenamefont {Grynberg}}]{Guidoni_1997}%
  \BibitemOpen
  \bibfield  {author} {\bibinfo {author} {\bibfnamefont {L.}~\bibnamefont
  {Guidoni}}, \bibinfo {author} {\bibfnamefont {C.}~\bibnamefont {Trich\'e}},
  \bibinfo {author} {\bibfnamefont {P.}~\bibnamefont {Verkerk}},\ and\ \bibinfo
  {author} {\bibfnamefont {G.}~\bibnamefont {Grynberg}},\ }\bibfield  {title}
  {\bibinfo {title} {Quasiperiodic optical lattices},\ }\href
  {https://doi.org/10.1103/PhysRevLett.79.3363} {\bibfield  {journal} {\bibinfo
   {journal} {Phys. Rev. Lett.}\ }\textbf {\bibinfo {volume} {79}},\ \bibinfo
  {pages} {3363} (\bibinfo {year} {1997})}\BibitemShut {NoStop}%
\bibitem [{\citenamefont {Guidoni}\ \emph {et~al.}(1999)\citenamefont
  {Guidoni}, \citenamefont {D\'epret}, \citenamefont {di~Stefano},\ and\
  \citenamefont {Verkerk}}]{Guidoni_1999}%
  \BibitemOpen
  \bibfield  {author} {\bibinfo {author} {\bibfnamefont {L.}~\bibnamefont
  {Guidoni}}, \bibinfo {author} {\bibfnamefont {B.}~\bibnamefont {D\'epret}},
  \bibinfo {author} {\bibfnamefont {A.}~\bibnamefont {di~Stefano}},\ and\
  \bibinfo {author} {\bibfnamefont {P.}~\bibnamefont {Verkerk}},\ }\bibfield
  {title} {\bibinfo {title} {Atomic diffusion in an optical quasicrystal with
  five-fold symmetry},\ }\href {https://doi.org/10.1103/PhysRevA.60.R4233}
  {\bibfield  {journal} {\bibinfo  {journal} {Phys. Rev. A}\ }\textbf {\bibinfo
  {volume} {60}},\ \bibinfo {pages} {R4233} (\bibinfo {year}
  {1999})}\BibitemShut {NoStop}%
\bibitem [{\citenamefont {Corcovilos}\ and\ \citenamefont
  {Mittal}(2019)}]{Corcovilos_2019}%
  \BibitemOpen
  \bibfield  {author} {\bibinfo {author} {\bibfnamefont {T.~A.}\ \bibnamefont
  {Corcovilos}}\ and\ \bibinfo {author} {\bibfnamefont {J.}~\bibnamefont
  {Mittal}},\ }\bibfield  {title} {\bibinfo {title} {Two-dimensional optical
  quasicrystal potentials for ultracold atom experiments},\ }\href
  {https://doi.org/10.1364/AO.58.002256} {\bibfield  {journal} {\bibinfo
  {journal} {Applied optics}\ }\textbf {\bibinfo {volume} {58}},\ \bibinfo
  {pages} {2256} (\bibinfo {year} {2019})}\BibitemShut {NoStop}%
\bibitem [{\citenamefont {Viebahn}\ \emph {et~al.}(2019)\citenamefont
  {Viebahn}, \citenamefont {Sbroscia}, \citenamefont {Carter}, \citenamefont
  {Yu},\ and\ \citenamefont {Schneider}}]{Viebahn_2019}%
  \BibitemOpen
  \bibfield  {author} {\bibinfo {author} {\bibfnamefont {K.}~\bibnamefont
  {Viebahn}}, \bibinfo {author} {\bibfnamefont {M.}~\bibnamefont {Sbroscia}},
  \bibinfo {author} {\bibfnamefont {E.}~\bibnamefont {Carter}}, \bibinfo
  {author} {\bibfnamefont {J.-C.}\ \bibnamefont {Yu}},\ and\ \bibinfo {author}
  {\bibfnamefont {U.}~\bibnamefont {Schneider}},\ }\bibfield  {title} {\bibinfo
  {title} {Matter-wave diffraction from a quasicrystalline optical lattice},\
  }\href {https://doi.org/10.1103/PhysRevLett.122.110404} {\bibfield  {journal}
  {\bibinfo  {journal} {Phys. Rev. Lett.}\ }\textbf {\bibinfo {volume} {122}},\
  \bibinfo {pages} {110404} (\bibinfo {year} {2019})}\BibitemShut {NoStop}%
\bibitem [{\citenamefont {Sbroscia}\ \emph {et~al.}(2020)\citenamefont
  {Sbroscia}, \citenamefont {Viebahn}, \citenamefont {Carter}, \citenamefont
  {Yu}, \citenamefont {Gaunt},\ and\ \citenamefont
  {Schneider}}]{Sbroscia_2020}%
  \BibitemOpen
  \bibfield  {author} {\bibinfo {author} {\bibfnamefont {M.}~\bibnamefont
  {Sbroscia}}, \bibinfo {author} {\bibfnamefont {K.}~\bibnamefont {Viebahn}},
  \bibinfo {author} {\bibfnamefont {E.}~\bibnamefont {Carter}}, \bibinfo
  {author} {\bibfnamefont {J.-C.}\ \bibnamefont {Yu}}, \bibinfo {author}
  {\bibfnamefont {A.}~\bibnamefont {Gaunt}},\ and\ \bibinfo {author}
  {\bibfnamefont {U.}~\bibnamefont {Schneider}},\ }\bibfield  {title} {\bibinfo
  {title} {Observing localization in a 2d quasicrystalline optical lattice},\
  }\href {https://doi.org/10.1103/PhysRevLett.125.200604} {\bibfield  {journal}
  {\bibinfo  {journal} {Phys. Rev. Lett.}\ }\textbf {\bibinfo {volume} {125}},\
  \bibinfo {pages} {200604} (\bibinfo {year} {2020})}\BibitemShut {NoStop}%
\bibitem [{\citenamefont {Yu}\ \emph {et~al.}(2024)\citenamefont {Yu},
  \citenamefont {Bhave}, \citenamefont {Reeve}, \citenamefont {Song},\ and\
  \citenamefont {Schneider}}]{Yu_2024}%
  \BibitemOpen
  \bibfield  {author} {\bibinfo {author} {\bibfnamefont {J.-C.}\ \bibnamefont
  {Yu}}, \bibinfo {author} {\bibfnamefont {S.}~\bibnamefont {Bhave}}, \bibinfo
  {author} {\bibfnamefont {L.}~\bibnamefont {Reeve}}, \bibinfo {author}
  {\bibfnamefont {B.}~\bibnamefont {Song}},\ and\ \bibinfo {author}
  {\bibfnamefont {U.}~\bibnamefont {Schneider}},\ }\bibfield  {title} {\bibinfo
  {title} {Observing the two-dimensional bose glass in an optical
  quasicrystal},\ }\href {https://doi.org/10.1038/s41586-024-07875-2}
  {\bibfield  {journal} {\bibinfo  {journal} {Nature}\ }\textbf {\bibinfo
  {volume} {633}},\ \bibinfo {pages} {338} (\bibinfo {year}
  {2024})}\BibitemShut {NoStop}%
\bibitem [{\citenamefont {Li}\ \emph {et~al.}(2019)\citenamefont {Li},
  \citenamefont {Chen},\ and\ \citenamefont {Fisher}}]{Li_2019}%
  \BibitemOpen
  \bibfield  {author} {\bibinfo {author} {\bibfnamefont {Y.}~\bibnamefont
  {Li}}, \bibinfo {author} {\bibfnamefont {X.}~\bibnamefont {Chen}},\ and\
  \bibinfo {author} {\bibfnamefont {M.~P.~A.}\ \bibnamefont {Fisher}},\
  }\bibfield  {title} {\bibinfo {title} {Measurement-driven entanglement
  transition in hybrid quantum circuits},\ }\href
  {https://doi.org/10.1103/PhysRevB.100.134306} {\bibfield  {journal} {\bibinfo
   {journal} {Phys. Rev. B}\ }\textbf {\bibinfo {volume} {100}},\ \bibinfo
  {pages} {134306} (\bibinfo {year} {2019})}\BibitemShut {NoStop}%
\bibitem [{\citenamefont {Lu}\ and\ \citenamefont {Grover}(2021)}]{Lu_2021}%
  \BibitemOpen
  \bibfield  {author} {\bibinfo {author} {\bibfnamefont {T.-C.}\ \bibnamefont
  {Lu}}\ and\ \bibinfo {author} {\bibfnamefont {T.}~\bibnamefont {Grover}},\
  }\bibfield  {title} {\bibinfo {title} {Spacetime duality between localization
  transitions and measurement-induced transitions},\ }\href
  {https://doi.org/10.1103/PRXQuantum.2.040319} {\bibfield  {journal} {\bibinfo
   {journal} {PRX Quantum}\ }\textbf {\bibinfo {volume} {2}},\ \bibinfo {pages}
  {040319} (\bibinfo {year} {2021})}\BibitemShut {NoStop}%
\bibitem [{\citenamefont {Shkolnik}\ \emph {et~al.}(2023)\citenamefont
  {Shkolnik}, \citenamefont {Zabalo}, \citenamefont {Vasseur}, \citenamefont
  {Huse}, \citenamefont {Pixley},\ and\ \citenamefont {Gazit}}]{Shkolnik_2023}%
  \BibitemOpen
  \bibfield  {author} {\bibinfo {author} {\bibfnamefont {G.}~\bibnamefont
  {Shkolnik}}, \bibinfo {author} {\bibfnamefont {A.}~\bibnamefont {Zabalo}},
  \bibinfo {author} {\bibfnamefont {R.}~\bibnamefont {Vasseur}}, \bibinfo
  {author} {\bibfnamefont {D.~A.}\ \bibnamefont {Huse}}, \bibinfo {author}
  {\bibfnamefont {J.}~\bibnamefont {Pixley}},\ and\ \bibinfo {author}
  {\bibfnamefont {S.}~\bibnamefont {Gazit}},\ }\bibfield  {title} {\bibinfo
  {title} {Measurement induced criticality in quasiperiodic modulated random
  hybrid circuits},\ }\href {https://doi.org/10.1103/PhysRevB.108.184204}
  {\bibfield  {journal} {\bibinfo  {journal} {Phys. Rev. B}\ }\textbf {\bibinfo
  {volume} {108}},\ \bibinfo {pages} {184204} (\bibinfo {year}
  {2023})}\BibitemShut {NoStop}%
\bibitem [{\citenamefont {Zabalo}\ \emph {et~al.}(2023)\citenamefont {Zabalo},
  \citenamefont {Wilson}, \citenamefont {Gullans}, \citenamefont {Vasseur},
  \citenamefont {Gopalakrishnan}, \citenamefont {Huse},\ and\ \citenamefont
  {Pixley}}]{Zabalo_2023}%
  \BibitemOpen
  \bibfield  {author} {\bibinfo {author} {\bibfnamefont {A.}~\bibnamefont
  {Zabalo}}, \bibinfo {author} {\bibfnamefont {J.~H.}\ \bibnamefont {Wilson}},
  \bibinfo {author} {\bibfnamefont {M.~J.}\ \bibnamefont {Gullans}}, \bibinfo
  {author} {\bibfnamefont {R.}~\bibnamefont {Vasseur}}, \bibinfo {author}
  {\bibfnamefont {S.}~\bibnamefont {Gopalakrishnan}}, \bibinfo {author}
  {\bibfnamefont {D.~A.}\ \bibnamefont {Huse}},\ and\ \bibinfo {author}
  {\bibfnamefont {J.}~\bibnamefont {Pixley}},\ }\bibfield  {title} {\bibinfo
  {title} {Infinite-randomness criticality in monitored quantum dynamics with
  static disorder},\ }\href {https://doi.org/10.1103/PhysRevB.107.L220204}
  {\bibfield  {journal} {\bibinfo  {journal} {Phys. Rev. B}\ }\textbf {\bibinfo
  {volume} {107}},\ \bibinfo {pages} {L220204} (\bibinfo {year}
  {2023})}\BibitemShut {NoStop}%
\bibitem [{\citenamefont {Matsubara}\ \emph {et~al.}(2025)\citenamefont
  {Matsubara}, \citenamefont {Yamamoto},\ and\ \citenamefont
  {Koga}}]{Matsubara_AAH}%
  \BibitemOpen
  \bibfield  {author} {\bibinfo {author} {\bibfnamefont {T.}~\bibnamefont
  {Matsubara}}, \bibinfo {author} {\bibfnamefont {K.}~\bibnamefont
  {Yamamoto}},\ and\ \bibinfo {author} {\bibfnamefont {A.}~\bibnamefont
  {Koga}},\ }\bibfield  {title} {\bibinfo {title} {Measurement-induced phase
  transitions for free fermions in a quasiperiodic potential},\ }\href
  {https://doi.org/10.1103/3zfd-3hqt} {\bibfield  {journal} {\bibinfo
  {journal} {Phys. Rev. B}\ }\textbf {\bibinfo {volume} {112}},\ \bibinfo
  {pages} {054309} (\bibinfo {year} {2025})}\BibitemShut {NoStop}%
\bibitem [{\citenamefont {Zhao}\ \emph {et~al.}(2026)\citenamefont {Zhao},
  \citenamefont {Huang}, \citenamefont {Zhang}, \citenamefont {Li},\ and\
  \citenamefont {Zhong}}]{Zhao_2026}%
  \BibitemOpen
  \bibfield  {author} {\bibinfo {author} {\bibfnamefont {Y.-J.}\ \bibnamefont
  {Zhao}}, \bibinfo {author} {\bibfnamefont {X.}~\bibnamefont {Huang}},
  \bibinfo {author} {\bibfnamefont {Y.-R.}\ \bibnamefont {Zhang}}, \bibinfo
  {author} {\bibfnamefont {H.-Z.}\ \bibnamefont {Li}},\ and\ \bibinfo {author}
  {\bibfnamefont {J.-X.}\ \bibnamefont {Zhong}},\ }\bibfield  {title} {\bibinfo
  {title} {Entanglement phases and phase transitions in monitored free fermion
  systems of localization},\ }\href {https://doi.org/10.1103/q1d9-943y}
  {\bibfield  {journal} {\bibinfo  {journal} {Phys. Rev. B}\ }\textbf {\bibinfo
  {volume} {113}},\ \bibinfo {pages} {064301} (\bibinfo {year}
  {2026})}\BibitemShut {NoStop}%
\bibitem [{\citenamefont {Singha}\ \emph {et~al.}(2026)\citenamefont {Singha},
  \citenamefont {Roy}, \citenamefont {Szyniszewski},\ and\ \citenamefont
  {Sharma}}]{Singha_2026}%
  \BibitemOpen
  \bibfield  {author} {\bibinfo {author} {\bibfnamefont {P.}~\bibnamefont
  {Singha}}, \bibinfo {author} {\bibfnamefont {N.}~\bibnamefont {Roy}},
  \bibinfo {author} {\bibfnamefont {M.}~\bibnamefont {Szyniszewski}},\ and\
  \bibinfo {author} {\bibfnamefont {A.}~\bibnamefont {Sharma}},\ }\bibfield
  {title} {\bibinfo {title} {Controlled zeno-induced localization of free
  fermions in a quasiperiodic chain},\ }\href
  {https://doi.org/10.1103/27g7-tvqh} {\bibfield  {journal} {\bibinfo
  {journal} {Phys. Rev. B}\ }\textbf {\bibinfo {volume} {113}},\ \bibinfo
  {pages} {235122} (\bibinfo {year} {2026})}\BibitemShut {NoStop}%
\bibitem [{\citenamefont {Yin}\ \emph {et~al.}(2026)\citenamefont {Yin},
  \citenamefont {Bo},\ and\ \citenamefont
  {Garc{\'\i}a-Garc{\'\i}a}}]{Yin_2026}%
  \BibitemOpen
  \bibfield  {author} {\bibinfo {author} {\bibfnamefont {C.}~\bibnamefont
  {Yin}}, \bibinfo {author} {\bibfnamefont {F.}~\bibnamefont {Bo}},\ and\
  \bibinfo {author} {\bibfnamefont {A.~M.}\ \bibnamefont
  {Garc{\'\i}a-Garc{\'\i}a}},\ }\bibfield  {title} {\bibinfo {title} {No
  measurement induced phase transition in the entanglement dynamics of
  monitored non-interacting one-dimensional fermions in a disordered or
  quasiperiodic potential},\ }\bibfield  {journal} {\bibinfo  {journal} {arXiv
  preprint arXiv:2605.10758}\ }\href
  {https://doi.org/10.48550/arXiv.2605.10758} {10.48550/arXiv.2605.10758}
  (\bibinfo {year} {2026})\BibitemShut {NoStop}%
\bibitem [{\citenamefont {Jagannathan}(2021)}]{Jagannathan_Rev}%
  \BibitemOpen
  \bibfield  {author} {\bibinfo {author} {\bibfnamefont {A.}~\bibnamefont
  {Jagannathan}},\ }\bibfield  {title} {\bibinfo {title} {The fibonacci
  quasicrystal: Case study of hidden dimensions and multifractality},\ }\href
  {https://doi.org/10.1103/RevModPhys.93.045001} {\bibfield  {journal}
  {\bibinfo  {journal} {Rev. Mod. Phys.}\ }\textbf {\bibinfo {volume} {93}},\
  \bibinfo {pages} {045001} (\bibinfo {year} {2021})}\BibitemShut {NoStop}%
\bibitem [{\citenamefont {Daley}(2014)}]{Daley_2014}%
  \BibitemOpen
  \bibfield  {author} {\bibinfo {author} {\bibfnamefont {A.~J.}\ \bibnamefont
  {Daley}},\ }\bibfield  {title} {\bibinfo {title} {Quantum trajectories and
  open many-body quantum systems},\ }\href
  {https://doi.org/10.1080/00018732.2014.933502} {\bibfield  {journal}
  {\bibinfo  {journal} {Advances in Physics}\ }\textbf {\bibinfo {volume}
  {63}},\ \bibinfo {pages} {77} (\bibinfo {year} {2014})}\BibitemShut {NoStop}%
\bibitem [{\citenamefont {Dalibard}\ \emph {et~al.}(1992)\citenamefont
  {Dalibard}, \citenamefont {Castin},\ and\ \citenamefont
  {M{\o}lmer}}]{Dalibard_1992}%
  \BibitemOpen
  \bibfield  {author} {\bibinfo {author} {\bibfnamefont {J.}~\bibnamefont
  {Dalibard}}, \bibinfo {author} {\bibfnamefont {Y.}~\bibnamefont {Castin}},\
  and\ \bibinfo {author} {\bibfnamefont {K.}~\bibnamefont {M{\o}lmer}},\
  }\bibfield  {title} {\bibinfo {title} {Wave-function approach to dissipative
  processes in quantum optics},\ }\href
  {https://doi.org/10.1103/PhysRevLett.68.580} {\bibfield  {journal} {\bibinfo
  {journal} {Phys. Rev. Lett.}\ }\textbf {\bibinfo {volume} {68}},\ \bibinfo
  {pages} {580} (\bibinfo {year} {1992})}\BibitemShut {NoStop}%
\bibitem [{\citenamefont {Wiseman}\ and\ \citenamefont
  {Milburn}(1993)}]{Wiseman_1993}%
  \BibitemOpen
  \bibfield  {author} {\bibinfo {author} {\bibfnamefont {H.}~\bibnamefont
  {Wiseman}}\ and\ \bibinfo {author} {\bibfnamefont {G.}~\bibnamefont
  {Milburn}},\ }\bibfield  {title} {\bibinfo {title} {Interpretation of quantum
  jump and diffusion processes illustrated on the bloch sphere},\ }\href
  {https://doi.org/10.1103/PhysRevA.47.1652} {\bibfield  {journal} {\bibinfo
  {journal} {Phys. Rev. A}\ }\textbf {\bibinfo {volume} {47}},\ \bibinfo
  {pages} {1652} (\bibinfo {year} {1993})}\BibitemShut {NoStop}%
\bibitem [{\citenamefont {Jagannathan}\ \emph {et~al.}(2007)\citenamefont
  {Jagannathan}, \citenamefont {Szallas}, \citenamefont {Wessel},\ and\
  \citenamefont {Duneau}}]{Jagannathan_2004}%
  \BibitemOpen
  \bibfield  {author} {\bibinfo {author} {\bibfnamefont {A.}~\bibnamefont
  {Jagannathan}}, \bibinfo {author} {\bibfnamefont {A.}~\bibnamefont
  {Szallas}}, \bibinfo {author} {\bibfnamefont {S.}~\bibnamefont {Wessel}},\
  and\ \bibinfo {author} {\bibfnamefont {M.}~\bibnamefont {Duneau}},\
  }\bibfield  {title} {\bibinfo {title} {Penrose quantum antiferromagnet},\
  }\href {https://doi.org/10.1103/PhysRevB.75.212407} {\bibfield  {journal}
  {\bibinfo  {journal} {Phys. Rev. B}\ }\textbf {\bibinfo {volume} {75}},\
  \bibinfo {pages} {212407} (\bibinfo {year} {2007})}\BibitemShut {NoStop}%
\bibitem [{\citenamefont {Szallas}\ and\ \citenamefont
  {Jagannathan}(2008)}]{Attila_2008}%
  \BibitemOpen
  \bibfield  {author} {\bibinfo {author} {\bibfnamefont {A.}~\bibnamefont
  {Szallas}}\ and\ \bibinfo {author} {\bibfnamefont {A.}~\bibnamefont
  {Jagannathan}},\ }\bibfield  {title} {\bibinfo {title} {Spin waves and local
  magnetizations on the penrose tiling},\ }\href
  {https://doi.org/10.1103/PhysRevB.77.104427} {\bibfield  {journal} {\bibinfo
  {journal} {Phys. Rev. B}\ }\textbf {\bibinfo {volume} {77}},\ \bibinfo
  {pages} {104427} (\bibinfo {year} {2008})}\BibitemShut {NoStop}%
\bibitem [{\citenamefont {Koga}\ and\ \citenamefont
  {Tsunetsugu}(2017)}]{Koga_2017}%
  \BibitemOpen
  \bibfield  {author} {\bibinfo {author} {\bibfnamefont {A.}~\bibnamefont
  {Koga}}\ and\ \bibinfo {author} {\bibfnamefont {H.}~\bibnamefont
  {Tsunetsugu}},\ }\bibfield  {title} {\bibinfo {title} {Antiferromagnetic
  order in the hubbard model on the penrose lattice},\ }\href
  {https://doi.org/10.1103/PhysRevB.96.214402} {\bibfield  {journal} {\bibinfo
  {journal} {Phys. Rev. B}\ }\textbf {\bibinfo {volume} {96}},\ \bibinfo
  {pages} {214402} (\bibinfo {year} {2017})}\BibitemShut {NoStop}%
\bibitem [{\citenamefont {Koga}(2020)}]{Koga_2020}%
  \BibitemOpen
  \bibfield  {author} {\bibinfo {author} {\bibfnamefont {A.}~\bibnamefont
  {Koga}},\ }\bibfield  {title} {\bibinfo {title} {Superlattice structure in
  the antiferromagnetically ordered state in the hubbard model on the
  ammann-beenker tiling},\ }\href {https://doi.org/10.1103/PhysRevB.102.115125}
  {\bibfield  {journal} {\bibinfo  {journal} {Phys. Rev. B}\ }\textbf {\bibinfo
  {volume} {102}},\ \bibinfo {pages} {115125} (\bibinfo {year}
  {2020})}\BibitemShut {NoStop}%
\bibitem [{\citenamefont {Matsubara}\ \emph {et~al.}(2024)\citenamefont
  {Matsubara}, \citenamefont {Koga},\ and\ \citenamefont
  {Coates}}]{Matsubara_3GMT}%
  \BibitemOpen
  \bibfield  {author} {\bibinfo {author} {\bibfnamefont {T.}~\bibnamefont
  {Matsubara}}, \bibinfo {author} {\bibfnamefont {A.}~\bibnamefont {Koga}},\
  and\ \bibinfo {author} {\bibfnamefont {S.}~\bibnamefont {Coates}},\
  }\bibfield  {title} {\bibinfo {title} {Ferromagnetically ordered states in
  the hubbard model on the ${H}_{00}$ hexagonal golden-mean tiling},\ }\href
  {https://doi.org/10.1103/PhysRevB.109.014413} {\bibfield  {journal} {\bibinfo
   {journal} {Phys. Rev. B}\ }\textbf {\bibinfo {volume} {109}},\ \bibinfo
  {pages} {014413} (\bibinfo {year} {2024})}\BibitemShut {NoStop}%
\bibitem [{\citenamefont {Koga}\ and\ \citenamefont
  {Matsubara}(2025)}]{Koga_honey}%
  \BibitemOpen
  \bibfield  {author} {\bibinfo {author} {\bibfnamefont {A.}~\bibnamefont
  {Koga}}\ and\ \bibinfo {author} {\bibfnamefont {T.}~\bibnamefont
  {Matsubara}},\ }\bibfield  {title} {\bibinfo {title} {Quasiperiodically
  modulated honeycomb lattices and their magnetic properties},\ }\href
  {https://doi.org/10.1103/7hsy-sct6} {\bibfield  {journal} {\bibinfo
  {journal} {Phys. Rev. B}\ }\textbf {\bibinfo {volume} {111}},\ \bibinfo
  {pages} {214422} (\bibinfo {year} {2025})}\BibitemShut {NoStop}%
\bibitem [{\citenamefont {Takemori}\ \emph {et~al.}(2020)\citenamefont
  {Takemori}, \citenamefont {Arita},\ and\ \citenamefont
  {Sakai}}]{Takemori_2020}%
  \BibitemOpen
  \bibfield  {author} {\bibinfo {author} {\bibfnamefont {N.}~\bibnamefont
  {Takemori}}, \bibinfo {author} {\bibfnamefont {R.}~\bibnamefont {Arita}},\
  and\ \bibinfo {author} {\bibfnamefont {S.}~\bibnamefont {Sakai}},\ }\bibfield
   {title} {\bibinfo {title} {Physical properties of weak-coupling
  quasiperiodic superconductors},\ }\href
  {https://doi.org/10.1103/PhysRevB.102.115108} {\bibfield  {journal} {\bibinfo
   {journal} {Phys. Rev. B}\ }\textbf {\bibinfo {volume} {102}},\ \bibinfo
  {pages} {115108} (\bibinfo {year} {2020})}\BibitemShut {NoStop}%
\bibitem [{\citenamefont {Sakai}\ \emph {et~al.}(2022)\citenamefont {Sakai},
  \citenamefont {Arita},\ and\ \citenamefont {Ohtsuki}}]{Sakai_2022}%
  \BibitemOpen
  \bibfield  {author} {\bibinfo {author} {\bibfnamefont {S.}~\bibnamefont
  {Sakai}}, \bibinfo {author} {\bibfnamefont {R.}~\bibnamefont {Arita}},\ and\
  \bibinfo {author} {\bibfnamefont {T.}~\bibnamefont {Ohtsuki}},\ }\bibfield
  {title} {\bibinfo {title} {Hyperuniform electron distributions controlled by
  electron interactions in quasicrystals},\ }\href
  {https://doi.org/10.1103/PhysRevB.105.054202} {\bibfield  {journal} {\bibinfo
   {journal} {Phys. Rev. B}\ }\textbf {\bibinfo {volume} {105}},\ \bibinfo
  {pages} {054202} (\bibinfo {year} {2022})}\BibitemShut {NoStop}%
\bibitem [{\citenamefont {Mirzhalilov}\ and\ \citenamefont
  {Oktel}(2020)}]{Murod_2020}%
  \BibitemOpen
  \bibfield  {author} {\bibinfo {author} {\bibfnamefont {M.}~\bibnamefont
  {Mirzhalilov}}\ and\ \bibinfo {author} {\bibfnamefont {M.~O.}\ \bibnamefont
  {Oktel}},\ }\bibfield  {title} {\bibinfo {title} {Perpendicular space
  accounting of localized states in a quasicrystal},\ }\href
  {https://doi.org/10.1103/PhysRevB.102.064213} {\bibfield  {journal} {\bibinfo
   {journal} {Phys. Rev. B}\ }\textbf {\bibinfo {volume} {102}},\ \bibinfo
  {pages} {064213} (\bibinfo {year} {2020})}\BibitemShut {NoStop}%
\bibitem [{\citenamefont {Oktel}(2022)}]{Oktel_2022}%
  \BibitemOpen
  \bibfield  {author} {\bibinfo {author} {\bibfnamefont {M.~O.}\ \bibnamefont
  {Oktel}},\ }\bibfield  {title} {\bibinfo {title} {Localized states in local
  isomorphism classes of pentagonal quasicrystals},\ }\href
  {https://doi.org/10.1103/PhysRevB.106.024201} {\bibfield  {journal} {\bibinfo
   {journal} {Phys. Rev. B}\ }\textbf {\bibinfo {volume} {106}},\ \bibinfo
  {pages} {024201} (\bibinfo {year} {2022})}\BibitemShut {NoStop}%
\bibitem [{\citenamefont {Keskiner}\ and\ \citenamefont
  {Oktel}(2022)}]{Akif_2022}%
  \BibitemOpen
  \bibfield  {author} {\bibinfo {author} {\bibfnamefont {M.~A.}\ \bibnamefont
  {Keskiner}}\ and\ \bibinfo {author} {\bibfnamefont {M.~O.}\ \bibnamefont
  {Oktel}},\ }\bibfield  {title} {\bibinfo {title} {Strictly localized states
  on the socolar dodecagonal lattice},\ }\href
  {https://doi.org/10.1103/PhysRevB.106.064207} {\bibfield  {journal} {\bibinfo
   {journal} {Phys. Rev. B}\ }\textbf {\bibinfo {volume} {106}},\ \bibinfo
  {pages} {064207} (\bibinfo {year} {2022})}\BibitemShut {NoStop}%
\bibitem [{\citenamefont {Hori}\ \emph
  {et~al.}(2024{\natexlab{a}})\citenamefont {Hori}, \citenamefont {Sugimoto},
  \citenamefont {Tohyama},\ and\ \citenamefont {Tanaka}}]{Hori_TSC}%
  \BibitemOpen
  \bibfield  {author} {\bibinfo {author} {\bibfnamefont {M.}~\bibnamefont
  {Hori}}, \bibinfo {author} {\bibfnamefont {T.}~\bibnamefont {Sugimoto}},
  \bibinfo {author} {\bibfnamefont {T.}~\bibnamefont {Tohyama}},\ and\ \bibinfo
  {author} {\bibfnamefont {K.}~\bibnamefont {Tanaka}},\ }\bibfield  {title}
  {\bibinfo {title} {Self-consistent study of topological superconductivity in
  two-dimensional quasicrystals},\ }\href
  {https://doi.org/10.1103/PhysRevB.110.144512} {\bibfield  {journal} {\bibinfo
   {journal} {Phys. Rev. B}\ }\textbf {\bibinfo {volume} {110}},\ \bibinfo
  {pages} {144512} (\bibinfo {year} {2024}{\natexlab{a}})}\BibitemShut
  {NoStop}%
\bibitem [{\citenamefont {Hori}\ \emph
  {et~al.}(2024{\natexlab{b}})\citenamefont {Hori}, \citenamefont {Sugimoto},
  \citenamefont {Hashizume},\ and\ \citenamefont {Tohyama}}]{Hori_Bose}%
  \BibitemOpen
  \bibfield  {author} {\bibinfo {author} {\bibfnamefont {M.}~\bibnamefont
  {Hori}}, \bibinfo {author} {\bibfnamefont {T.}~\bibnamefont {Sugimoto}},
  \bibinfo {author} {\bibfnamefont {Y.}~\bibnamefont {Hashizume}},\ and\
  \bibinfo {author} {\bibfnamefont {T.}~\bibnamefont {Tohyama}},\ }\bibfield
  {title} {\bibinfo {title} {Multifractality and hyperuniformity in
  quasicrystalline bose–hubbard models with and without disorder},\ }\href
  {https://doi.org/10.7566/JPSJ.93.114005} {\bibfield  {journal} {\bibinfo
  {journal} {Journal of the Physical Society of Japan}\ }\textbf {\bibinfo
  {volume} {93}},\ \bibinfo {pages} {114005} (\bibinfo {year}
  {2024}{\natexlab{b}})}\BibitemShut {NoStop}%
\bibitem [{\citenamefont {Yamamoto}\ and\ \citenamefont
  {Inoue}(2024)}]{Yamamoto_2024}%
  \BibitemOpen
  \bibfield  {author} {\bibinfo {author} {\bibfnamefont {S.}~\bibnamefont
  {Yamamoto}}\ and\ \bibinfo {author} {\bibfnamefont {T.}~\bibnamefont
  {Inoue}},\ }\bibfield  {title} {\bibinfo {title} {Magnon confinement on the
  two-dimensional penrose lattice: Perpendicular-space analysis of the dynamic
  structure factor},\ }\href {https://doi.org/10.3390/cryst14080702} {\bibfield
   {journal} {\bibinfo  {journal} {Crystals}\ }\textbf {\bibinfo {volume}
  {14}},\ \bibinfo {pages} {702} (\bibinfo {year} {2024})}\BibitemShut
  {NoStop}%
\bibitem [{\citenamefont {Sakai}\ \emph {et~al.}(2017)\citenamefont {Sakai},
  \citenamefont {Takemori}, \citenamefont {Koga},\ and\ \citenamefont
  {Arita}}]{Sakai_2017}%
  \BibitemOpen
  \bibfield  {author} {\bibinfo {author} {\bibfnamefont {S.}~\bibnamefont
  {Sakai}}, \bibinfo {author} {\bibfnamefont {N.}~\bibnamefont {Takemori}},
  \bibinfo {author} {\bibfnamefont {A.}~\bibnamefont {Koga}},\ and\ \bibinfo
  {author} {\bibfnamefont {R.}~\bibnamefont {Arita}},\ }\bibfield  {title}
  {\bibinfo {title} {Superconductivity on a quasiperiodic lattice:
  Extended-to-localized crossover of cooper pairs},\ }\href
  {https://doi.org/10.1103/PhysRevB.95.024509} {\bibfield  {journal} {\bibinfo
  {journal} {Phys. Rev. B}\ }\textbf {\bibinfo {volume} {95}},\ \bibinfo
  {pages} {024509} (\bibinfo {year} {2017})}\BibitemShut {NoStop}%
\bibitem [{\citenamefont {Matsubara}\ and\ \citenamefont
  {Yamamoto}(2026)}]{zenodo202607}%
  \BibitemOpen
  \bibfield  {author} {\bibinfo {author} {\bibfnamefont {T.}~\bibnamefont
  {Matsubara}}\ and\ \bibinfo {author} {\bibfnamefont {K.}~\bibnamefont
  {Yamamoto}},\ }\bibfield  {title} {\bibinfo {title} {Dataset for
  "measurement-induced spatially nonuniform fluctuations of the local particle
  number and their crossover in a quasiperiodic free-fermion chain"},\ }\href
  {https://doi.org/10.5281/zenodo.21188162} {10.5281/zenodo.21188162} (\bibinfo
  {year} {2026})\BibitemShut {NoStop}%
\bibitem [{\citenamefont {Goldman}\ and\ \citenamefont
  {Kelton}(1993)}]{Goldman_1993}%
  \BibitemOpen
  \bibfield  {author} {\bibinfo {author} {\bibfnamefont {A.~I.}\ \bibnamefont
  {Goldman}}\ and\ \bibinfo {author} {\bibfnamefont {R.~F.}\ \bibnamefont
  {Kelton}},\ }\bibfield  {title} {\bibinfo {title} {Quasicrystals and
  crystalline approximants},\ }\href
  {https://doi.org/10.1103/RevModPhys.65.213} {\bibfield  {journal} {\bibinfo
  {journal} {Rev. Mod. Phys.}\ }\textbf {\bibinfo {volume} {65}},\ \bibinfo
  {pages} {213} (\bibinfo {year} {1993})}\BibitemShut {NoStop}%
\bibitem [{\citenamefont {Godreche}\ and\ \citenamefont
  {Luck}(1990)}]{Godreche_1990}%
  \BibitemOpen
  \bibfield  {author} {\bibinfo {author} {\bibfnamefont {C.}~\bibnamefont
  {Godreche}}\ and\ \bibinfo {author} {\bibfnamefont {J.~M.}\ \bibnamefont
  {Luck}},\ }\bibfield  {title} {\bibinfo {title} {Multifractal analysis in
  reciprocal space and the nature of the fourier transform of self-similar
  structures},\ }\href {https://doi.org/10.1088/0305-4470/23/16/024} {\bibfield
   {journal} {\bibinfo  {journal} {Journal of Physics A: Mathematical and
  General}\ }\textbf {\bibinfo {volume} {23}},\ \bibinfo {pages} {3769}
  (\bibinfo {year} {1990})}\BibitemShut {NoStop}%
\end{thebibliography}%

\end{document}